\documentclass{ptephy_v1}

\preprintnumber{LFTC-20-5/57} 

\usepackage{amsmath}  
\usepackage{amsfonts}
\usepackage{amssymb}
\usepackage{setspace}
\usepackage{graphicx} 
\usepackage{xspace}
\usepackage{cancel}
\usepackage{url}

\usepackage{color}
\usepackage{ulem}
\usepackage{cancel}

\def\K0bar{\overline{K^0}}

\def\qbar{{\bar{q}}}

\def\d0bar{{\bar{D}^0}}

\def\qbar{\bar{q}}
\def\Qbar{\bar{Q}}

\def\be{\begin{equation}}
\def\ee{\end{equation}}
\def\bea{\begin{eqnarray}}
\def\eea{\end{eqnarray}}
\def\nn{\nonumber}

\begin{document}

\title{
Magnetic moments of the octet, decuplet, low-lying charm, and low-lying bottom baryons 
in a nuclear medium
}

\author{K.~Tsushima}

\affil{Laborat\'orio de F\'{\i}sica Te\'orica e Computacional-LFTC, 
Universidade Cidade de S\~ao Paulo  
and 
Universidade Cruzeiro do Sul, 
01506-000, S\~ao Paulo, SP, Brazil
\email{kazuo.tsushima@gmail.com, kazuo.tsushima@cruzeirodosul.edu.br}}


\begin{abstract}
We study the magnetic moments of the octet, decuplet, low-lying charm, and 
low-lying bottom baryons with nonzero light quarks in symmetric nuclear matter     
using the quark-meson coupling (QMC) model, which satisfies the constraint  
for the allowed maximum change (swelling) of the in-medium nucleon size derived from 
the $y$-scaling data for $^3$He$(e,e')$ and $^{56}$Fe$(e,e')$.  
This is the first study to estimate the in-medium      
magnetic moments of the low-lying charm and bottom baryons with nonzero light quarks.
The present QMC model also satisfies the expected allowed maximum enhancement  
of the nucleon magnetic moments in nuclear matter. 
Moreover, it has been proven that the calculated in-medium to free 
proton electromagnetic form factor (EMFF) ratios calculated within the QMC model,  
reproduce well the proton EMFF super ratio extracted from $^4{\rm He}(\vec{e},e'\vec{p})^3{\rm H}$ 
at Jefferson Laboratory (JLab).
The medium modifications of the magnetic moments are estimated by evaluating  
the in-medium to free space baryon magnetic moment ratios to compensate the MIT bag deficiency to 
describe the free space octet baryon magnetic moments, where ratios are often 
measured directly in experiments even without knowing the absolute values, such as the free and 
bound proton electromagnetic form factors, as well as the European Muon Collaboration (EMC) effect 
to extract the structure function $F_2$ ratio of the bound to free nucleons by the corresponding 
cross section ratio. We also present the results calculated with the different  
current quark mass values for the strange and bottom quarks to see 
the possible impact. 
Furthermore, for a practical use, we give the explicit density dependent 
parametrizations for the vector potentials of the baryons and light-$(u, d)$ quarks, 
as well as for the effective masses of the baryons treated in this study, and of the mesons,   
$\omega,\rho,K,K^*,\eta,\eta',D,D^*,B$, and $B^*$.
\end{abstract}


\maketitle

\section{Introduction}

To study the properties of hadrons containing both the heavy quarks (charm ($c$) and bottom ($b$)) 
and the isoboublet light quarks (up ($u$) and down ($d$)) are particularly interesting, 
since the heavy quarks can be regarded as static color/gluon sources, while the isodoublet 
light quarks surrounding them can be regarded as interacting strongly with other hadrons.
(Hereafter, we will simply denote the isodoublet light quarks $u$ and $d$ as light quarks, 
but not the strange quark $s$, otherwise stated.)
This gives an alternative picture for the structure of heavy hadrons additional to 
that of the hadrons composed of purely light quarks. In heavy hadrons with light quarks, 
the light quarks contribute to their masses by dynamical symmetry breaking. 
Thus, when the heavy hadrons with nonzero light quarks  
are produced in nuclei, e.g., in the future antiProton ANnihilations at DArmstadt 
(PANDA) experiment, we can advance in our understanding of hadron structure~\cite{PANDA:2021ozp}.
The physics of PANDA aims at explicitly to produce the heavy hadrons in nuclei among the other 
physics issues. Of course, to study the in-medium properties of hadrons with strange ($s$) quarks,  
such as (strange) hyperons are also very interesting and important, in connection with the 
strange-hypernuclei, and the ''hyperon puzzle'' in neutron star structure. 

In astrophysical laboratories such as neutron stars and compact stars, as well as in dense nuclear 
medium produced in heavy ion (HI) collisions, the effects of the strong magnetic field on the 
hadron properties were  
studied~\cite{Chakrabarty:1997ef,Leinson:1998yr,Broderick:2000pe,Cardall:2000bs,Mao:2001fq,
Mao:2001cv,Yue:2006xt, Sinha:2008wb,Yue:2009jh,Ryu:2010zzb, 
Sinha:2010fm,Rabhi:2011ej,PenaArteaga:2011wm,Ryu:2012hf,Maruyama:2012hf,deLima:2013dda}.
Furthermore, the effects of the baryon (anomalous) magnetic moments on the (proto-)neutron star 
structure under very strong magnetic filed were   
investigated in Refs.~\cite{Mao:2001fq,Ryu:2010zzb,Ryu:2012hf,Maruyama:2012hf}.
Moreover, the effects of charm quark in a dense medium, namely the stability of charm star was  
studied in Refs.~\cite{Kettner:1994zs,Glendenning,Jimenez:2019kji}. 

In this study we focus on the modifications of baryon magnetic moments in 
a nuclear medium, of the octed, decuplet, the low-lying charm, and the low-lying bottom baryons 
with nonzero light quarks. 
By this, we can study the in-medium electromagnetic (EM) interactions 
of the light and heavy baryons, the differences in the medium modifications,    
and the roles of the light and heavy quarks in a nuclear medium. 

We studied in Ref.~\cite{Tsushima:2018goq} the strong interaction properties   
for the octet, low-lying charm, and low-lying bottom baryons with nonzero light quarks 
in symmetric nuclear matter by the quark-meson coupling (QMC) 
model invented by Guichon~\cite{Guichon:1987jp}. 
In this study, we extend further to include the decuplet baryons,  
and proceed to study the modifications of the magnetic 
moments in a nuclear medium, which have potential impacts on 
the studies of the neutron star and magnetar structure.
In particular, the $\Delta$ baryon properties in a nuclear medium 
have collected renewed     
interests~\cite{Chen:2007kxa,Chen:2009am,Kolomeitsev:2016ptu,
Reichert:2019lny,Reichert:2020uxs,Li:2019tjx,Motta:2019ywl,Li:2020ias,
Thapa:2021kfo,Dexheimer:2021sxs}.

The QMC model has been successfully applied for various studies, e.g.,            
for the properties of finite 
(hyper)nuclei~\cite{Guichon:1995ue,Saito:1996sf,Saito:1996yb,Stone:2016qmi,Tsushima:1997rd,
Tsushima:1997cu,Guichon:2008zz,Tsushima:2002ua,Tsushima:2002sm,Tsushima:2003dd}, hadron properties 
in medium~\cite{Saito:1997ae,Tsushima:1997df,Tsushima:1998qw,Tsushima:1998qp,Tsushima:1998ru,
Sibirtsev:1999jr,Tsushima:2002cc}, nuclear 
reactions~\cite{Sibirtsev:1999js,Shyam:2008ny,Tsushima:2009zh,Shyam:2011aa,Chatterjee:2012ja,
Tsushima:2012pt,Shyam:2016bzq,Shyam:2016uxa,Shyam:2018iws}, and neutron star 
structure~\cite{RikovskaStone:2006ta,Katayama:2012ge,Miyatsu:2011bc,Whittenbury:2013wma,
Thomas:2013sea}.
(See Refs.~\cite{Saito:2005rv,Krein:2017usp,Guichon:2018uew} for reviews.)
It should be emphasized that, the predicted $^{15}_\Xi$C hypernucleus single-particle energies  
by the QMC model~\cite{Guichon:2008zz,Shyam:2019laf}, were indeed observed very closely    
in the experiments~\cite{Yoshimoto:2021ljs}. This may give some confidence for the 
treatment of the strange quark sector in the QMC model to be explained below.

Self-consistent exchange of the Lorentz-scalar-isoscalar $\sigma$-,  
Lorentz-vector-isoscalar $\omega$-, and Lorentz-vector-isovector $\rho$-meson fields,  
coupled directly to the confined, relativistically moving light $u$ and $d$ quarks, 
is the key mechanism of the QMC model.   
This mechanism, though simple, is known to achieve the novel saturation properties 
of nuclear matter---the saturation is achieved due to  
the quark structure of nucleon and quark dynamics. 
All the relevant coupling constants between the light quarks 
and the $\sigma$-, $\omega$-, and $\rho$-meson fields in any hadrons 
can be treated as the same as those in the nucleon, once the coupling constants 
are determined/constrained by the fit to the nuclear matter saturation constraints.

The physics behind the simple picture of the QMC model is associated with 
the dynamical symmetry breaking, although the usual QMC model such as the present one, 
does not have explicit chiral symmetry due to the lack of 
pion (Goldstone boson) field in the model. (See Refs.~\cite{Nagai:2008ai,Choi:2021odz} 
for chiral quark-meson coupling (CQMC) model, which explicitly incorporates  
pion field using the cloudy bag model instead of the MIT bag model, 
to be consistent with chiral symmetry.)
Namely, the fact that the light-quark condensates are expected to reduce the magnitude 
faster than those of the strange and heavier quarks as nuclear density 
increases, and it is supported by: 
(i) study of the in-medium strange and light quark condensates in connection 
with the $\phi$-meson mass shift in nuclear 
matter in the QCD sun-rule approach~\cite{Gubler:2014pta,Gubler:2018ctz,Gubler.private}, 
(ii) study made based on the NJL model~\cite{Tsushima:1991fe,Maruyama:1992ab}  
for the strange and light quark condensates in medium, and (iii) the result that the heavy quark 
condensates are proportional to the gluon condensate that is obtained by 
the operator product expansion~\cite{Shifman:1978bx} and also by a world-line effective 
action-based study~\cite{Antonov:2012ud}, together with the model independent 
result that the magnitude of gluon condensate at nuclear matter saturation density 
decreases only about 5\% by the quantum chromodynamics (QCD) trace anomaly and Hellman-Feynman 
theorem~\cite{Cohen:1991nk}.  
The light quark condensates are associated with dynamical chiral symmetry breaking  
and partial restoration of chiral symmetry in nuclear medium, 
where the latter is reflected to the reduction of the light quark condensates.
Although the QMC model Lagrangian does not have chiral symmetry, the model incorporates 
the expected facts of the density dependence of the in-medium quark condensates 
phenomenologically. That is, the model approximates that the $\sigma$-, $\omega$-, 
and $\rho$-meson fields couple directly only to the light quarks, 
but neither to the strange nor heavier quarks. 
Although one can consider the couplings of strange quark 
with $\phi$-meson field and some other types of the Lorentz scalar-meson fields, 
these would introduce unconstrained coupling constants which cannot be determined/constrained 
by the nuclear matter saturation properties, where the nuclear matter saturation properties
are the basic properties for calibrating any reasonable nuclear matter models.

As mentioned already, to study the in-medium properties of heavy baryons 
with nonzero light quarks are very important to understand dynamical symmetry 
breaking, its partial restoration, and the roles of the light quarks in medium. 
These phenomena can provide us with additional information on the origin 
of (dynamical) masses of hadrons and ''normal'' (not ''dark'') matter,  
which we can observe directly in our universe. 
Because of the importance, many studies have been made for the heavy baryon hypernuclei 
as well as the properties of heavy baryons with nonzero light quarks in a nuclear 
medium~\cite{Tyapkin,Dover,Gibson,Bando1,Bando2,bcexp1,bcexp2,Tsushima:2002cc,Tsushima:2002ua,
Tsushima:2002sm,Tsushima:2003dd,Tan:2004mt,Wang:2011hta,Wang:2011yj,Wang:epjc72,Ghosh:2014oia,
Hosaka:2016ypm,Azizi:2016dmr,Azizi:2018dtb,Er:2019hhk,Azizi:2019yoq,Ohtani:2017wdc,Carames:2018xek,
Yasui:2018sxz,Vidana:2019amb,Abu-Shady:2019fir}. 

In effective models such as the present model, the (current) quark mass values 
do not have a direct connection with QCD, but we may regard the quark mass value effect 
as a model parameter dependence. 
Then, we also present the results calculated with the different 
values of the current quark masses for the strange and bottom quarks in this study.
Although the magnetic moments of the octet~\cite{Ryu:2008st,Ramalho:2012pu,Singh:2017mxj} and 
decuplet~\cite{Singh:2020nwp} baryons in medium were studied, 
there exist no studies for the low-lying charm or bottom baryons with 
nonzero light quarks in a nuclear medium, while some studies in free space were made   
by symmetry-based quark models~\cite{Lichtenberg:1976fi,Johnson:1976mv,Franklin:1981rc}, 
the MIT bag model~\cite{Bose:1980vy,Bernotas:2012nz,Bernotas:2013eia,Simonis:2018rld}, 
QCD sum rules~\cite{Aliev:2015axa,Aliev:2015qea}, 
a relativistic three-quark model~\cite{Faessler:2006ft}, 
nonrelativistic hyper central model~\cite{Patel:2007gx}, 
an independent-quark model based on Dirac equation~\cite{Barik:1983ics}, 
and a relativistic potential model~\cite{Jena:1986xs}.

As for the in-medium modification of the bound nucleon size, or the bound nucleon electromagnetic  
form factors (EMFFs) which is directly associated with the present study, 
the constraint on the allowed maximum change (swelling) of the 
bound nucleon size was derived by the $y$-scaling data for 
$^3$He$(e,e')$~\cite{Sick:1985ygc} and $^{56}$Fe$(e,e')$~\cite{Sick:1986ns}. 
It was concluded that the allowed maximum increase of the bound nucleon size is, 
3 - 6\% in $^3$He, and 2 - 3\% in $^{56}$Fe. 
More precise analysis was performed by McKeown, and 
the upper limit of 3.6\% in $^3$He was obtained~\cite{Mckeown:1986kn}. 
The larger upper limit values may possibly be obtained by different methods. 
In Ref.~\cite{Morgenstern:2001jt} the relative change of the proton charge radius 
of $13 \pm 4$\% in a heavy nucleus was suggested, based on the analysis  
for the longitudinal Coulomb response function (Coulomb sum rule) using  
the effective momentum approximation, where the quenching of the longitudinal response function    
was reinterpreted assuming the dipole expression for the proton 
charge form factor. However, the derived value seems to be more indirect 
than that obtained by the $y$-scaling data analysis.  

The bound nucleon size (charge and magnetic radii) was discussed in the past  
in connection with the predictions of the bound nucleon EMFFs 
in the QMC model~\cite{Lu:1997mu,Lu:1998tn} as well as in the CQMC model~\cite{Nagai:2008ai}. 
In Ref.~\cite{Lu:1998tn} it was stated that the 10\% decrease of the 
bag constant at the normal nuclear matter density $\rho_0$ (= 0.15 fm$^{-3}$) 
is already quite large, and this would severely reduce the bound nucleon EMFFs. 
Using the average baryon density estimated by the 
QMC model, $<\rho_B(^3{\rm He})> \simeq 0.35\rho_0$ and 
$<\rho_B(^{56}{\rm Fe})> \simeq 0.71\rho_0$, the $y$-scaling-based results 
lead the allowed maximum change (swelling) of the nucleon size 
at $\rho_0$ respectively to, 8.5 - 17.1\% and 4.2 - 8.5\%. 
Since the MIT bag model for both the matter (charge) radius and magnetic moment of proton  
are proportional to the bag radius, the bag radius as well as 
the nucleon magnetic moment at $\rho_0$ cannot be enhanced  
more than 17.1\% if one takes the $y$-scaling result properly.  
Although the range, 4.2 - 8.2\% derived from the $^{56}$Fe data is expected 
to be more proper to extrapolate to the normal 
nuclear matter density $\rho_0$, we allow the larger range in the discussion.  
There are studies made for the in-medium octet baryon magnetic 
moments~\cite{Ryu:2008st}, and the impact of them on neutron stars under strong 
magnetic field~\cite{Ryu:2010zzb}, and the authors reported about a 25\% enhancement 
of the nucleon magnetic moment, and about a 20\% increase of the nucleon bag radius 
at density 0.17 fm$^{-3}$ (1.133$\rho_0 = 1.133 \times 0.15$ fm$^{-3}$).  
The results seem to give too large enhancement, which are originated from the density 
dependent bag constant using the modified quark-meson coupling (MQMC) model.
In this study, we use the standard QMC model, and indeed the results will turn out to satisfy 
the even the tighter $y$-scaling data constraint. 
Thus, the present results are expected to be constrained by the allowed in-medium modifications 
of the octet baryon magnetic moments, as well as for those of the decuplet, low-lying charm and 
bottom baryons with nonzero light quarks.

Although it is known that the MIT bag model has a deficiency to produce   
the magnitude of the free space octet baryon magnetic moments, 
we can focus on the in-medium to free space ratios, as many experiments directly 
measure ratios to extract meaningful physical quantities, such as 
to extract the free proton~\cite{Jones:1999rz,Gayou:2001qd} 
as well as the bound proton~\cite{Strauch:2002wu,Paolone:2010qc,Malov:2000rh} 
EMFF ratios by measuring simultaneously the transverse ($P_t$) and the longitudinal 
($P_l$) recoil proton polarization to extract the proton electric ($G^p_E$) over magnetic 
($G^p_M$) form factor ratio $G^p_E/G^p_M$. 
Indeed, the super ratio, $[G^p_E/G^p_M(^4{\rm He})]/[G^p_E/G^p_M(^1{\rm H})]$,  
calculated using the in-medium to free EMFF ''ratios'' predicted by the QMC model, 
reproduce well the data as shown in Refs.~\cite{Strauch:2002wu,Dieterich:2000mu,Paolone:2010qc}.
Note that, the meson cloud contributions for the total medium modifications of EMFFs 
are at the order of a few tens percent, namely, the order of 0.2 $\sim$ 0.3 \% for the total 
modifications of about 10\%. Furthermore, the relativistic kinematic factors are canceled 
out in the QMC model calculated EMFF ratios. 
Because the in-medium modifications apply directly for the light quarks in the QMC model  
and the reproduced data are for the proton composed of purely light quarks, 
as far as the effects of light quarks are concerned, 
we may have some confidence for the in-medium to free EMFF ratios 
for the strange and heavy baryons, where the light quark medium modifications are  
mainly responsible for the medium modifications of these baryons.
Recall, also the successful predictions for the $^{15}_\Xi$C hypernucleus 
single-particle energies, associated with the strange quark sector.  

Furthermore, in Ref.~\cite{Jones:1999rz} it is stated that for the method of 
measuring the ratio, ''Neither the beam polarization nor the polarimeter 
analyzing power needs to be known, which results in small systematic uncertainties.'' 
The examples for the experiments mentioned above clearly demonstrate that the usefulness of 
measuring/calculating ratios of physical quantities. 
Another example is to extract the structure function ($F_2$) ratio of the bound 
to free nucleons, [$F_2^{\rm bound\hspace{1ex}nucleon}/F_2^{\rm free\hspace{1ex}nucleon}$],  
by measuring the corresponding cross section ratio 
(e.g. see Ref.~\cite{Bodek:1983qn,Seely:2009gt}), 
known as the ''EMC'' effect (ratios)~\cite{EMC}.

Concerning the magnetic moments of heavy baryons with $c$ and/or $b$ quarks/quark,  
some ambiguities arise in constructing the flavor-spin wave functions~\cite{Franklin:1981rc}. 
These are associated with the so-called ''quark order'',  
originating from the fact that the spin, isospin, SU(3) flavor symmetry,   
and the Pauli principle cannot help much. The possible different quark orders in the 
flavor-spin wave functions yield different results for the calculated magnetic moments.
This is concerned for the $\Xi_{c,b}$ baryons in the present study.
For these $\Xi_{c,b}$ baryons, the two lightest quarks ($u$ and $s$) or ($d$ and $s$) 
are taken as the first two quark antisymmetric pair denoted as $[u,s]$ or $[d,s]$ in the wave 
functions~\cite{Yu:2018yxl}, and thus the magnetic moments of $\Xi_{c,b}$ are given by 
$\mu_{c,b}$~\cite{Franklin:1981rc,Bernotas:2012nz} 
and nearly the same as those of the $\Lambda_{c,b}$.
Thus, the free as well as the in-medium magnetic moments of the $\Xi_{c,b}$ 
are almost similar to those of the $\Lambda_{c,b}$ baryons as will be 
shown explicitly later.
Then, although some issues exist for the quark order as discussed in Ref.~\cite{Franklin:1981rc},   
we may assume the natural quark order as realized for the octet baryon sector, and 
study the magnetic moments and the transition 
magnetic moments of the octet, decuplet, low-lying charm, and low-lying bottom baryons 
with nonzero light quarks in symmetric nuclear matter.
Note that, the ''1-2 quark order'' is supported in Ref.~\cite{Franklin:1981rc} 
as the best quark ordering for flavor-degenerate baryons for the masses.
Furthermore, we discuss possible ambiguities originating from 
the MIT bag model artifact for the transition magnetic moments.
This is the first study to estimate the in-medium      
magnetic moments of the low-lying charm and bottom baryons 
with nonzero light quarks.

As an important side remark, we would like to emphasize that, 
the explicit density dependent parametrizations 
are given for the vector potentials of the baryons and light-$(u, d)$ quarks, 
as well as for the effective masses (Lorentz-scalar-isoscalar mean field potentials) 
of the low-lying baryons treated in this study, 
and of the mesons, $\omega,\rho,K,K^*,\eta,\eta',D,D^*,B$, and $B^*$, for a practical use.

\section{The QMC model}
\label{qmc}

In this section we outline the QMC model following Refs.~\cite{Saito:2005rv,Krein:2017usp}.
Since the Hartree-Fock approximation in the QMC model    
gives very similar results with those of the Hartree approximation~\cite{Krein:1998vc}, 
we use the Hartree approximation to be consistent with Ref.~\cite{Tsushima:2018goq}.
(See Ref.~\cite{Whittenbury:2013wma} for the Hartree-Fock approximation in the QMC model 
applied for studying the neutron star structure with hyperons.)

Furthermore, the one-gluon-exchange based color-magnetic interaction between the quarks, 
which enhances the in-medium mass splittings between the $\Sigma$ and $\Lambda$ 
baryons~\cite{Guichon:2008zz} as well as the $N$ and $\Delta$ baryons~\cite{Motta:2019ywl} 
playing an important role in the studies of hypernuclei and neutron star structure, 
is not included in this study. 

In a practical aspect, we would like to emphasize that the explicit density dependent 
parametrizations for the Lorentz-vector potentials of the baryons, as well as for the 
effective masses (Lorentz-scalar-isoscalar potentials) of the low-lying baryons and mesons except 
for pion, will be given in this section.

\subsection{Nuclear matter}

Using the Born-Oppenheimer approximation, a relativistic effective Lagrangian density 
for a ''hypernucleus'' in the QMC model 
may be given by~\cite{Tsushima:1997cu,Tsushima:2003dd,Saito:2005rv,Krein:2017usp} 
(to consider the nuclear matter limit, we call the following configuration as 
a ''hypernucleus'' using a terminology ''hyperon ($Y$)'' for each baryon  
indicated in Eq.~(\ref{eq:LagY})), 
\begin{eqnarray}
{\cal L}^{HY}_{QMC} &=& {\cal L}^N_{QMC} + {\cal L}^Y_{QMC},
\label{eq:LagYQMC} \\
{\cal L}^N_{QMC} &=&  \overline{\psi}_N(\vec{r})
[ i \gamma \cdot \partial - m_N^* (\sigma) \nn\\
& & - (\, g_\omega \omega(\vec{r})
+ g_\rho \dfrac{\tau^N_3}{2} b(\vec{r})
+ \dfrac{e}{2} (1+\tau^N_3) A^0(\vec{r}) \,) \gamma_0
] \psi_N(\vec{r}) \quad \nn \\
& & - \dfrac{1}{2}[ (\nabla \sigma(\vec{r}))^2 +
m_{\sigma}^2 \sigma(\vec{r})^2 ]
+ \dfrac{1}{2}[ (\nabla \omega(\vec{r}))^2 + m_{\omega}^2
\omega(\vec{r})^2 ] \nn \\
 & & + \dfrac{1}{2}[ (\nabla b(\vec{r}))^2 + m_{\rho}^2 b(\vec{r})^2 ]
+ \dfrac{1}{2} (\nabla A^0(\vec{r}))^2, 
\label{eq:LagN} \\
{\cal L}^Y_{QMC} &=&
\overline{\psi}_Y(\vec{r})
\left[ i \gamma \cdot \partial
- m_Y^*(\sigma)
- (\, g^Y_\omega \omega(\vec{r})
+ g^Y_\rho I^Y_3 b(\vec{r})
+ e Q_Y A^0(\vec{r}) \,) \gamma_0
\right] \psi_Y(\vec{r}), 
\nn\\
&&\hspace{-4ex}(Y=\Lambda,\Sigma^{0,\pm},\Xi^{0,-},\Delta^{0,\pm,++},\Sigma^{*\,0,\pm},\Xi^{*\,0,-},
\Lambda^+_c,\Sigma_c^{0,+,++},\Xi_c^{0,+},\Lambda_b^0,\Sigma_b^{0,\pm},\Xi_b^{0,-}),
\label{eq:LagY}
\end{eqnarray}
where, the quasi-particles moving in single-particle orbits are three-quark 
clusters with the quantum numbers of a nucleon, a $\Delta$ baryon, a strange, a charm or 
a bottom ''hyperon'' when expanded to the same order in 
velocity~\cite{Guichon:1995ue,Saito:1996sf,Tsushima:1997cu,Tsushima:2002ua,
Tsushima:2003dd,Tsushima:2002cc}.  
In the above $\psi_N(\vec{r})$ [$\psi_Y(\vec{r})$] 
is the nucleon [$\Delta$ baryon, hyperon (strange, charm or bottom baryon)] field. 
The mean-meson fields represented by, $\sigma, \omega$, and $b$    
are the Lorentz-scalar-isoscalar, Lorentz-vector-isoscalar, and third component of  
the Lorentz-vector-isovector fields, respectively, while $A^0$ is the Coulomb field.
Hereafter, the quantities in medium will be denoted by an asterisk, $^*$. 
Note that, in the Lagrangian density Eq.~(\ref{eq:LagY}), the phenomenologically   
introduced effective Pauli potentials, which contain also the channel coupling effects   
at the baryon level for the $\Lambda, \Sigma$ and $\Xi$~\cite{Tsushima:1997cu},  
are not written explicitly. The potentials were needed to reproduce the observed 
lowest single-particle energy in $^{208}_\Lambda$Pb-hypernucleus, 
as well as the energy difference between the $\Lambda$ and $\Sigma$-hyperons obtained in the 
G-matrix calculation in nuclear matter~\cite{Tsushima:1997cu},  
in addition to the effective Lagrangian density Eq.~(\ref{eq:LagY}).
But these repulsive ``Pauli'' (vector) potentials will be included in the explicit 
density dependent parametrizations,  
while the one-gluon exchange interaction in medium,  
that yields to agree with the fact of ''no experimental observation of the $\Sigma$ hypernuclei'',  
will not be included, where the interaction was introduced in the latest version of the QMC 
model~\cite{Guichon:2008zz}. Thus, in order to agree with this fact, we will 
include a phenomenological repulsive vector potential for the $\Sigma$, in such a way 
that it yields the $\Sigma$ total potential of $\simeq +30$ MeV at $\rho_0$, and the corresponding 
parametrizations for the vector potential will also be given. 

The coupling constants of the hyperon appearing in Eq.~(\ref{eq:LagY}) are,   
$g^Y_\omega = (n_q/3) g_\omega$, and $g^Y_\rho \equiv g_\rho = g_\rho^q$, 
with $n_q$ being the number of valence light quarks in the hyperon $Y$ 
($n_q = 3$ for $N$), where $g_\omega$ and $g_\rho$ 
appearing in Eq.~(\ref{eq:LagN}) are the $\omega$-$N$ and $\rho$-$N$ 
coupling constants, respectively. 
$I^Y_3$ and $Q_Y$ are the third component of the hyperon isospin operator and the electric
charge in units of the positron charge, $e$, respectively.
The couplings between the meson fields and quarks, as already mentioned,  
reflect the fact that the magnitude of the light-quark condensates is expected to reduce 
faster than those of the strange and heavier quarks as baryon (nuclear) density increases.

The $\sigma$-field dependent $\sigma$-$N$ [$\sigma$-$Y$] 
coupling strength for the nucleon $N$ [hyperon $Y$],  
$g^N_\sigma(\sigma)$ [$g^Y_\sigma(\sigma)$] implicitly in Eq.~(\ref{eq:LagN}) 
[Eq.~(\ref{eq:LagY})], is defined by  
\bea
& &m_{N,Y}^*(\sigma) \equiv m_{N,Y} - g^{N,Y}_\sigma(\sigma)
\sigma(\vec{r}),  
\hspace{2ex} (Y = \Lambda,\Sigma,\Xi,\Delta,\Sigma^*,\Xi^*,  
\Lambda_c,\Sigma_c,\Xi_c,\Lambda_b,\Sigma_b,\Xi_b), 
\label{effmass}
\eea
where $m_N$ [$m_Y$] is the free nucleon [hyperon] mass. 
Note that, the dependence of the coupling strengths on the 
scalar field $\sigma$, must be calculated self-consistently within the quark
model~\cite{Guichon:1987jp,Guichon:1995ue,Tsushima:1997cu,Tsushima:2002ua,
Tsushima:2002sm,Tsushima:2002cc} (the MIT bag model in the present case).
This characterizes the QMC model differently from quantum hadrodynamics 
(QHD)~\cite{Walecka:1974qa,Serot:1984ey}, as well as from other  
naive symmetry-based approaches. Namely, although in such approaches,  
$g^Y_\sigma(\sigma)/g^N_\sigma(\sigma)$ 
may be $2/3$ or $1/3$ depending on the number of the light quarks $n_q$ in the hyperon $Y$  
in free space (means $\sigma = 0$), this may not be true any more in a nuclear medium.
(Even in free space this is not true, since the bag radii of the nucleon and hyperon 
are not exactly the same~\cite{Chodos:1974je,Tsushima:1997cu}.)

For the later convenience, we define $C_{N,Y}(\sigma) \equiv 
S_{N,Y}(\sigma) / S_{N,Y}(\sigma=0)$,  
and $S_{N,Y}(\sigma)$ in connection with $m^*_{N,Y}$~\cite{Tsushima:2018goq}, 
by denoting $q (\equiv u, d)$ the light quarks as, 
\bea
\frac{d m_{N,Y}^*(\sigma)}{d \sigma}
&=& - n_q g_{\sigma}^q \int_{bag} d^3y\, 
{\overline \psi}_q(\vec{y}) \psi_q(\vec{y})
\nn\\
&\equiv& - n_q g_\sigma^q S_{N,Y}(\sigma)  
= - \left[ n_q g_\sigma^q S_{N,Y} (\sigma=0)\right]
\left( \dfrac{S_{N,Y}(\sigma)}{\left[ n_q g_\sigma^q S_{N,Y} (\sigma=0)\right]} \right) \nn \\
&\equiv& - \left[ n_q g_\sigma^q S_{N,Y} (\sigma=0)\right]\, C_{N,Y}(\sigma) 
= - \dfrac{d}{d \sigma}
\left[ g^{N,Y}_\sigma(\sigma) \sigma \right],
\label{Ssigma}
\eea
where $g^q_\sigma$ is the (light-quark)-$\sigma$ coupling constant, 
and $\psi_q$ is the light-quark ground state wave function in $N$ or 
$Y$ immersed in a nuclear medium, where the in-medium masses $m^*_{N,Y}$ 
are associated with the {\it quark scalar charge}.
The $\sigma$-$N$ and $\sigma$-$Y$ coupling constants in free space (i.e., $\sigma=0$) 
are defined by, 
\begin{equation}
g^{N,Y}_\sigma \equiv g_\sigma^{N,Y}(\sigma = 0) \equiv n_q\, g^q_\sigma\, S_{N,Y} (\sigma = 0). 
\label{sigmacc}
\end{equation}
Note that, the values of $S_N(\sigma)$ and $S_Y(\sigma)$ in Eq.~(\ref{Ssigma}) are 
different, since the light-quark ground state wave functions 
in $N$ and $Y$ are different in free space as well as in medium. 
Since the light quarks in any hadrons are expected to feel 
the same scalar and vector potentials as those in the nucleon, 
one can systematically study the hadron properties in medium 
using the same (light-quark)-meson coupling constants, which are constrained 
by the nuclear matter saturation properties. 
This is one of the big advantages of the QMC model.

Next, we consider the rest frame of symmetric nuclear matter, a spin and isospin 
saturated, infinitely large system with only the strong interaction. 
In this case, the self-consistent effect from the embedded one hyperon 
to the (nuclear matter + one hyperon) system can be neglected,  
although for a hypernucleus, the effect is self-consistently included 
together with the influence of the Pauli potentials and channel couplings   
as $1/A$ effect, with $A$ being the total baryon number of the hypernucleus 
(see Refs.~\cite{Tsushima:2018goq,Guichon:1987jp,Tsushima:1997cu,Saito:2005rv} for details). 
Thus, the quark-meson coupling constants that are determined by the saturation properties 
of symmetric nuclear matter (without a hyperon), 
as well as the total energy per nucleon, will not be affected in 
the system of the (nuclear matter + one hyperon).

The Dirac equations for the quarks and antiquarks  
in nuclear matter, in a bag of a hadron, $h$, ($q = u$ or $d$, and 
$Q \equiv s,c$ or $b$, hereafter),  
are given by  ($x=(t,\vec{x})$ and for $|\vec{x}|\le$ 
bag 
radius)~\cite{Tsushima:1997df,Tsushima:1998ru,Sibirtsev:1999jr,Tsushima:2002cc,Sibirtsev:1999js}, 
\begin{eqnarray}
\left[ i \gamma \cdot \partial_x -
(m_q - V^q_\sigma)
\mp \gamma^0
\left( V^q_\omega +
\dfrac{1}{2} V^q_\rho
\right) \right] 
\left( \begin{array}{c} \psi_u(x)  \\
\psi_{\bar{u}}(x) \\ \end{array} \right) &=& 0,
\label{Diracu}\\
\left[ i \gamma \cdot \partial_x -
(m_q - V^q_\sigma)
\mp \gamma^0
\left( V^q_\omega -
\dfrac{1}{2} V^q_\rho
\right) \right]
\left( \begin{array}{c} \psi_d(x)  \\
\psi_{\bar{d}}(x) \\ \end{array} \right) &=& 0,
\label{Diracd}\\
\left[ i \gamma \cdot \partial_x - m_{Q} \right] \psi_{Q} (x) = 0, ~~~~~~~~  
\left[ i \gamma \cdot \partial_x - m_{Q} \right] \psi_{\overline{Q}} (x) &=& 0,  
\label{DiracQ}
\end{eqnarray}
where, the mean field potentials are defined by, 
$V^q_\sigma \equiv g^q_\sigma \sigma$, 
$V^q_\omega \equiv g^q_\omega \omega$, and
$V^q_\rho \equiv g^q_\rho b$,
with $g^q_\sigma$, $g^q_\omega$, and
$g^q_\rho$ being the corresponding quark-meson coupling constants. 
We assume SU(2) symmetry, $m_{u,\bar{u}}=m_{d,\bar{d}} \equiv m_{q,\bar{q}}$. 
The Lorentz-scalar ''effective quark masses'' 
are defined by, $m^*_{u,\bar{u}}=m^*_{d,\bar{d}}=m^*_{q,\bar{q}} \equiv 
m_{q,\bar{q}}-V^q_{\sigma}$, and thus $m_q^*$ is dominated by 
$-V^q_{\sigma}$ as baryon density increases, and can be negative.
Note that, $m_Q=m_Q^*$, since the $\sigma$ field does not couple 
to the heavier quarks $Q=s,c,b$.
Furthermore, since $\rho$-meson mean field becomes zero, $V^q_{\rho}=0$, in Eqs.~(\ref{Diracu}) 
and~(\ref{Diracd}) in symmetric nuclear matter in the Hartree approximation,      
we will ignore it.

The same mean fields $\sigma$ and $\omega$ for the quarks  
in Eqs.~(\ref{Diracu}) and~(\ref{Diracd}), satisfy self-consistently 
the following equations at the nucleon level, with  
$m_N^*(\sigma)$ to be calculated by Eq.~(\ref{hmass}):
\begin{eqnarray}
{\omega} &=& \dfrac{g_\omega}{m_\omega^2} \rho_B 
\equiv \dfrac{g_\omega}{m_\omega^2} 
\dfrac{4}{(2\pi)^3}\int d^3{k}\ \theta (k_F - |\vec{k}|),  
\label{omgf}\\
{\sigma} &=& \dfrac{g_\sigma^N}{m_\sigma^2}C_N({\sigma}) \rho_s
\equiv \dfrac{g_\sigma^N }{m_\sigma^2}C_N({\sigma}) 
\dfrac{4}{(2\pi)^3}\int d^3{k} \ \theta (k_F - |\vec{k}|)
\dfrac{m_N^*(\sigma)}{\sqrt{m_N^{* 2}(\sigma)+\vec{k}^2}}, 
\label{sigf}
\end{eqnarray}
where $k_F$ is the nucleon Fermi momentum.

Because of the underlying quark structure of the nucleon used to calculate
$m^*_N(\sigma)$ in the nuclear medium, $C_N(\sigma)$ decreases as $\sigma$ increases,
whereas in the usual point-like nucleon, $C_N(\sigma) = 1$.
It is this variation of $C_N(\sigma)$ 
(or equivalently $\sigma$-dependence of the coupling as $g_\sigma^N (\sigma(\rho_B))$), 
that yields a novel saturation mechanism for nuclear matter---$\sigma$-dependence originates 
from the quark structure of the nucleon. 
The important dynamics which originates from the quark structure
of the nucleon, is included in $C_N(\sigma)$. 
This $C_N(\sigma)$ also yields three-body or density dependent effective 
forces at the nucleon level~\cite{Guichon:2018uew,Guichon:2006er}.  
As a consequence, the QMC model gives the nuclear incompressibility of $K \simeq 280$~MeV 
with the free space inputs $m_q = 5$ MeV and nucleon bag radius $0.8$ fm~\cite{Saito:2005rv}. 
The value is in contrast to a naive version of 
QHD~\cite{Walecka:1974qa,Serot:1984ey},
that results in much larger value, $K \simeq 500$~MeV, where
the empirically extracted value falls in the range $K = 200 - 300$ MeV.
(See Ref.~\cite{Dutra:2012mb} for details.)

Once the self-consistency equation for the ${\sigma}$ field 
Eq.~(\ref{sigf}) is solved, one can calculate the total energy per nucleon:
\begin{equation}
E^\mathrm{tot}/A=\dfrac{4}{(2\pi)^3 \rho_B}\int d^3{k} \
\theta (k_F - |\vec{k}|) \sqrt{m_N^{* 2}(\sigma)+
\vec{k}^2}+\dfrac{m_\sigma^2 {\sigma}^2}{2 \rho_B}+
\dfrac{g_\omega^2 \rho_B}{2m_\omega^2} .
\label{toten}
\end{equation}
The parameters appearing in the Lagrangian density   
Eqs.~(\ref{eq:LagYQMC})-(\ref{eq:LagY}) and also above are, 
$m_\omega = 783$ MeV, $m_\rho = 770$ MeV, $m_\sigma = 550$ MeV 
and $e^2/4\pi = 1/137.036$~\cite{Guichon:1995ue,Saito:1996sf}.
The coupling constants, $g^N_{\sigma} \equiv g_\sigma$, $g^N_\omega \equiv g_{\omega}$, 
and $g^N_{\rho} \equiv g_\rho$ at the nucleon level, are determined by 
the fit to the binding energy of 15.7~MeV at the saturation density $\rho_0$ = 0.15 fm$^{-3}$
($k_F^0$ = 1.305 fm$^{-1}$) for symmetric nuclear matter, as well as  
$g_\rho$ to the symmetry energy of~35 MeV.
The corresponding quark-meson coupling constants determined, 
and the current quark mass values (inputs), 
respectively denoted by ''Set I`` and ''Set II``, are listed  
in Tables~\ref{coupcc1} and~\ref{coupcc2}, where in the Set I, the 
current quark mass values of the strange ($m_s$) and bottom ($m_b$) quarks 
are different form those in Ref.~\cite{PDG}---Particle Data Group (PDG).
Although the (current) quark mass values in effective models do not have direct connection  
with those in QCD, we use the latest values quoted in Ref.~\cite{PDG} for the Set II, 
where one may see a model parameter dependence by the Set I and Set II results.
Note that, the current quark mass values except for the $u$ and $d$ quarks,  
do not influence the nuclear matter saturation properties, thus the relevant quark-meson 
coupling constants, $g^q_\sigma, g^q_\omega$, and $g^q_\rho$, are the same  
in Tables~\ref{coupcc1} and~\ref{coupcc2}.
In the past including Ref.~\cite{Tsushima:2018goq}, 
the use of the strange quark current 
mass value of $m_s = 250$ MeV was motivated by the success   
in reproducing the light hadron masses in the MIT bag model 
with $m_s = 279$ MeV~\cite{DeGrand:1975cf}.  
In the present study we also use the different values for the strange and bottom quark 
current masses, respectively $m_s = 93$ MeV and $m_b = 4180$ MeV (Set II) 
given in Ref.~\cite{PDG}.

%
\begin{table}[htb]
\begin{center}
\caption{
(Set I) Current quark mass values (inputs), quark-meson coupling constants 
and the bag constant $B_p$~\cite{Tsushima:2018goq}, obtained with 
the inputs: free nucleon bag radius $R_N=0.8$ fm, empirical values 
$E^{\rm tot}/A - m_N = -15.7$ MeV ($m_N = 939$ MeV) at the saturation density  
$\rho_0=0.15$ fm$^{-3}$, and the symmetry energy, 35 MeV.
}
\label{coupcc1}
\vspace{1ex}
\begin{tabular}{r|r||l|l}
\hline
\hline
$m_{u,d}$ &5    MeV &$g^q_\sigma$ &5.69\\
$m_s$     &250  MeV &$g^q_\omega$ &2.72\\
$m_c$     &1270 MeV &$g^q_\rho$   &9.33\\
$m_b$     &4200 MeV &$B_p^{1/4}$  &170 MeV\\
\hline
\hline
\end{tabular}
\end{center}
\end{table}
%
\begin{table}[htb]
\begin{center}
\caption{
(Set II) Same as the caption of Table~\ref{coupcc1}, but the current quark mass values for the 
strange ($m_s$) and bottom ($m_b$) quarks are from Ref.~\cite{PDG}.
}
\label{coupcc2}
\vspace{1ex}
\begin{tabular}{r|r||l|l}
\hline
\hline
$m_{u,d}$ &5    MeV &$g^q_\sigma$ &5.69\\
$m_s$     &93  MeV  &$g^q_\omega$ &2.72\\
$m_c$     &1270 MeV &$g^q_\rho$   &9.33\\
$m_b$     &4180 MeV &$B_p^{1/4}$  &170 MeV\\
\hline
\hline
\end{tabular}
\end{center}
\end{table}

The corresponding coupling constant values at the nucleon level are, 
$g_\sigma^2/4\pi = (g_\sigma^N)^2/4\pi = 5.39$ 
(see Eq.~(\ref{sigmacc}) with $S_N(0) = 0.4827$, where Ref.~\cite{Tsushima:2018goq} 
mistakenly gave the value for finite nuclei),    
$g_\omega^2/4\pi = (g_\omega^N)^2/4\pi = (3g^q_\omega)^2/4\pi = 5.30$,  
and $g_\rho^2/4\pi = (g_\rho^N)^2/4\pi = (g^q_\rho)^2/4\pi = 6.93$.

The mass of a hadron $h$ in symmetric nuclear matter, $m^*_h$ (free mass is $m_h$), 
is calculated, together with the mass stability condition with respect to the in-medium 
bag radius at a given density:
\begin{eqnarray}
m_h^* &=& \sum_{j=q,\qbar,Q,\Qbar} 
\dfrac{ n_j\Omega_j^* - z_h}{R_h^*}
+ \frac{4}{3}\pi R_h^{* 3} B_p,\quad
\hspace{2ex}\frac{d m_h^*}{d R^*_h} = 0,
\label{hmass}
\end{eqnarray}
where $\Omega_q^*=\Omega_{\bar{q}}^*
=[(x_q^*)^2 + (R_h^* m_q^*)^2]^{1/2}\,(q=u,d)$, with
$m_q^*=m_q{-}g^q_\sigma \sigma=m_q-V^q_\sigma$,
$\Omega_Q^*=\Omega_{\Qbar}^*=[(x_Q^*)^2 + (R_h^* m_Q)^2]^{1/2}\,(Q=s,c,b)$,
and $x_{q,Q}^*$ are the lowest mode bag eigenfrequencies.
$B_p$ is the bag constant (assumed to be independent of density), 
$n_{q,Q}$ $[n_{\qbar,\Qbar}]$ are the lowest mode valence 
quark [antiquark] numbers of each quark flavor $q=(u,d),Q=(s,c,b)$ 
in the hadron $h$, while $z_h$ parametrizes the sum of the
center-of-mass and gluon fluctuation effects,   
which is assumed to be independent of density~\cite{Guichon:1995ue}. 
The bag constant $B_p = {\rm (170\, MeV)}^4$ is determined 
by the free nucleon mass $m_N = 939$ MeV, free nucleon bag radius $R_N = 0.8$ fm, 
and $m_q = 5$ MeV, which are considered to be the standard input values in the QMC 
model~\cite{Saito:2005rv}.
Recall that, the quark-meson coupling constants, $g^q_\sigma$, $g^q_\omega$
and $g^q_\rho$, have already been determined by the nuclear matter saturation properties, 
we can use the same coupling constants which are empirically constrained, 
for the light quarks in any hadrons.

\begin{figure}[htb]
\vspace{3ex}
\begin{center}
\includegraphics[scale=0.26]{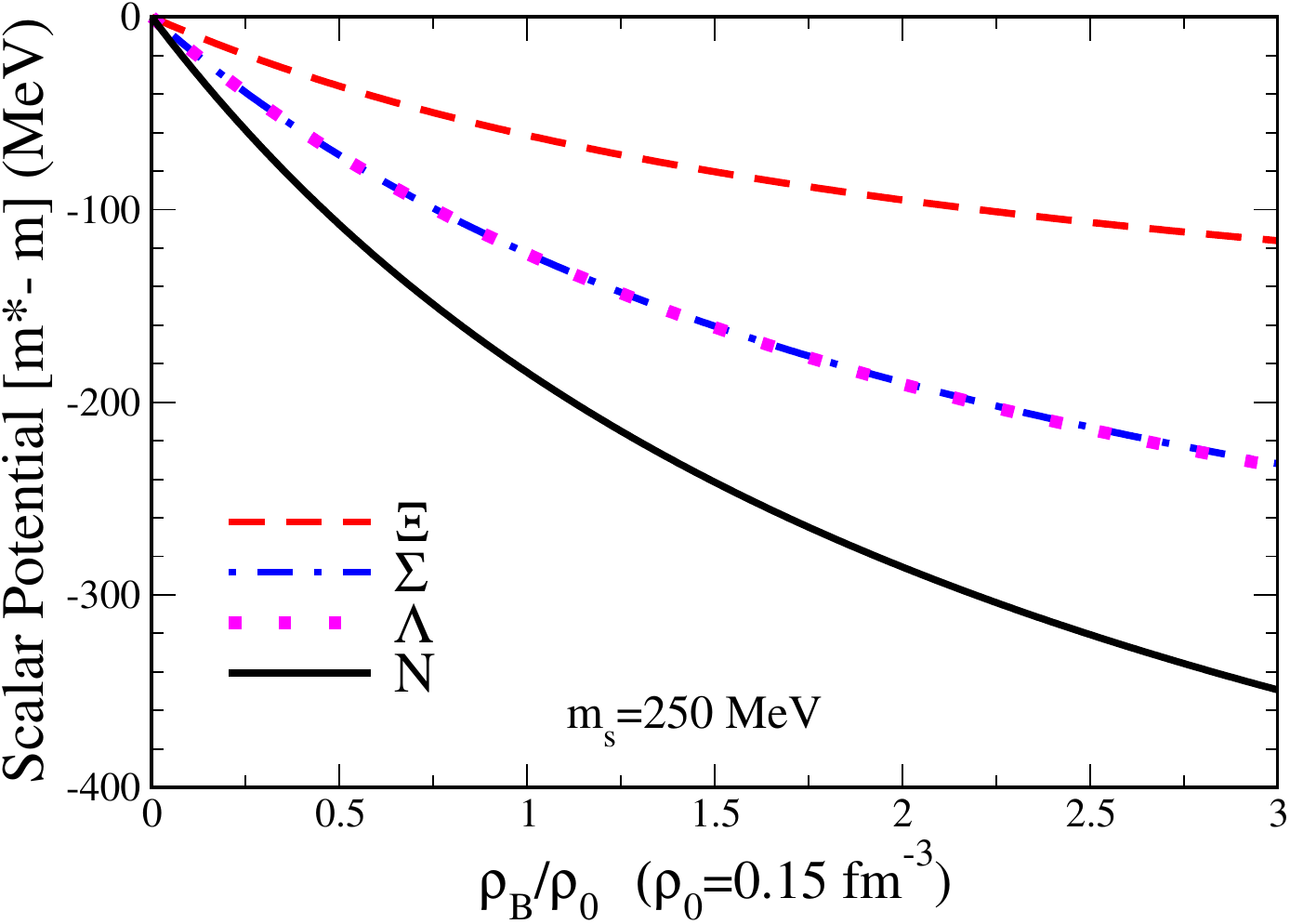}
\hspace{3ex}
\includegraphics[scale=0.26]{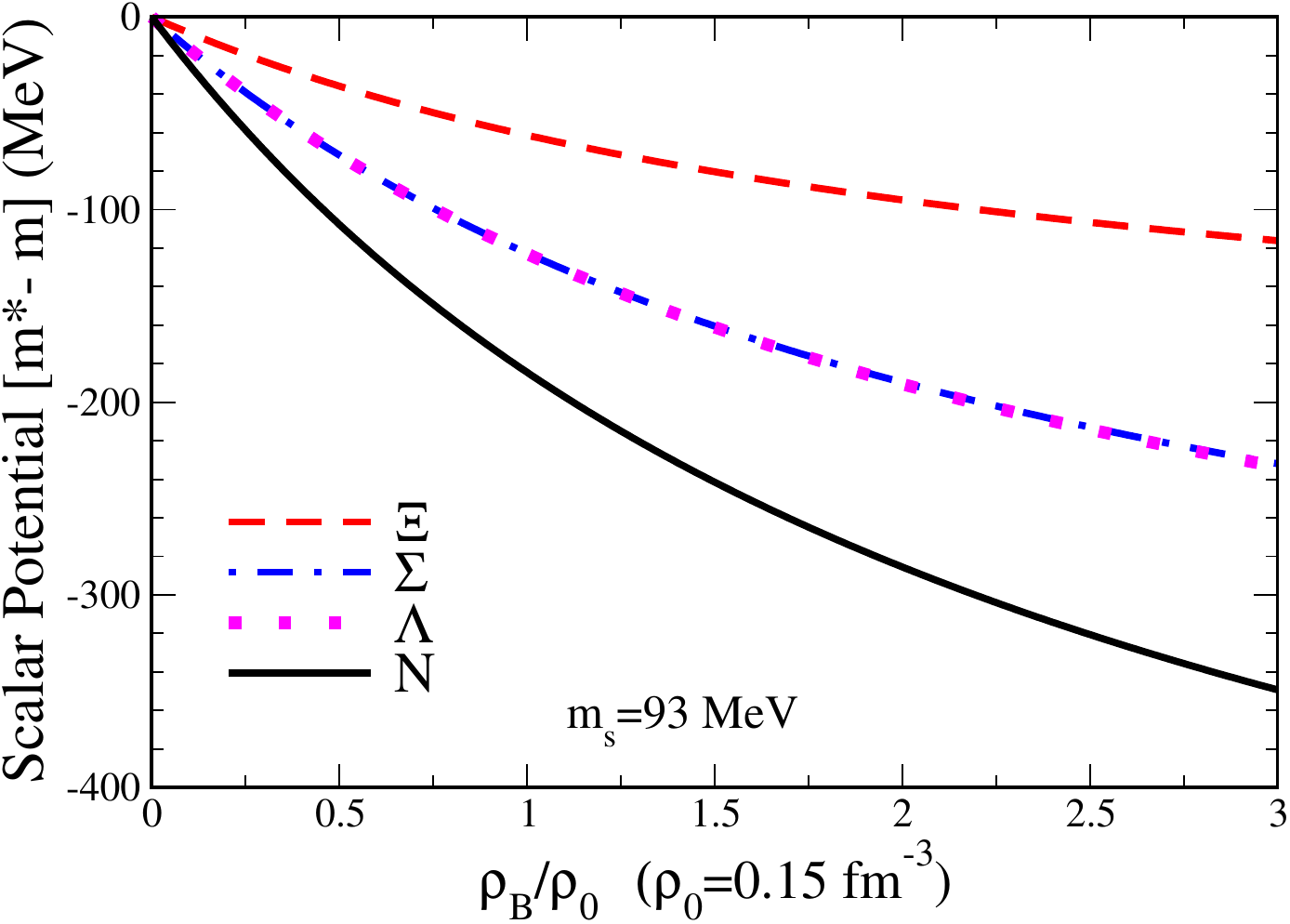}
\vspace{6ex}
\\
\includegraphics[scale=0.26]{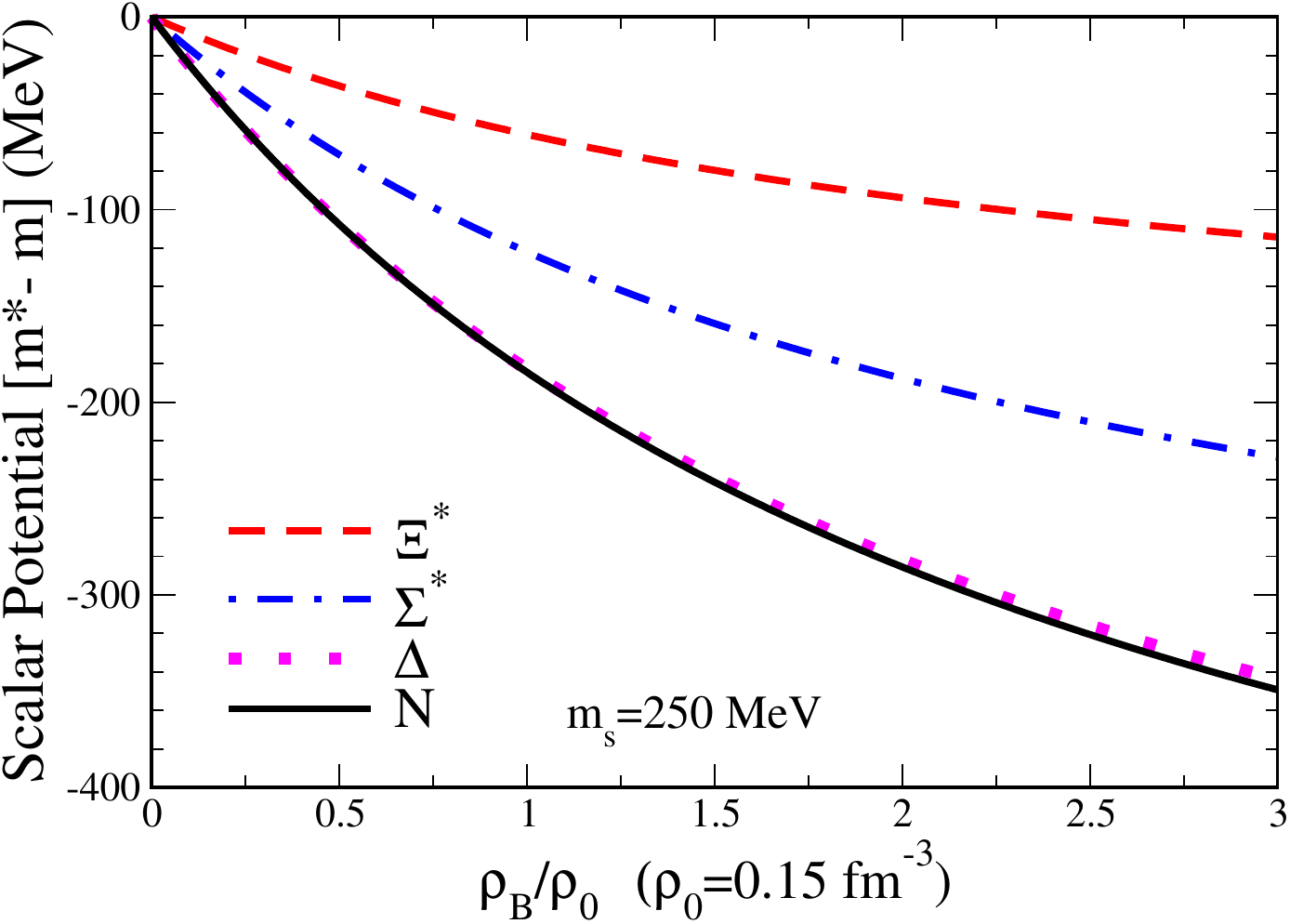}
\hspace{3ex}
\includegraphics[scale=0.26]{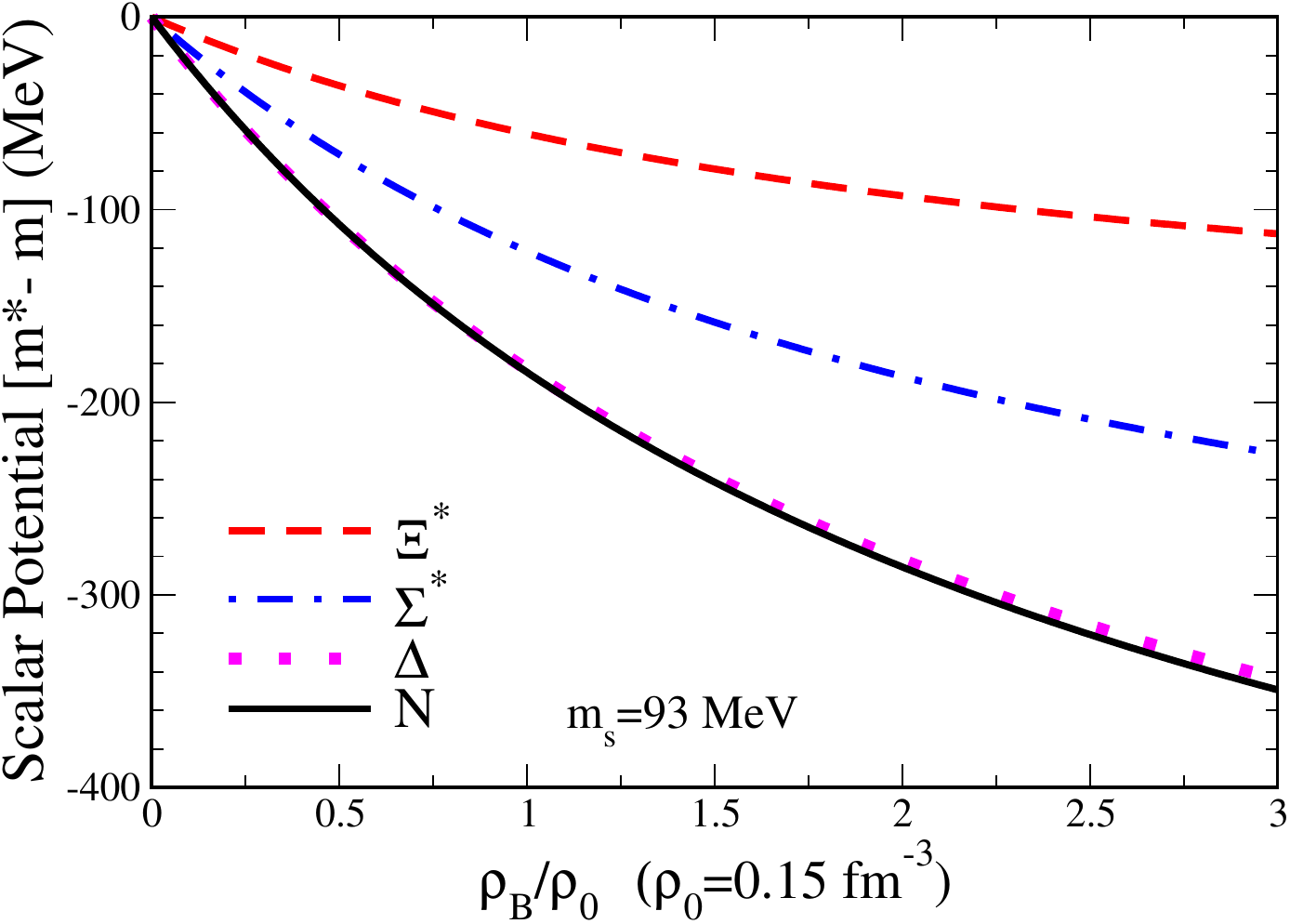}
\vspace{6ex}
\\
\includegraphics[scale=0.26]{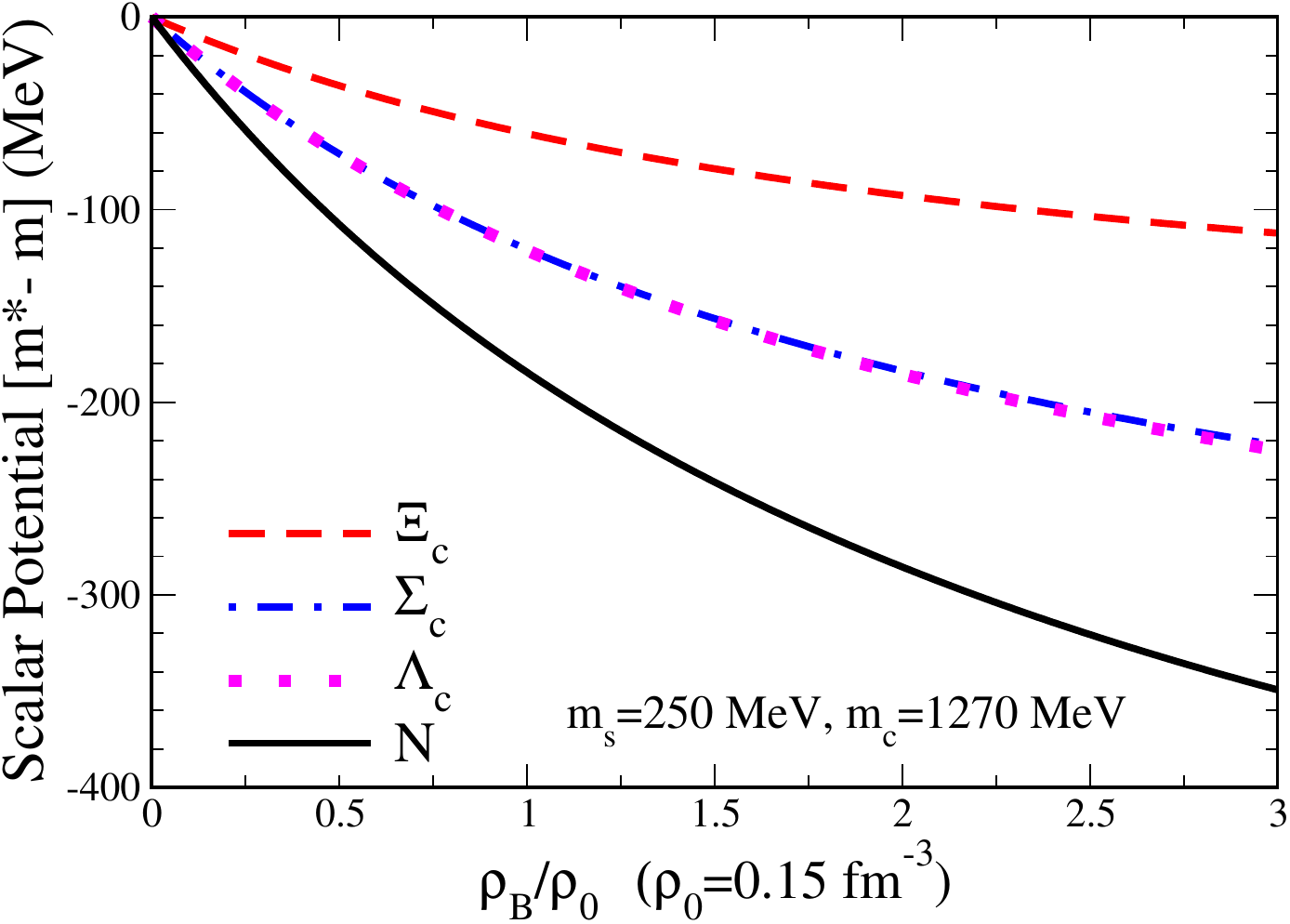}
\hspace{3ex}
\includegraphics[scale=0.26]{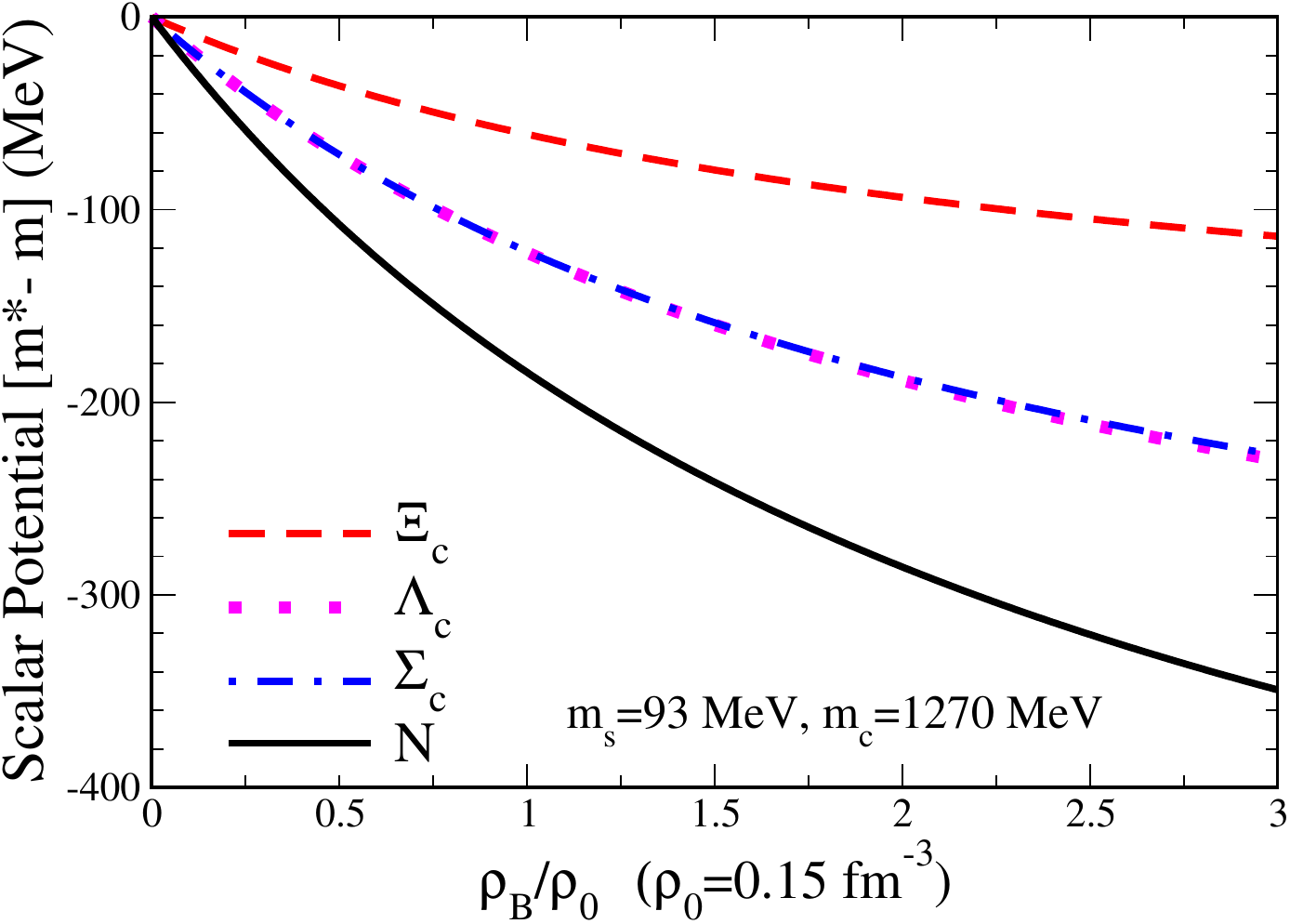}
\vspace{6ex}
\\
\includegraphics[scale=0.26]{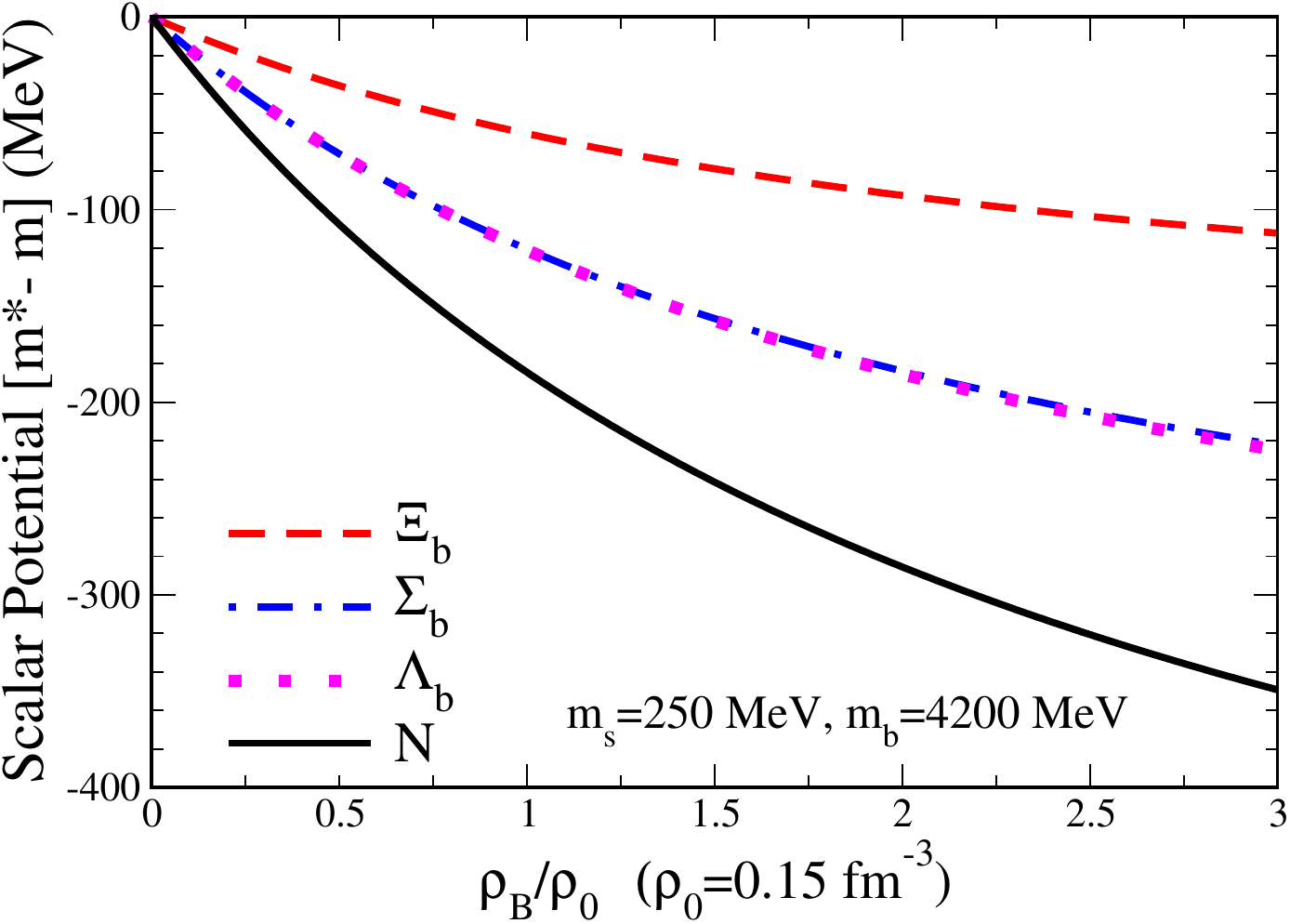}
\hspace{3ex}
\includegraphics[scale=0.26]{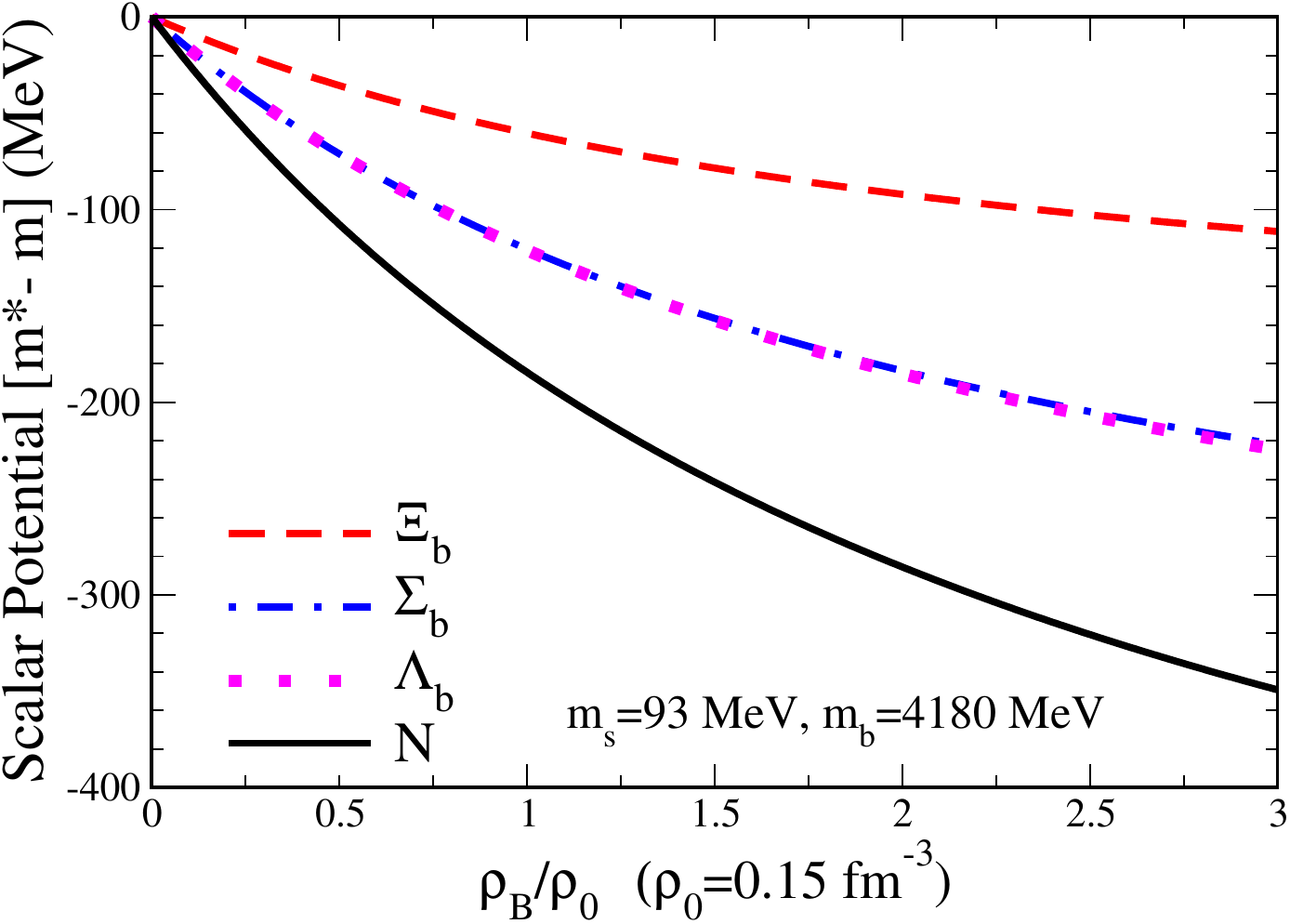}
\caption{\label{VBs} 
Density dependence of the baryon scalar potentials, $[m^* - m]$, 
for the octet, decuplet, low-lying charm, and low-lying bottom baryons, 
for the Set I (left panel) and Set II (right panel).
}
\end{center}
\vspace{2ex}
\end{figure}

\begin{figure}[htb]
\vspace{3ex}
\begin{center}
\includegraphics[scale=0.26]{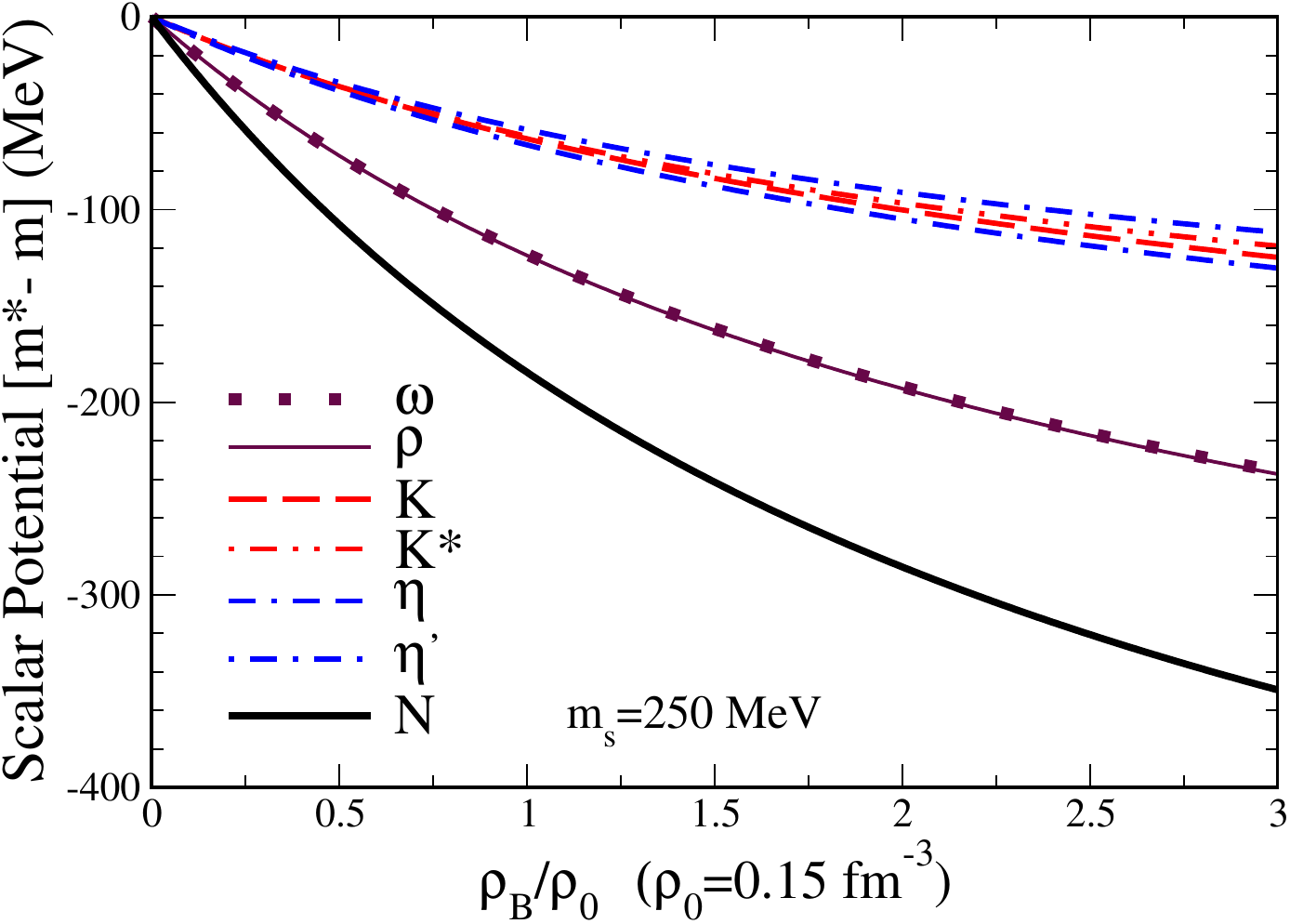}
\hspace{3ex}
\includegraphics[scale=0.26]{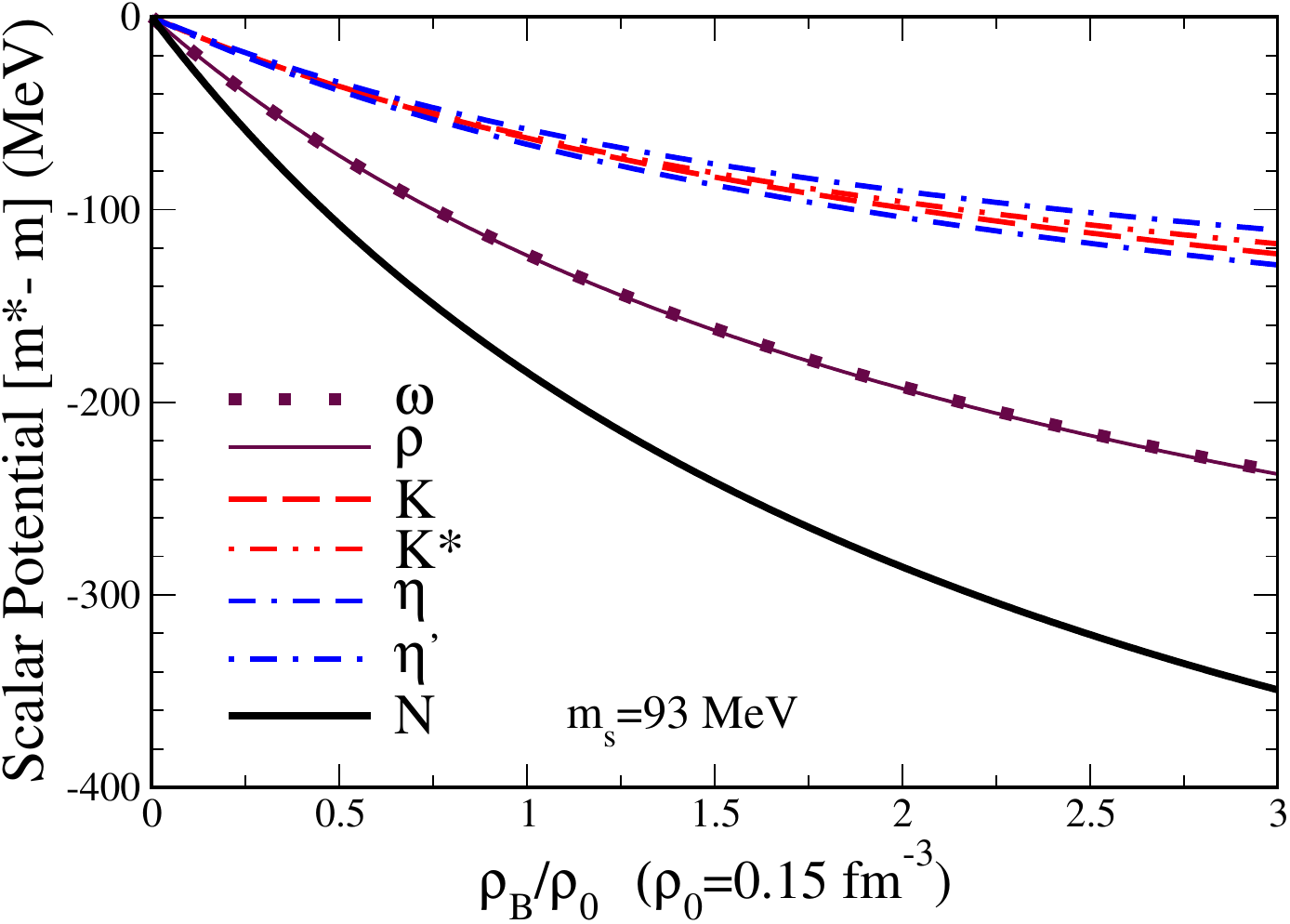}
\vspace{6ex}
\\
\includegraphics[scale=0.26]{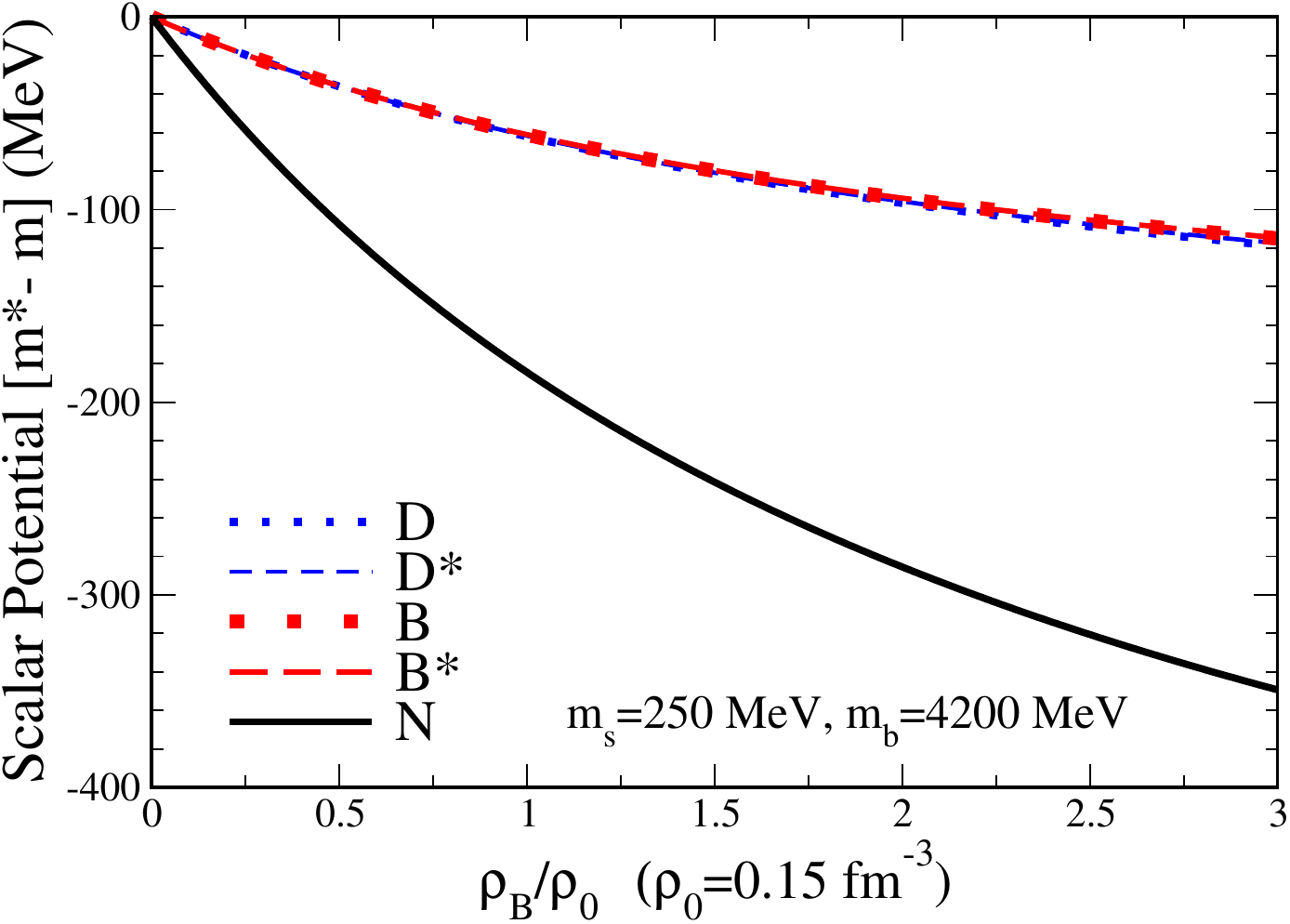}
\hspace{3ex}
\includegraphics[scale=0.26]{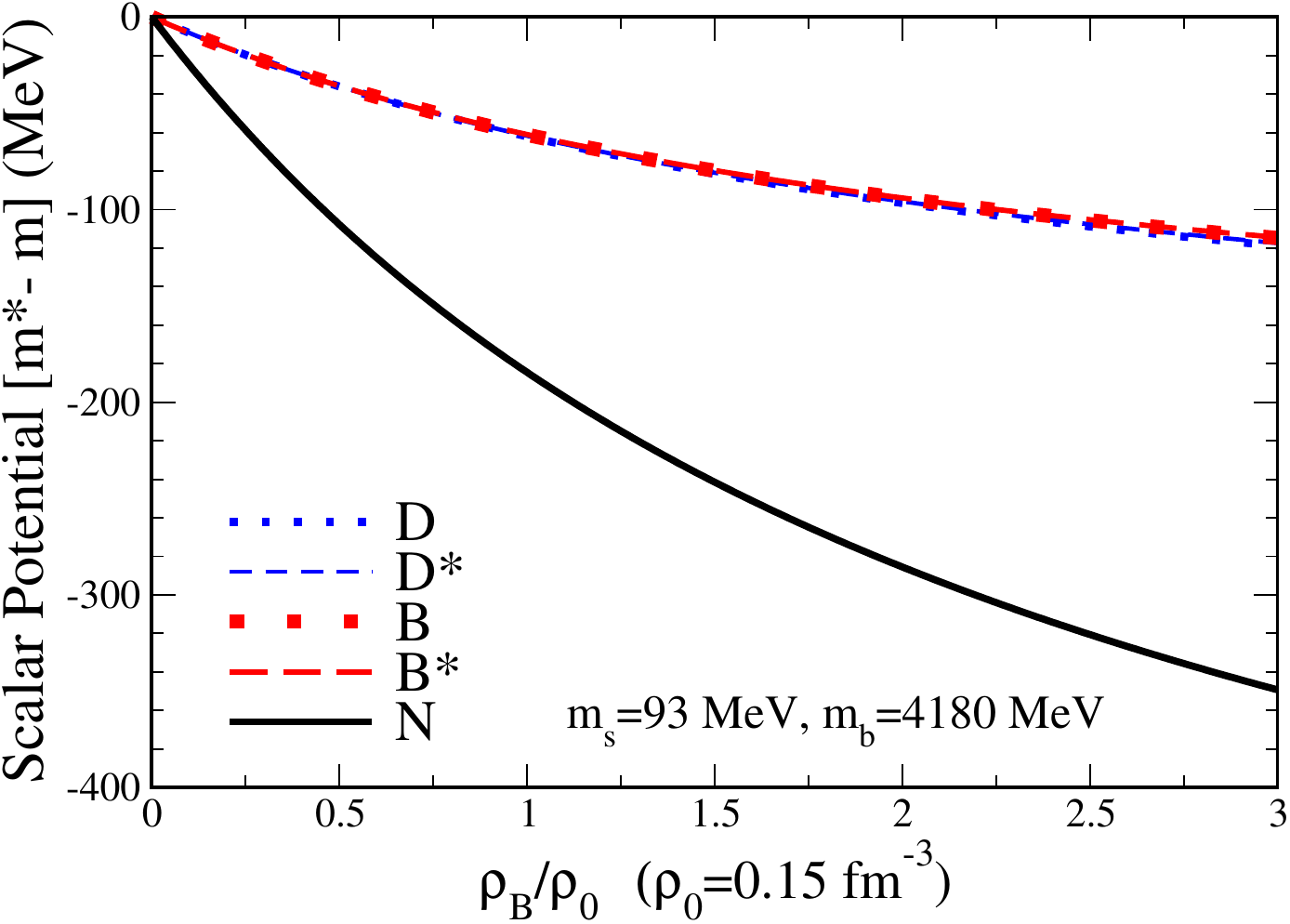}
\caption{\label{VMs} 
Density dependence of the meson scalar potentials, $[m^* - m]$, 
for the light and strange mesons (upper panel), 
and charm and bottom mesons (lower panel) for   
the Set I (left panel) and Set II (right panel). 
Note that, for the $\eta$ and $\eta'$ mesons, the pseudoscalar octet(8)-singlet(1) mixing 
angle of $\theta_P = -11.3^\circ$ from the linear mass formula~\cite{PDG} is used.
This makes the effective mass of $\eta$ ($\eta'$) lighter (heavier) than that 
of $\eta_8$ ($\eta_1$), where the possible density dependence of the mixing angle 
for $\theta_P$ is ignored.
}
\end{center}
\vspace{2ex}
\end{figure}

The calculated effective baryon and meson scalar potentials 
$[m_B^*-m_B]$ and $[m_M^*-m_M]$  are shown in Figs.~\ref{VBs} and~\ref{VMs}, 
respectively by simply denoted by $[m^*-m]$, for both  
the Set I (left panel) and Set II (right panel). 
One can notice that the ''light quark number counting rule'', namely the amount of the 
Lorentz-scalar potential is proportional to the number of light quarks in the hadron, 
is realized well for both the baryon and meson cases. 
(The $\eta$ and $\eta'$ meson cases will be discussed later.)

The ground state wave function of the quark $q$ or $Q$ in the hadron $h$ 
immersed in the nuclear medium, satisfies the boundary condition at the bag surface,
\begin{equation}
j_0 (x_{q,Q}^*) =  \beta^{h*}_{q,Q}\, j_1 (x_{q,Q}^*),
\label{boundary}
\end{equation} 
where $j_{\,0,1}$ are the spherical Bessel functions, and 
\begin{equation}
\beta_{q}^{h*} = \sqrt{\frac{\Omega_{q}^{*} - m_{q}^{*} R_h^{*}}{\Omega_{q}^{*} 
+ m_{q}^{*} R_h^{*}}}, \hspace{5ex}
\beta_{Q}^{h*} = \sqrt{\frac{\Omega_{Q}^{*} - m_{Q} R_h^{*}}{\Omega_{Q}^{*} 
+ m_{Q} R_h^{*}}}.
\label{eq:beta}
\end{equation}
The ground state quark wave functions $\psi_{q,Q}^{B\,*}(\vec{r})$ 
in a baryon $B$ in symmetric nuclear 
matter are given by replacing $h \to B$ in the above, 
\be
\psi_{q,Q}^{B\,*}(\vec{r}) 
= N_{q,Q}^{B\,*} 
\left( \begin{array}{c}
j_0(x_{q,Q}^*\, r/R^*_B) \\
i\,\beta_{q,Q}^{B*}\, {\vec \sigma}\cdot\hat{r}\, j_1(x_{q,Q}^*\, r/R^*_B)
\end{array} \right) \frac{\chi_s}{\sqrt{4\pi}}, 
\label{wf}
\ee
\noindent
with
\be
(N_{q,Q}^{B\,*})^{-2} = 2(R^*_B)^3 j_0^2(x_{q,Q}^*)
\left[\Omega_{q,Q}^* (\Omega_{q,Q}^* - 1) + m_{q,Q}^* R^*_B/2\right]/x_{q,Q}^{*\,2},
\label{norm}
\ee
where $r=|\vec{r}|, \hat{r}=\vec{r}/r, m^*_Q = m_Q$ as already mentioned, 
and $\chi_s$ is the Pauli spinor.

\subsection{Density dependent parametrizations}

\begin{figure}[htb]
\vspace{3ex}
\begin{center}
\includegraphics[scale=0.3]{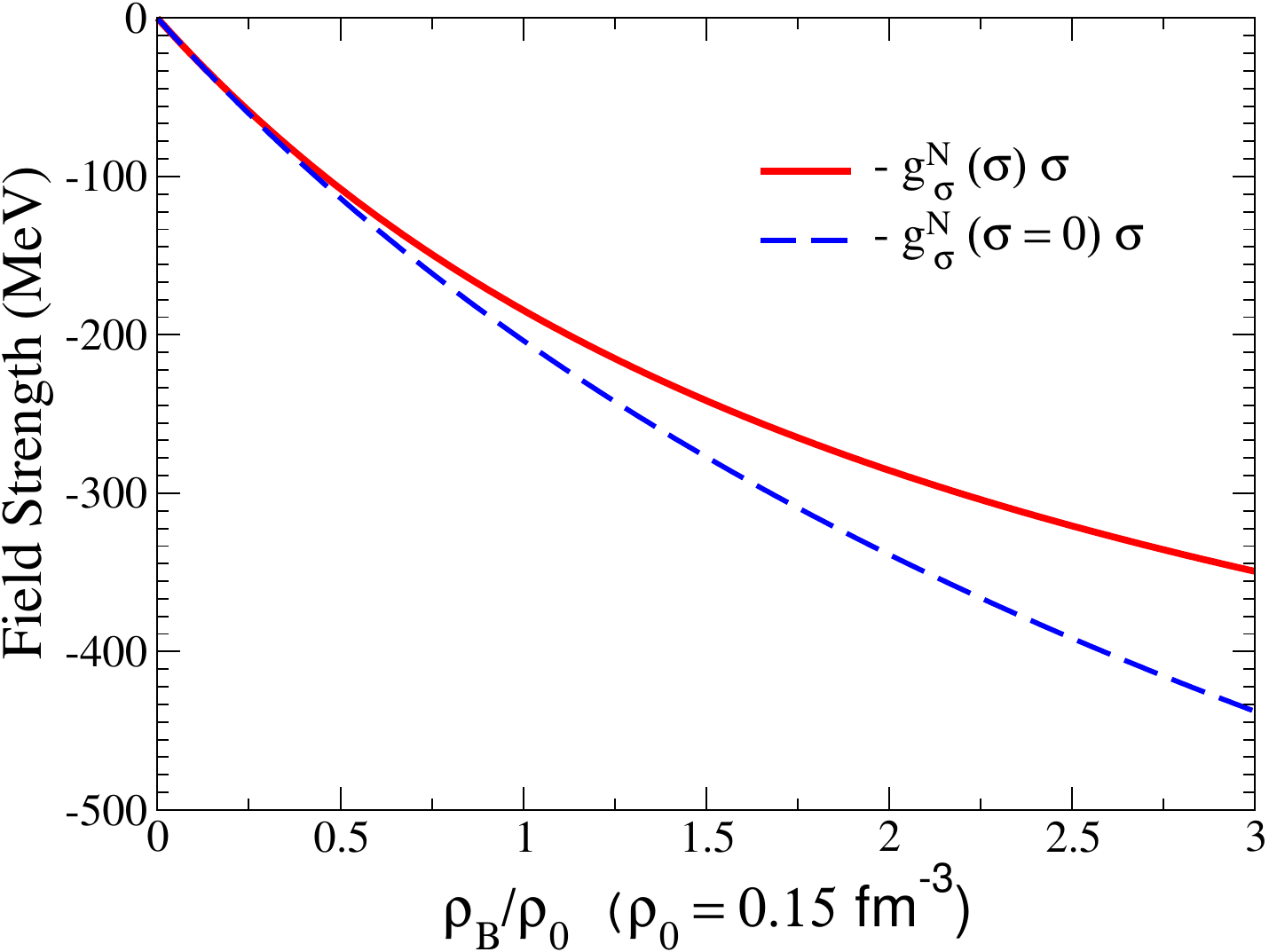}
\hspace{3ex}
\includegraphics[scale=0.3]{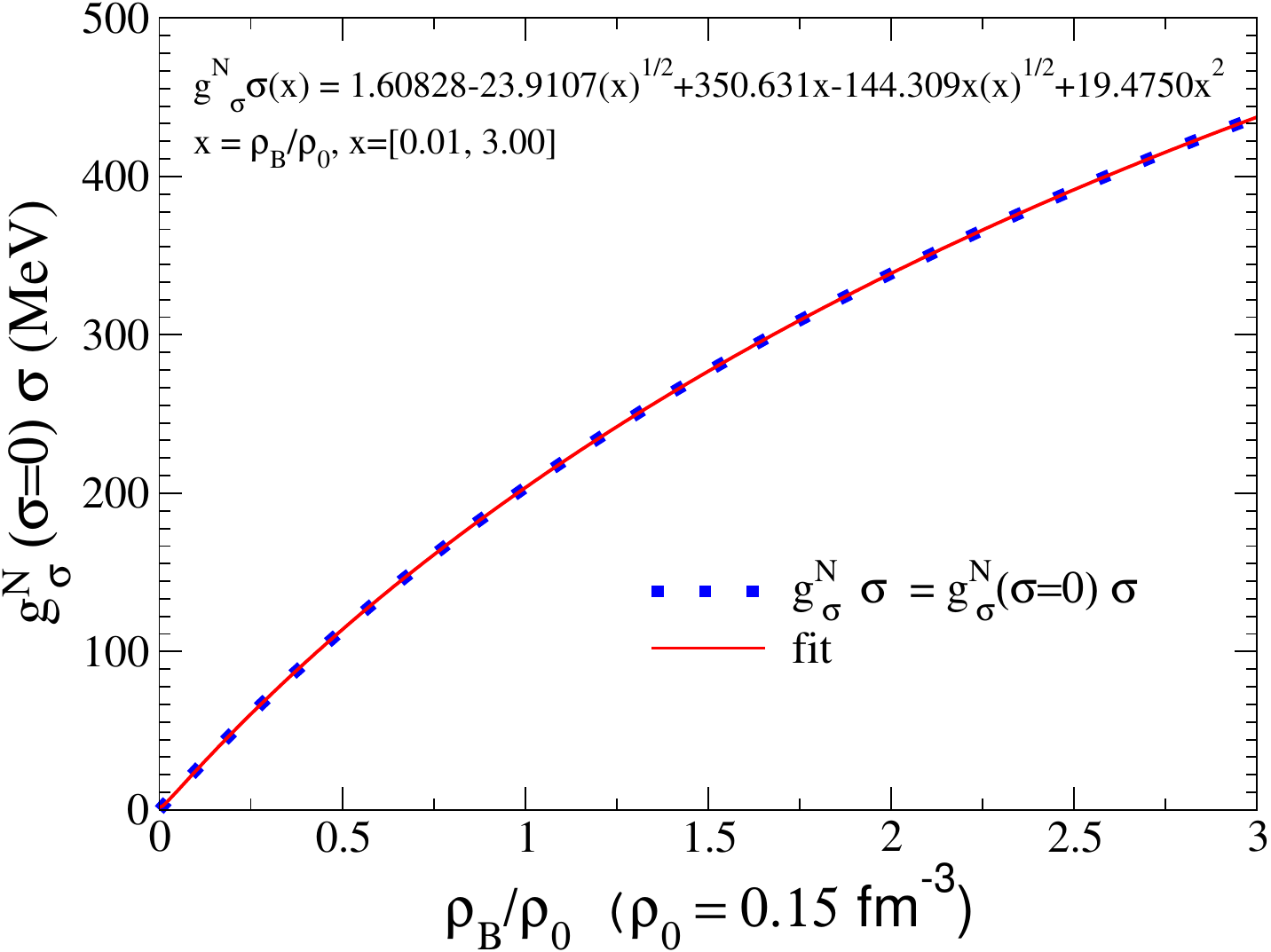}
\caption{\label{gNss} 
Density dependence of the $-g^N_\sigma (\sigma) \sigma$ 
and $-g^N_\sigma (\sigma=0)\sigma=-g^N_\sigma\sigma$ (left panel) 
and the fit result for $g^N_\sigma 
(\sigma=0)\sigma=g^N_\sigma\sigma$ (right panel).
}
\end{center}
\vspace{2ex}
\end{figure}

In connection with the Lorentz-scalar potentials 
[$m_B^* - m_B] = -g^{B=N,Y}_\sigma (\sigma)\, \sigma$ shown in Fig.~\ref{VBs} 
(or equivalently the effective baryon masses $m_B^*$),  
it has been found that the function $C_B({\sigma})\, 
(B = N,\Lambda,\Sigma,\Xi,\Delta,\Sigma^*,\Xi^*,\Lambda_c,\Sigma_c,\Xi_c,\Lambda_b,\Sigma_b,\Xi_b)$
appearing in the last line in Eq.~(\ref{Ssigma}) can be parameterized as a linear
form in the $\sigma$ field, $g^N_\sigma\sigma=g^N_\sigma(\sigma=0)\sigma$ shown in 
Fig.~\ref{gNss}, for a practical use~\cite{Guichon:1995ue,Saito:1996sf,Tsushima:1997cu}, 
\begin{equation}
C_B ({\sigma}) = 1 - a_B
\times (g^N_{\sigma} {\sigma}),\hspace{1em}
(B = N,\Lambda,\Sigma,\Xi,\Delta,\Sigma^*,\Xi^*,\Lambda_c,\Sigma_c,\Xi_c,\Lambda_b,\Sigma_b,\Xi_b), 
\label{cynsigma}
\end{equation}
where we compare with the $\sigma$ dependent coupling case, $g^N_\sigma(\sigma)\sigma$,  
and without the dependent case, $g^N_\sigma \sigma = g^N_\sigma (\sigma=0) \sigma$ 
in the left panel of Fig.~\ref{gNss}.

%
The values obtained for $a_B$ are listed in Tables~\ref{slopeI} and~\ref{slopeII} respectively 
for the Set I and Set II.
Note that, for the antibaryon $\overline{B}$, $a_{\overline{B}} = a_B$,  
and also $m^*_{\overline{B}} = m^*_B$ 
with $n_q \to n_{\overline{q}}$ below.
This parameterization works very well up to
about three times of normal nuclear matter density $3 \rho_0$.
Then, the effective mass of baryon $B$ in nuclear matter, where $m^*_B - m_B$ are  
shown in Fig.~\ref{VBs}, is well approximated by (the accuracy will be discussed in the last part of 
this subsection):
\bea
m^*_B &\simeq& m_B - \dfrac{n_q}{3} g^N_\sigma 
\left[1-\dfrac{a_B}{2}(g^N_\sigma {\sigma})\right]\sigma, 
\nonumber\\
&=& m_B - \dfrac{n_q}{3} \left[ (g^N_\sigma\sigma) - \dfrac{a_B}{2}(g^N_\sigma\sigma)^2 \right], 
\nonumber\\
&&\hspace{15ex}(B = 
N,\Lambda,\Sigma,\Xi,\Delta,\Sigma^*,\Xi^*,\Lambda_c,\Sigma_c,\Xi_c,\Lambda_b,\Sigma_b,\Xi_b),
\label{MBstar}
\eea
with $n_q$ being the valence light-quark number in the baryon $B$,  
where $g^N_\sigma\sigma=g^N_\sigma(\sigma=0)\sigma$ (in MeV) can be fitted (parametrized) 
well as a function of $x$ for $0 \le x=\rho_B/\rho_0 \le 3.0$ ($\rho_0=0.15$ fm$^{-3}$) as,
\bea
(g^N_\sigma\sigma) (x)\hspace{-1ex} &=&\hspace{-1ex} 1.60828 -23.9107\, \sqrt{x} + 350.631\, x - 
144.309\, x\sqrt{x} + 19.4750\, x^2\hspace{1ex}(x>0),  
\nonumber\\
(g^N_\sigma\sigma)(x)\hspace{-1ex} &=&\hspace{-1ex} 0 \hspace{1ex}(x=0),
\label{sigmax}
\eea
with $\chi^2/{\rm d.o.f} = 0.06653/301$, where $0.01 \le x$ is in practice, 
and the fitted result is shown in the right panel of Fig.~\ref{gNss}. Then, using 
Eq.~(\ref{MBstar}) and the $a_B$ values given 
in Tables~\ref{slopeI} and~\ref{slopeII}, one can obtain the corresponding 
effective mass values $m^*_B$ for a given baryon density $\rho_B$ (fm$^{-3}$), 
in particular, that for $N$, $m^*_N$ with $a_N=9.15\times10^{-4}$ MeV$^{-1}$ 
and $n_q = 3$ may be useful.

Furthermore, the Lorentz-vector-isoscalar $\omega$ mean field potential $V^B_\omega$ [$V^h_v$]    
(in MeV) for a baryon $B$ [hadron $h$], and 
the $\omega$ potential for the light quarks 
($q=u,d$), $V^q_\omega$ (MeV) can also be parametrized using 
$x=\rho_B/\rho_0$ ($\rho_0=0.15$ fm$^{-3}$) as, 
\bea
V^B_\omega(x) &=& b_B\, x,   
\\
V^q_\omega(x) &=& = 41.77\, x
\\
V_v^h(x) &=& V^h_\omega 
= (n_q-n_{\overline{q}}) V^q_\omega 
= (n_q-n_{\overline{q}})\times 41.77\, x\hspace{2ex} ({\rm except\, for\, the\, 
baryon\, octet}), 
\label{Vvpot}
\eea
where the values of $b_B$ are the same for both the Set I and Set II, 
but given in Tables~\ref{slopeI} and~\ref{slopeII}, respectively.
Note that, for $\Lambda, \Sigma$ and $\Xi$ hyperons, phenomenologically 
introduced quark-based ''Pauli potentials''~\cite{Tsushima:1997cu} 
are included in $b_B$ assuming the same for both the Set I and Set II. 
As for the $\Sigma$ total potential in a nonrelativistic sense to apply for the upper 
component of the Dirac spinor, $[m^*_\Sigma - m_\Sigma] + V^\Sigma_v$ at $\rho_0$,   
the above parametrizations give attractive potential of $\simeq -22$ MeV  
for both the Set I and Set II. 
If we want to agree with the ''no experimental observation of the $\Sigma$ hypernuclei'', 
we need to introduce the alternative, phenomenological parametrizations to yield 
e.g., $[m^*_\Sigma - m_\Sigma] + V^\Sigma_v \simeq + 30$ MeV 
at $\rho_0$, and we give in Tables~\ref{slopeI} and~\ref{slopeII} 
for this case by denoting $\tilde{b}_\Sigma$.

To be complete, we give also the Lorentz-vector-isovector mean field potential 
(in MeV) as a function of $y \equiv \rho_3/\rho_0=(\rho_p-\rho_n)/\rho_0$ with the 
isospin-third component of the hadron $h$, $I^h_3$, 
\be
I^h_3 V^h_\rho(y) = I^h_3 \times 84.61\, y,
\label{Vrhopot}
\ee
where, the expression given in Ref.~\cite{Saito:2005rv} wrongly contained 
a factor 1/2.

Similarly, by defining $n^M_q \equiv (n_q+n_{\overline{q}})$, 
and using $a_M = a_{\overline{M}}$, 
the effective masses of the low-lying mesons $m^*_M$ 
($M=\omega,\rho,K,K^*,\eta,\eta',D,D^*,B,B^*$) are given below, 
except for pion, which is the (nearly) Goldstone boson and difficult in describing 
consistently in naive independent/additive quark models 
(pion mass is not expected to be modified up to about normal nuclear 
matter density~\cite{Kienle,Vogl,Meissner}): 
\bea
m^*_M &\simeq& m_M - \dfrac{n^M_q}{3} g^N_\sigma 
\left[1-\dfrac{a_M}{2}(g^N_\sigma {\sigma})\right]\sigma, 
\nonumber\\
&=& m_M - \dfrac{n^M_q}{3} \left[ (g^N_\sigma\sigma) - 
\dfrac{a_M}{2}(g^N_\sigma\sigma)^2 \right]  
\nonumber\\
&&\hspace{8ex}(M=\omega,\rho,K,K^*,\eta,\eta',D,D^*,B,B^*,\hspace{1ex} 
{\rm with}\hspace{1ex} n^M_q\to1\, {\rm for}\, \eta\, {\rm and}\, \eta').
\label{MMstar}
\eea
In the above, $n^M_q = n^{\eta,\eta'}_q \to 1$ 
for the $\eta$ and $\eta'$ mesons may be verified from 
Fig.~\ref{VMs}, where the pseudoscalar octet(8)-singlet(1) mixing 
angle of $\theta_P = -11.3^\circ$ from the linear mass formula~\cite{PDG} is used.
This makes the effective mass of $\eta$ ($\eta'$) lighter (heavier) than that 
of $\eta_8$ ($\eta_1$), neglecting the possible density dependence of the mixing angle 
$\theta_P$, as also shown explicitly in Ref.~\cite{Tsushima:1998qp,Tsushima:1998qw,Saito:2005rv} 
by the effective mass ratios. 
The $n^{\eta,\eta'}_q \to 1$ for the $\eta$ and $\eta'$ 
reflects the fact that the numerator of the total 
energy for the $\eta$ and 
$\eta'$ in Eq.~(\ref{hmass}) becomes nearly the sum as $\simeq (\Omega_q^* + \Omega_s^*)$  
when the flavor octet and singlet wave functions are used with     
the mixing effect of $\theta_P = -11.3^\circ$,  
which is different from the $\phi$-$\omega$ ideal-mixing case.
Note that the $\omega$ and $\rho$ mesons appearing above are the those of the SU(6) quark model, 
and should not be confused with the $\omega$ and $\rho$ (mean) fields in the QMC model.
The obtained slope parameters $a_M$ are listed in 
Tables~\ref{slopeI} and~\ref{slopeII} respectively 
for the Set I and Set II. 
Note that the $\omega$ mean filed potential for the kaon ($K$) in the QMC model 
is necessary to be modified as,
\be
V^K_\omega (x) \simeq 1.96 \times  41.77\, x,  
\ee
to reproduce the empirically extracted repulsive 
$K^+$ total potential~\cite{Tsushima:1997df} for both the Set I and Set II.
This gives the total $K^+$ potential of $\simeq + 19$ MeV at $\rho_0$ for the both sets.

We comment briefly the accuracy of the parametrizations given above.
Since the vector potentials proportional to the baryon density $\rho_B$ or $\rho_3=\rho_p-\rho_n$  
the parametrizations are simple and should be good. For the effective masses of the 
baryons and mesons, the quality of the parametrizations with $a_B$, $a_M$ and 
$(g^N_\sigma\sigma)(x)$ of Eq.~(\ref{sigmax}) are all well within the 1.0\% deviations from the 
calculated results for $0.01 \le \rho_B/\rho_0 \le 3.0$, except that 
the $m^*_\eta$ has a maximum of 1.7\% 
deviation ($\simeq 7$ MeV) from the original result.  
Thus, for the practical purposes, one can comfortably use the given parametrizations 
for the effective masses (Lorentz-scalar-isoscalar potentials) 
of the baryons and mesons for a given 
baryon density for $0.01 \le \rho_B/\rho_0 \le 3.0$, as well as for the Lorentz-vector-isoscalar   
and Lorentz-vector-isovector mean field potentials.
Recall that the $\eta$ and $\eta'$ cases are subject to the mixing of the 
octet and the singlet states, and also $n_q \to 1$ is applied for them.
However, the observed maximum deviation of 1.7\% from the parametrizations for them  
are surprisingly good.

%
\begin{table}[htb]
\begin{center}
\caption{(Set I) Effective mass slope parameter $a_B$ [Eq.~(\ref{MBstar})],  
and the vector potential parameter $b_B$ [Eq.~(\ref{Vvpot})], 
for $B = N,\Lambda,\Sigma,\Xi,\Delta, \Sigma^*, 
\Xi^*, \Lambda_c,\Sigma_c,\Xi_c,\Lambda_b,\Sigma_b,\Xi_b$, and 
the effective mass slope parameter $a_M$ [Eq.~(\ref{MMstar})] for 
$M = \omega,\rho,K,K^*,\eta,\eta',D,D^*,B,B^*$ 
Note that the tiny differences in values of $a_B$ from those in  
Refs.~\cite{Saito:2005rv,Krein:2017usp}, are due to the differences 
in the number of data points for calculating $a_B$, 
but such differences in $a_B$ give negligible effects.   
Concerning the $\Sigma$ vector potential, the alternative parametrization 
$\tilde{b}_\Sigma$ yields the total potential 
of $[m^*_\Sigma - m_\Sigma] + V^\Sigma_v \simeq +30$ MeV at $\rho_0$.
}
\label{slopeI}
\hspace*{-7ex}
\begin{tabular}[t]{c|c||c|c||c|c||c|c}
\hline
\hline
$a_B$ &$\times 10^{-4}$ MeV$^{-1}$ &$a_B$ &$\times 10^{-4}$ MeV$^{-1}$ 
&$a_B$ &$\times 10^{-4}$ MeV$^{-1}$ &$a_B$ &$\times 10^{-4}$ MeV$^{-1}$\\
\hline
$a_N$         &9.15  &$a_\Delta$      &10.08   &--- &---                &--- &--- \\
$a_\Lambda$   &9.35  &---             &---     &$a_{\Lambda_c}$ &9.90   &$a_{\Lambda_b}$ &10.78 \\
$a_\Sigma$    &9.59  &$a_{\Sigma^*}$  &10.15   &$a_{\Sigma_c}$  &10.34  &$a_{\Sigma_b}$  &11.22 \\
$a_{\Xi}$     &9.52  &$a_{\Xi^*}$     &10.15   &$a_{\Xi_c}$     &9.99   &$a_{\Xi_b}$     &10.83 \\
\hline
\hline
$b_B$ &MeV &$b_B$ &MeV &$b_B$ &MeV &$b_B$ &MeV\\
\hline
$b_N$         &125.30  &$b_\Delta$      &125.30  &--- &---                &--- &--- \\
$b_\Lambda$   &92.57   &---             &---     &$b_{\Lambda_c}$ &83.54    &$b_{\Lambda_b}$ &83.54 
\\
$b_\Sigma$    &100.12  &$b_{\Sigma^*}$  &83.54   &$b_{\Sigma_c}$  &83.54   &$b_{\Sigma_b}$  &83.54 
\\
$\tilde{b}_\Sigma$    &152.42 & & & & & &  
\\
$b_{\Xi}$     &46.29   &$b_{\Xi^*}$     &41.77   &$b_{\Xi_c}$     &41.77   &$b_{\Xi_b}$     &41.77 
\\
\hline
\hline
$a_M$ &$\times 10^{-4}$ MeV$^{-1}$ &$a_M$ &$\times 10^{-4}$ MeV$^{-1}$ 
&$a_M$ &$\times 10^{-4}$ MeV$^{-1}$ &$a_M$ &$\times 10^{-4}$ MeV$^{-1}$\\
\hline
$a_\omega$ &8.73  &$a_K$              &6.66   &$a_D$     &8.61   &$a_B$      &9.92 \\
$a_\rho$   &8.70  &$a_{K^*}$          &8.60   &$a_{D^*}$ &9.09   &$a_{B^*}$  &10.04\\
---  &---         &$a_\eta(n^\eta_q\to 1)$    &7.03   &---       &---   &---         &--- \\
---  &---         &$a_{\eta'}(n^{\eta'}_q\to 1)$ &8.81   &---       &---   &---         &--- \\
\hline 
\hline
\end{tabular}
\end{center}
\end{table}
%

%
\begin{table}[htb]
\begin{center}
\caption{(Set II) See the caption of Table~\ref{slopeI}, but for the parameters Set II.
}
\label{slopeII}
\hspace*{-7ex}
\begin{tabular}[t]{c|c||c|c||c|c||c|c}
\hline
\hline
$a_B$ &$\times 10^{-4}$ MeV$^{-1}$ &$a_B$ &$\times 10^{-4}$ MeV$^{-1}$ 
&$a_B$ &$\times 10^{-4}$ MeV$^{-1}$ &$a_B$ &$\times 10^{-4}$ MeV$^{-1}$\\
\hline
$a_N$         &9.15   &$a_\Delta$      &10.08   &--- &---                &--- &--- \\
$a_\Lambda$   &9.68   &---             &---     &$a_{\Lambda_c}$ &9.90    &$a_{\Lambda_b}$ &10.82 \\
$a_\Sigma$  &9.91   &$a_{\Sigma^*}$  &10.44   &$a_{\Sigma_c}$  &10.34   &$a_{\Sigma_b}$  &11.27 \\
$a_\Xi$     &10.15  &$a_{\Xi^*}$     &10.71   &$a_{\Xi_c}$     &10.28   &$a_{\Xi_b}$     &11.13 \\
\hline
\hline
$b_B$ &MeV &$b_B$ &MeV &$b_B$ &MeV &$b_B$ &MeV\\
\hline
$b_N$         &125.30  &$b_\Delta$      &125.30  &--- &---                &--- &--- \\
$b_\Lambda$   &92.57   &---             &---     &$b_{\Lambda_c}$ &83.54    &$b_{\Lambda_b}$ &83.54 
\\
$b_\Sigma$  &100.12  &$b_{\Sigma^*}$  &83.54   &$b_{\Sigma_c}$  &83.54   &$b_{\Sigma_b}$  &83.54 
\\
$\tilde{b}_\Sigma$    &152.00 & & & & & &
\\
$b_\Xi$     &46.29   &$b_{\Xi^*}$     &41.77   &$b_{\Xi_c}$     &41.77   &$b_{\Xi_b}$     &41.77 
\\
\hline
\hline
$a_M$ &$\times 10^{-4}$ MeV$^{-1}$ &$a_M$ &$\times 10^{-4}$ MeV$^{-1}$ 
&$a_M$ &$\times 10^{-4}$ MeV$^{-1}$ &$a_M$ &$\times 10^{-4}$ MeV$^{-1}$\\
\hline
$a_\omega$ &8.73  &$a_K$              &7.24   &$a_D$     &8.61   &$a_B$      &9.97 \\
$a_\rho$   &8.70  &$a_{K^*}$          &8.99   &$a_{D^*}$ &9.09   &$a_{B^*}$  &10.10\\
---  &---         &$a_\eta(n^\eta_q\to 1)$    &7.54   &---       &---   &---         &--- \\
---  &---         &$a_{\eta'}(n^{\eta'}_q\to 1)$ &9.20   &---       &---   &---         &--- \\
\hline 
\hline
\end{tabular}
\end{center}
\end{table}

\section{Baryon magnetic moments in symmetric nuclear matter}

In Ref.~\cite{Tsushima:2018goq} we obtained the MIT bag model wave functions 
in symmetric nuclear matter for the octet, low-lying charm, 
and low-lying bottom baryons with nonzero light quarks in the Set I current quark mass 
values (Table~\ref{coupcc1}). 
In this study we extend further to include the decuplet baryons for the Set I 
as well as for the Set II, and calculate the wave functions of the octet, 
decuplet, low-lying charm, and low-lying bottom baryons with nonzero light quarks 
also for the Set II (Table~\ref{coupcc2}).
Below, we calculate the magnetic moments of these baryons and some transition 
magnetic moments in symmetric nuclear matter using the MIT bag (QMC model) wave functions 
in the Set I (Table~\ref{coupcc1}) and Set II (Table~\ref{coupcc2}).

First, we discuss the magnetic moment of an octet baryon $B$, $\mu_B$, in free space (vacuum).
For the octet baryons $B(q_1,q_2,q_3)$ specifying by this the quark order, 
the familiar SU(6) flavor-spin wave functions are constructed based on the isospin, 
spin, and Pauli principle, and consistent with    
the ''1-2 quark order'' for the quarks $q_1$ and $q_2$, 
namely, the first two quarks $q_1$ and $q_2$ are the closest 
in mass~\cite{Franklin:1981rc}.
Good examples may be the wave functions of $\Lambda$ 
and $\Sigma^0$ baryons. 
The first two quarks $(q_1,q_2) = (u,d)$,   
are antisymmetric in the $\Lambda$, while symmetric in the $\Sigma^0$.
Similar arguments may not necessarily be applicable for 
the baryons with $c$ and/or $b$ quarks with nonzero light quarks, 
in particular, for the baryons such as $B(q_1,q_2,q_3)\, (q_1=q, q_2=Q \ne q_3=Q')$, 
since isospin symmetry and Pauli principle cannot help.
As discussed in Ref.~\cite{Franklin:1981rc}, 
different assignments for the quarks $q_1, q_2$ and $q_3$ 
in $B(q_1,q_2,q_3)\, (q_{1,2,3}=q,Q)$ are possible in some cases  
without violating the Pauli principle,  
and the different assignments give different results 
for the calculated magnetic moments.
This is concerned for the $\Xi_{c,b}$ baryons in the present study.
Although the first two quark pairs $(u,s)$ or $(d,s)$ are antisymmetric in the low-lying 
$\Xi_{c,b}$ baryons~\cite{Yu:2018yxl}, 
these cases agree with the assumption of the ''1-2 quark order'' for these heavy 
baryons---as a natural assumption, the same as that for the octet baryons. 
Despite some discussions were made in Ref.~\cite{Franklin:1981rc} for the quark order, 
we take the quarks $q_1$ and $q_2$ are the closest in mass in the wave functions.
This cases, the $\Xi_{c,b}$ wave functions have similar structure with 
those of the $\Lambda_{c,b}$ baryons. 
Note that, the ''1-2 quark order'' is supported in Ref.~\cite{Franklin:1981rc} 
as the best quark ordering for flavor-degenerate baryons for the masses.

The magnetic moment $\mu_B$ of the baryon $B(q_1,q_2,q_3)$  
in an impulse approximation (independent quark picture), 
is expressed in terms of the $j$-th quark magnetic moments $\mu_j\hspace{1ex}(j=1,2,3)$:  
\bea
\mu_B &=& \frac{1}{3} \left( 2\mu_1 + 2\mu_2 - \mu_3 \right) \hspace{3ex} 
(B \ne \Lambda, \Lambda_{c,b} \ne {\rm decuplet}), 
\label{muB}
\\
\mu_B &=& \mu_3 \hspace{3ex} (B = \Lambda, \Lambda_{c,b}, \Xi_{c,b}), 
\label{mu3}
\\
\mu_B &=& \mu_1 + \mu_2 + \mu_3 \hspace{3ex} (B = {\rm decuplet})
\eea
where we ignore any possible opposite parity state mixing, isospin mixing, 
and flavor mixing. Note that in Eq.~(\ref{mu3}) the quark order $q_1$ and $q_2$ 
is implied in such a way that they are coupled to isospin-0 and spin-0 pair, 
and the notation $B(q_1,q_2,q_3)$ indicates this (in Tables~\ref{Bmag1} and~\ref{Bmag2}). 
We give explicit expressions for the baryon magnetic 
moments and the transition magnetic moments in the second columns  
in Tables~\ref{Bmag1} and~\ref{Bmag2}, respectively for the Set I and Set II.
\begin{table}[htb]
\begin{center}
\caption{
Magnetic moments and transition magnetic moments in free space 
and in symmetric nuclear matter calculated by the Set I (Table~\ref{coupcc1}).
The free space results are given in the third column in nuclear magneton, $e/2m_N$ 
($e$ the positron charge and $m_N$ the free nucleon mass, 939 MeV), with experimental data
from Ref.~\cite{PDG} when exist (fourth column),   
while the in-medium to free space ratios,  
$\mu^*_B(\rho_B)/\mu_B$ ($\mu_B \equiv 
\mu_B(\rho_B=0)$) and $|\mu^*_{BB'}(\rho_B)/\mu_{BB'}|$ ($\mu_{BB'} \equiv \mu_{BB'}(\rho_B=0)$),  
for $\rho_B = (\rho_0, 2\rho_0, 3\rho_0)$ with $\rho_0 = 0.15$ fm$^{-3}$, 
are given in the (fifth, sixth, seventh) column.
The central values and errors for the free nucleon magnetic 
moments are shown by rounded inside the Table, but the correct values are shown in 
the footnote of the Table.
The expressions in the second column are calculated using the     
flavor-spin wave functions with the ''1-2 quark order'' for the baryon $B(q_1,q_2,q_3)$,  
namely, the quarks $q_1$ and $q_2$ are taken to be the closest in 
mass~\cite{Franklin:1981rc}, which is concerned for the $\Xi_{c,b}$ baryons below.
}
\label{Bmag1}
\hspace*{-9ex}
{\small
\begin{tabular}{l|c|cc||ccc}
\hline
\hline
$B(q_1,q_2,q_3)$ &$\mu_B$ (expression) &$\mu_B$ &Expt.  
&$\mu_B^*(\rho_0)/\mu_B$ &$\mu_B^*(2\rho_0)/\mu_B$ &$\mu_B^*(3\rho_0)/\mu_B$\\
\hline
$p(uud)$             &$(4\mu_u-\mu_d)/3$          &1.535  &2.793$\pm$0.000$^{1)}$ 
&1.077  &1.103 &1.111\\
$n(ddu)$             &$(4\mu_d-\mu_u)/3$          &-1.023 &-1.913$\pm$0.000$^{2)}$
&1.077  &1.103 &1.111\\
\hline
$\Lambda (uds)$      &$\mu_s$                     &-0.429 &-0.613$\pm$0.004&0.997  &0.991 &0.985\\
$\Sigma^+(uus)$      &$(4\mu_u-\mu_s)/3$          &1.557  &2.458$\pm$0.010 &1.086  &1.133 &1.162\\
$\Sigma^0(uds)$      &$(2\mu_u+2\mu_d-\mu_s)/3$   &0.499  &&1.067  &1.102 &1.123\\
$\Sigma^-(dds)$      &$(4\mu_d-\mu_s)/3$          &-0.560 &-1.160$\pm$0.025 &1.121  &1.189 &1.231\\
$\Xi^0(ssu)$         &$(4\mu_s-\mu_u)/3$          &-0.929 &-1.250$\pm$0.014 &1.035  &1.055 &1.067\\
$\Xi^-(ssd)$         &$(4\mu_s-\mu_d)/3$          &-0.405 &-0.6507$\pm$0.0025 
&0.956  &0.927 &0.907\\
\hline
$\Delta^{++}(uuu)$   &$3\mu_u$            &3.341  &3.7 - 9.8         &1.099 &1.151 &1.181\\
$\Delta^+   (uud)$   &$2\mu_u+\mu_d$      &1.671  &$2.7^{+1.0}_{-1.3}$ $^{3)}$ 
&1.099 &1.151 &1.181\\
$\Delta^0   (udd)$   &$2\mu_d+\mu_u$       &0      &(SU(2) symmetry)     &--- &--- &---\\
$\Delta^-   (ddd)$   &$3 \mu_d$            &-1.671 &                 &1.099 &1.151 &1.181\\
$\Sigma^{*+}(uus)$   &$2\mu_u+\mu_s$       &1.781  &                 &1.128 &1.201 &1.246\\
$\Sigma^{*0}(uds)$   &$\mu_u+\mu_d+\mu_s$  &0.102  &                 &1.571 &1.906 &2.120\\
$\Sigma^{*-}(dds)$   &$2\mu_d+\mu_s$       &-1.577 &                 &1.071 &1.110 &1.133\\
$\Xi^{*0}(ssu)$      &$2\mu_s+\mu_u$       &0.203  &                 &1.576 &1.924 &2.154\\
$\Xi^{*-}(ssd)$      &$2\mu_s+\mu_d$       &-1.473 &                 &1.038 &1.060 &1.073\\
\hline
$\Lambda_c^+(udc)$      &$\mu_c$                     &0.423  &&0.999 &0.998 &0.996\\
$\Sigma_c^{++}(uuc)$    &$(4\mu_u-\mu_c)/3$          &1.378  &&1.115 &1.179 &1.219\\
$\Sigma_c^+(udc)$       &$(2\mu_u+2\mu_d-\mu_c)/3$   &0.238  &&1.166 &1.261 &1.319\\
$\Sigma_c^0(ddc)$       &$(4\mu_d-\mu_c)/3$          &-0.903 &&1.087 &1.136 &1.167\\
$\Xi_c^+(usc)$          &$\mu_c$                     &0.424  &&1.000 &0.999 &0.998\\
$\Xi_c^0(dsc)$          &$\mu_c$                     &0.424  &&1.000 &0.999 &0.998\\
\hline
$\Lambda_b^0(udb)$      &$\mu_b$                     &-0.073 &&1.000 &1.000 &1.000\\
$\Sigma_b^+(uub)$       &$(4\mu_u-\mu_b)/3$          &1.675  &&1.111 &1.175 &1.214\\
$\Sigma_b^0(udb)$       &$(2\mu_u+2\mu_d-\mu_b)/3$   &0.437  &&1.107 &1.167 &1.205\\
$\Sigma_b^-(ddb)$       &$(4\mu_d-\mu_b)/3$          &-0.801 &&1.117 &1.183 &1.224\\
$\Xi_b^0(usb)$          &$\mu_b$                     &-0.073 &&1.000 &1.000 &1.000\\
$\Xi_b^-(dsb)$          &$\mu_b$                     &-0.073 &&1.000 &1.000 &1.000\\
\hline
\hline
Transition    &$|\mu_{B B'}|$ (expression) &$|\mu_{B B'}|$ & 
&$\mu_{B B'}^*(\rho_0)/\mu_{B B'}$ 
&$\mu_{B B'}^*(2\rho_0)/\mu_{BB'}$ 
&$\mu_{BB'}^*(3\rho_0)/\mu_{BB'}$\\
\hline
$\Sigma^0 \to \Lambda$           &$|(\mu_u-\mu_d)/\sqrt{3}|$ &0.868 &1.61$\pm$0.08 
&1.085 &1.129 &1.154\\
$\Sigma_c^+ \to \Lambda_c^+$     &$|(\mu_u-\mu_d)/\sqrt{3}|$ &0.899 &&1.086 &1.128 &1.151\\
$\Sigma_b^0 \to \Lambda_b^0$     &$|(\mu_u-\mu_d)/\sqrt{3}|$ &0.983 &&1.095 &1.143 &1.169\\
\hline
\hline  
\end{tabular}
} 
\end{center}
{\small
$^{1)} \mu_p^{\rm Expt.} = 2.7928473446 \pm 0.0000000008$.
\hspace{2ex}
$^{2)} \mu_n^{\rm Expt.} = -1.9130427 \pm 0.0000005$. 
\\
$^{3)}$ Theoretical uncertainties are not included~\cite{PDG}.
\hfill}
\end{table}
%

\begin{table}[htb]
\begin{center}
\caption{
See the caption of Table~\ref{Bmag1}, but the results are calculated with the 
different current quark mass values for the strange and bottom quarks by the Set II 
(Table~\ref{coupcc2}).
}
\label{Bmag2}
\hspace*{-9ex}
{\small
\begin{tabular}{l|c|cc||ccc}
\hline
\hline
$B(q_1,q_2,q_3)$ &$\mu_B$ (expression) &$\mu_B$ &Expt.  
&$\mu_B^*(\rho_0)/\mu_B$ &$\mu_B^*(2\rho_0)/\mu_B$ &$\mu_B^*(3\rho_0)/\mu_B$\\
\hline
$p(uud)$             &$(4\mu_u-\mu_d)/3$          &1.535  &2.793$\pm$0.000$^{1)}$
&1.077  &1.103 &1.111\\
$n(ddu)$             &$(4\mu_d-\mu_u)/3$          &-1.023 &-1.913$\pm$0.000$^{2)}$
&1.077  &1.103 &1.111\\
\hline
$\Lambda (uds)$      &$\mu_s$                     &-0.500 &-0.613$\pm$0.004 &0.996  &0.990 &0.983\\
$\Sigma^+(uus)$      &$(4\mu_u-\mu_s)/3$          &1.628  &2.458$\pm$0.010  &1.088  &1.137 &1.166\\
$\Sigma^0(uds)$      &$(2\mu_u+2\mu_d-\mu_s)/3$   &0.535  &                 &1.066  &1.103 &1.123\\
$\Sigma^-(dds)$      &$(4\mu_d-\mu_s)/3$          &-0.559 &-1.160$\pm$0.025 &1.113  &1.204 &1.250\\
$\Xi^0(ssu)$         &$(4\mu_s-\mu_u)/3$          &-1.068 &-1.250$\pm$0.014 &1.035  &1.054 &1.066\\
$\Xi^-(ssd)$         &$(4\mu_s-\mu_d)/3$          &-0.509 &-0.6507$\pm$0.0025 
&0.960  &0.934 &0.915\\
\hline
$\Delta^{++}(uuu)$   &$3\mu_u$            &3.341  &3.7 - 9.8   &1.099 &1.151 &1.181\\
$\Delta^+   (uud)$   &$2\mu_u+\mu_d$      &1.671  &$2.7^{+1.0}_{-1.3}$ $^{3)}$ 
&1.099 &1.151 &1.181\\
$\Delta^0   (udd)$   &$2\mu_d+\mu_u$       &0      &(SU(2) symmetry)     &--- &--- &---\\
$\Delta^-   (ddd)$   &$3\mu_d$             &-1.671 &           &1.099 &1.151 &1.181\\
$\Sigma^{*+}(uus)$   &$2\mu_u+\mu_s$       &1.767  &           &1.137 &1.216 &1.265\\
$\Sigma^{*0}(uds)$   &$\mu_u+\mu_d+\mu_s$  &0.040  &           &2.558 &3.486 &4.085\\
$\Sigma^{*-}(dds)$   &$2\mu_d+\mu_s$       &-1.687 &           &1.070 &1.109 &1.132\\
$\Xi^{*0}(ssu)$      &$2\mu_s+\mu_u$       &0.083  &           &2.566 &3.515 &4.139\\
$\Xi^{*-}(ssd)$      &$2\mu_s+\mu_d$       &-1.686 &           &1.037 &1.058 &1.071\\
\hline
$\Lambda_c^+(udc)$      &$\mu_c$                     &0.423  &&0.999 &0.998 &0.996\\
$\Sigma_c^{++}(uuc)$    &$(4\mu_u-\mu_c)/3$          &1.378  &&1.115 &1.179 &1.219\\
$\Sigma_c^+(udc)$       &$(2\mu_u+2\mu_d-\mu_c)/3$   &0.238  &&1.166 &1.261 &1.319\\
$\Sigma_c^0(ddc)$       &$(4\mu_d-\mu_c)/3$          &-0.903 &&1.087 &1.136 &1.167\\
$\Xi_c^+(usc)$          &$\mu_c$                     &0.426  &&1.000 &0.999 &0.998\\
$\Xi_c^0(dsc)$          &$\mu_c$                     &0.426  &&1.000 &0.999 &0.998\\
\hline
$\Lambda_b^0(udb)$      &$\mu_b$                     &-0.074 &&1.000 &1.000 &1.000\\
$\Sigma_b^+(uub)$       &$(4\mu_u-\mu_b)/3$          &1.681  &&1.112 &1.175 &1.215\\
$\Sigma_b^0(udb)$       &$(2\mu_u+2\mu_d-\mu_b)/3$   &0.439  &&1.107 &1.168 &1.206\\
$\Sigma_b^-(ddb)$       &$(4\mu_d-\mu_b)/3$          &-0.804 &&1.117 &1.184 &1.225\\
$\Xi_b^0(usb)$          &$\mu_b$                     &-0.074 &&1.000 &1.000 &1.000\\
$\Xi_b^-(dsb)$          &$\mu_b$                     &-0.074 &&1.000 &1.000 &1.000\\
\hline
\hline
Transition    &$|\mu_{B B'}|$ (expression) &$|\mu_{B B'}|$ &
&$\mu_{B B'}^*(\rho_0)/\mu_{B B'}$ 
&$\mu_{B B'}^*(2\rho_0)/\mu_{BB'}$ 
&$\mu_{BB'}^*(3\rho_0)/\mu_{BB'}$\\
\hline
$\Sigma^0 \to \Lambda$           &$|(\mu_u-\mu_d)/\sqrt{3}|$ &0.901 &1.61$\pm$0.08 
&1.089 &1.136 &1.163\\
$\Sigma_c^+ \to \Lambda_c^+$     &$|(\mu_u-\mu_d)/\sqrt{3}|$ &0.899 &&1.086 &1.128 &1.151\\
$\Sigma_b^0 \to \Lambda_b^0$     &$|(\mu_u-\mu_d)/\sqrt{3}|$ &0.988 &&1.095 &1.144 &1.170\\
\hline
\hline   
\end{tabular}
}
\end{center}
{\small
$^{1)} \mu_p^{\rm Expt.} = 2.7928473446 \pm 0.0000000008$.
\hspace{2ex}
$^{2)} \mu_n^{\rm Expt.} = -1.9130427 \pm 0.0000005$. 
\\
$^{3)}$ Theoretical uncertainties are not included~\cite{PDG}.
\hfill}
\end{table}
%

\begin{table}[htb]
\begin{center}
\caption{
Comparison of the free-space baryon magnetic moments (in nuclear magneton) by various models 
and lattice QCD simulations~\cite{Leinweber:1990dv,Leinweber:1992hy,Lee:2005ds} that are different 
from the MIT bag model, focusing on 
heavier baryons since the data for the octet baryons exist, where the model uncertainties (errors) 
are not given below even when 
quoted in some references. 
For some experimental values, see Tables~\ref{Bmag1} and~\ref{Bmag2}.
}
\label{Bmagvac}
\hspace*{-6ex}

{\small
\begin{tabular}{l|cccccccccc}
\hline
\hline
$B(q_1,q_2,q_3)$ &Set I &Set II 
&\cite{Singh:2017mxj,Singh:2020nwp} &\cite{Aliev:2015axa}
&\cite{Faessler:2006ft} &\cite{Patel:2007gx} &\cite{Barik:1983ics} &\cite{Jena:1986xs}
&\cite{Leinweber:1990dv,Leinweber:1992hy} &\cite{Lee:2005ds}\\
\hline
$p(uud)$        
&1.535 &1.535 &2.56 &--- &--- &--- &2.8732 &2.886 &2.3 &3.04\\
$n(ddu)$      
&-1.023 &-1.023 &-1.93 &--- &--- &--- &-1.9154 &-1.924 &-1.3 &-1.84\\
\hline
$\Lambda (uds)$      
&-0.429 &-0.500 &-0.55 &--- &--- &--- &-0.5512 &-0.580 &-0.40 &-0.70\\
$\Sigma^+(uus)$        
&1.557  &1.628 &2.60 &--- &--- &--- &2.7377 &2.758 &1.9 &2.87\\
$\Sigma^0(uds)$                       
&0.499  &0.535 &-1.48 &--- &--- &--- &0.8222 &0.834 &0.54 &0.76\\
$\Sigma^-(dds)$       
&-0.560 &-0.559 &-1.26 &--- &--- &--- &-1.0932 &-1.089 &-0.87 &-1.48\\
$\Xi^0(ssu)$          
&-0.929 &-1.068 &-1.32 &--- &--- &--- &-1.3734 &-1.414 &-0.95 &-1.37\\
$\Xi^-(ssd)$          
&-0.405 &-0.509 &-0.57 &--- &--- &--- &-0.4157 &-0.452 &-0.41 &-0.82\\
\hline
$\Delta^{++}(uuu)$    
&3.341 &3.341 &5.267 &--- &--- &--- &--- &--- &4.91 &5.24\\
$\Delta^+   (uud)$    
&1.671 &1.671 &2.430 &--- &--- &--- &--- &--- &2.46 &0.97\\
$\Delta^0   (udd)$         
       &0 &0 &-0.408 &--- &--- &--- &--- &--- &0.00 &-0.035\\
$\Delta^-   (ddd)$    
&-1.671 &-1.671 &-3.245 &--- &--- &--- &--- &--- &-2.46 &-2.98\\
$\Sigma^{*+}(uus)$    
&1.781  &1.767 &3.208 &--- &--- &--- &--- &--- &2.55 &1.27\\
$\Sigma^{*0}(uds)$    
&0.102  &0.040 &0.188 &--- &--- &--- &--- &--- &0.27 &0.33\\
$\Sigma^{*-}(dds)$    
&-1.577 &-1.687 &-2.105 &--- &--- &--- &--- &--- &-2.02 &-1.88\\
$\Xi^{*0}(ssu)$            
&0.203  &0.083 &0.508 &--- &--- &--- &--- &--- &0.46 &0.16\\
$\Xi^{*-}(ssd)$       
&-1.473 &-1.686 &-1.805 &--- &--- &--- &--- &--- &-1.68 &-0.62\\
\hline
$\Lambda_c^+(udc)$      
&0.423 &0.423 &--- &--- &0.42 &0.385 &0.341 &0.352 &--- &---\\
$\Sigma_c^{++}(uuc)$    
&1.378 &1.378 &--- &2.4 &1.76 &2.279 &2.44 &2.448 &--- &---\\
$\Sigma_c^+(udc)$       
&0.238 &0.238 &--- &0.5 &0.36 &0.501 &0.525 &0.524 &--- &---\\
$\Sigma_c^0(ddc)$       
&-0.903 &-0.903 &--- &-1.5 &-1.04 &-1.015 &-1.391 &-1.400 &--- &---\\
$\Xi_c^+(usc)$         
&0.424  &0.426 &--- &0.8 &0.41 &0.711 &0.796 &0.779 &--- &---\\
$\Xi_c^0(dsc)$          
&0.424  &0.426 &--- &-1.2 &0.39 &-0.966 &-1.12 &-1.145 &--- &---\\
\hline
$\Lambda_b^0(udb)$      
&-0.073 &-0.074 &--- &--- &-0.06 &-0.064 &--- &--- &--- &---\\
$\Sigma_b^+(uub)$       
&1.675  &1.681 &--- &2.4 &2.07 &2.229 &2.575 &2.586 &--- &---\\
$\Sigma_b^0(udb)$       
&0.437  &0.439 &--- &0.6 &0.53 &0.592 &0.659 &0.662 &--- &---\\
$\Sigma_b^-(ddb)$       
&-0.801 &-0.804 &--- &-1.3 &-1.01 &-1.047 &-1.256 &-1.261 &--- &---\\
$\Xi_b^0(usb)$          
&-0.073 &-0.074 &--- &0.7 &-0.06 &0.766 &0.93 &0.917 &--- &---\\
$\Xi_b^-(dsb)$          
&-0.073 &-0.074 &--- &-1.2 &-0.06 &-0.902 &-0.985 &-1.006 &--- &---\\
\hline
\hline  
\end{tabular}
} 
\end{center}
\end{table}
%

Next, as an example, we discuss the magnetic moment of 
a light quark $q$ in a baryon $B$ in the MIT bag model. 
For a heavy quark $Q$ in $B$, one may replace $q \to Q$ with 
the corresponding quantities.
The free space magnetic moment of the light quark $\mu_q$ with the charge $e_q$ in 
the baryon $B$ is given by,  
\bea
\mu_q  
\equiv e_q \eta_q
\equiv e_q\, \left[ (N_q^B)^2  
\int_0^{R_B} dr\, r^2\, \frac{2r}{3}\, j_0(x_q r/R_B)\, \beta_q^B j_1(x_q r/R_B) \right],   
\label{eta}
\eea
where, $R_B$ is the bag radius of the baryon $B$.
With the expressions given in the second columns of Tables~\ref{Bmag1} and~\ref{Bmag2}, 
it is straightforward to calculate the magnetic moments of those baryons 
in free space as well as in symmetric nuclear matter.

For the transition magnetic moments in free space, 
$B=(\Sigma^0,\Sigma_c^+,\Sigma_b^0) \to B'=(\Lambda,\Lambda_c^+,\Lambda_b^0)$,  
denoted respectively by  
$\mu_{BB'}=(\mu_{\Sigma^0\Lambda}, \mu_{\Sigma_c^+\Lambda_c^+}, \mu_{\Sigma_b\Lambda_b})$, 
some discussions are in order.  
In a rigorous calculation in the MIT bag model~\cite{Chodos:1974je}, 
the bag radius difference for the initial and final baryons arises.
This means that the same flavor spectator quark wave functions  
in the initial and final baryons are slightly different. 
Also, the integral upper limit is restricted to the common bag radius 
to be shown in Eq.~(\ref{etatrans}), and the defects for the spectator 
quark wave function overlaps arise. 
However, we ignore these subtle points due to the MIT bag model, which are 
expected to give negligible effects (indeed to be shown later), 
and approximate each spectator quark wave function overlap to be unity. 
By these approximations, the moduli of the free space transition magnetic moments 
(signs are not known) may be calculated by,
\bea
| \mu_{\Sigma^0 \Lambda} | 
= | \mu_{\Sigma_c^+ \Lambda_c^+} |
= | \mu_{\Sigma_b \Lambda_b} |
= \frac{1}{\sqrt{3}} \left| \mu_u - \mu_d \right|
\equiv \frac{1}{\sqrt{3}} \left| e_u \tilde{\eta}_u - e_d \tilde{\eta}_d \right|, 
\label{mutrans}
\eea
where $\tilde{\eta}_{u,d}$ in the last in Eq.~(\ref{mutrans})  
for $B=(\Sigma^0,\Sigma_c^+,\Sigma_b^0) \to B'=(\Lambda,\Lambda_c^+,\Lambda_b^0)$ are 
defined by:
\bea
\tilde{\eta}_q &\equiv& (N_{q}^{B'}N_q^B) \int_0^{{\rm min}(R_{B'},R_B)} dr\, r^2\, \frac{r}{3}\, 
\nn\\ 
& &\hspace{12ex} \times \left[ 
j_0(x'_{q} r/R_{B'})\, \beta_q^B j_1(x_{q} r/R_{B}) 
+ \beta_q^{B'} j_1(x'_{q} r/R_{B'})\, j_0(x_{q} r/R_{B}) 
\right]. 
\label{etatrans}
\eea
The above expression may arise ambiguities due to the so-called 
MIT bag model artifact, similar to those discussed in the  
weak-interaction vector charge 
calculation, namely, a naive integration over the common bag radius leads to 
a violation of the Ademoll-Gatto theorem~\cite{Ademollo:1964sr} 
as discussed in Ref.~\cite{Guichon:2017gbe}. 
Some discussions on this issue will be given later. 

In the third columns in Tables~\ref{Bmag1} and~\ref{Bmag2} we 
give the free space magnetic moments and transition magnetic moments, respectively calculated 
with the Set I (Table~\ref{coupcc1}) and Set II (Table~\ref{coupcc2}).
The experimental data are also given in fourth columns, denoted by ''Expt.'', but the values and 
errors for the nucleons are shown by rounded in the Tables, where the precise data are 
given in the footnote of each Table.
As already mentioned, the calculated magnitude of the octet baryon magnetic moments in free space  
do not reproduce well the data. Our focus of this study is on the density dependence of them, 
relative to the free space ones, which still can give some relevant and useful information.
To give some idea on the free-space baryon magnetic moments calculated in the present approach, 
we compare in Table~\ref{Bmagvac} the free-space baryon magnetic moments obtained by various 
models and lattice QCD simulations. 

Recall that, the in-medium to free proton EMFF ratios predicted by the QMC model, 
reproduce well the extracted proton EMFF super ratio 
$[G^p_E/G^p_M(^4{\rm He})]/[G^p_E/G^p_M(^1{\rm H})]$  
from the $^4{\rm He}(\vec{e},e'\vec{p})^3{\rm H}$ reaction  
measured at Jefferson Laboratory (JLab)~\cite{Strauch:2002wu,Dieterich:2000mu,Paolone:2010qc}.
The ratios of the in-medium to free space magnetic moments are given 
for three baryon densities, ($\rho_0, 2\rho_0, 3\rho_0$)     
in the (fifth, sixth, seventh) columns in Tables~\ref{Bmag1} and~\ref{Bmag2}, 
respectively for the Set I and Set II.

Here, we focus on the in-medium nucleon magnetic moments.
As already mentioned, the constraint for the allowed change (swelling) of 
the in-medium nucleon size, thus for the in-medium nucleon 
bag radii and magnetic moments, are allowed to increase no more than 17.1\% at $\rho_0$, 
if we take properly the $y$-scaling analysis results.
The present QMC model results give a 7.7\% enhancement of the 
nucleon magnetic moments at $\rho_0$, and well within the constraint.

One can notice that the most of the ratios become larger as the baryon density increases, 
except for $\Lambda, \Xi^-, \Lambda_c^+, \Lambda_b, \Xi^{+,0}_c$ and $\Xi^{0,-}_b$.
In particular, the large enhancement for the $\Sigma^{*0}$ and $\Xi^{*0}$ may be noted 
in the both, Set I and II. This enhancement is due to the small magnitudes of 
these magnetic moments in free space as can been seen in Tables~\ref{Bmag1} and~\ref{Bmag2}.
Between the corresponding results given in Tables~\ref{Bmag1} and~\ref{Bmag2}, 
one can also see the effect of the different 
strange quark mass values, respectively $m_s = 250$ and $m_s = 93$ MeV.
The absolute value of the strange quark magnetic moment $|\mu_s|$ becomes smaller 
$0.429 \to 0.275$ as the $m_s$ value changes $250 \to 93$ MeV. 
Then all changes involving the strange quark magnetic moment 
can be understood in Table~\ref{Bmag2} as follows. Namely, the 
corresponding magnetic moments in Table~\ref{Bmag2} are less influenced 
by the strange quark magnetic moment $\mu_s < 0$ 
than those in Table~\ref{Bmag1}.
On the other hand, the use of $m_b=4200$ or $m_b=4180$ MeV gives negligible difference.

The tiny decreases (or no changes up to the digits shown) observed for the magnetic moments of 
$\Lambda, \Xi^-, \Lambda_c^+, \Lambda_b, \Xi^{+,0}_c$ and $\Xi^{0,-}_b$ as baryon density 
increases can be understood as follows. The total magnetic moments of them are $\mu_s, \mu_c$ 
or $\mu_b$ for the $\Lambda$ and the corresponding $\Lambda$-like baryons, 
or $(4 \mu_s -\mu_{u,d})/3$ for $\Xi^{0,-}$. In the QMC model, the $s, c$ and $b$ quarks do not 
couple the mean fields $\sigma$ nor $\omega$. Thus, their wave functions are not directly modified 
as the first order interactions of $\sigma$ and $\omega$ mean fields.
However, the mean fields, in particular $\sigma$ field  
couples to the light quarks $u$ and $d$ in these baryons, and  
the effective masses of these baryons decrease (or they get the attractive Lorentz scalar 
potentials) as baryon density increases.
Accordingly, the bag radii of these baryons decrease very slightly as baryon density 
increases~\cite{Tsushima:2018goq} as the consequence of the simultaneous mass stability condition 
Eq.~(\ref{hmass}). Since the baryon (quark) magnetic moment is roughly 
proportional to the bag radius, and the wave functions of the $s,c,$ and $b$ 
quarks are not modified as the first order, the $s,c$ and $b$ quark magnetic moments 
are slightly modified as the second order effect of the bag radius change, and thus, the 
in-medium quark magnetic moments $\mu^*_{s,c,b}$ decrease very slightly due to the 
bag radius decrease. For the $\Xi^{0,-}$ cases are similar, because the $\mu^*_s$ dominates.
Thus, the magnetic moments of $\Lambda, \Xi^-, \Lambda_c^+, \Lambda_b, \Xi^{+,0}_c$ and 
$\Xi^{0,-}_b$ decrease very slightly as baryon density increases.

\begin{figure}[htb]
\vspace{3ex}
\begin{center}
\includegraphics[scale=0.3]{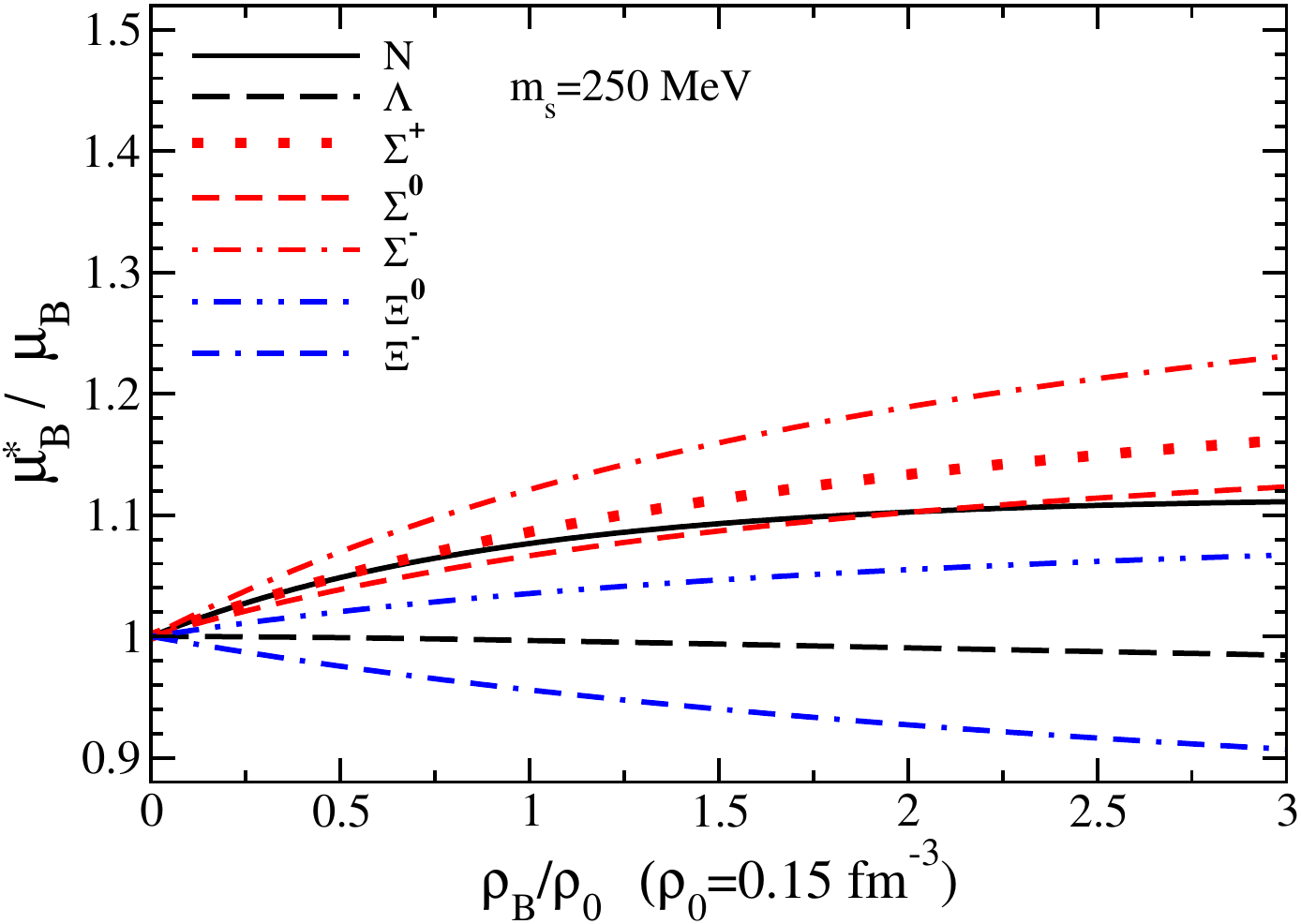}
\hspace{3ex}
\includegraphics[scale=0.3]{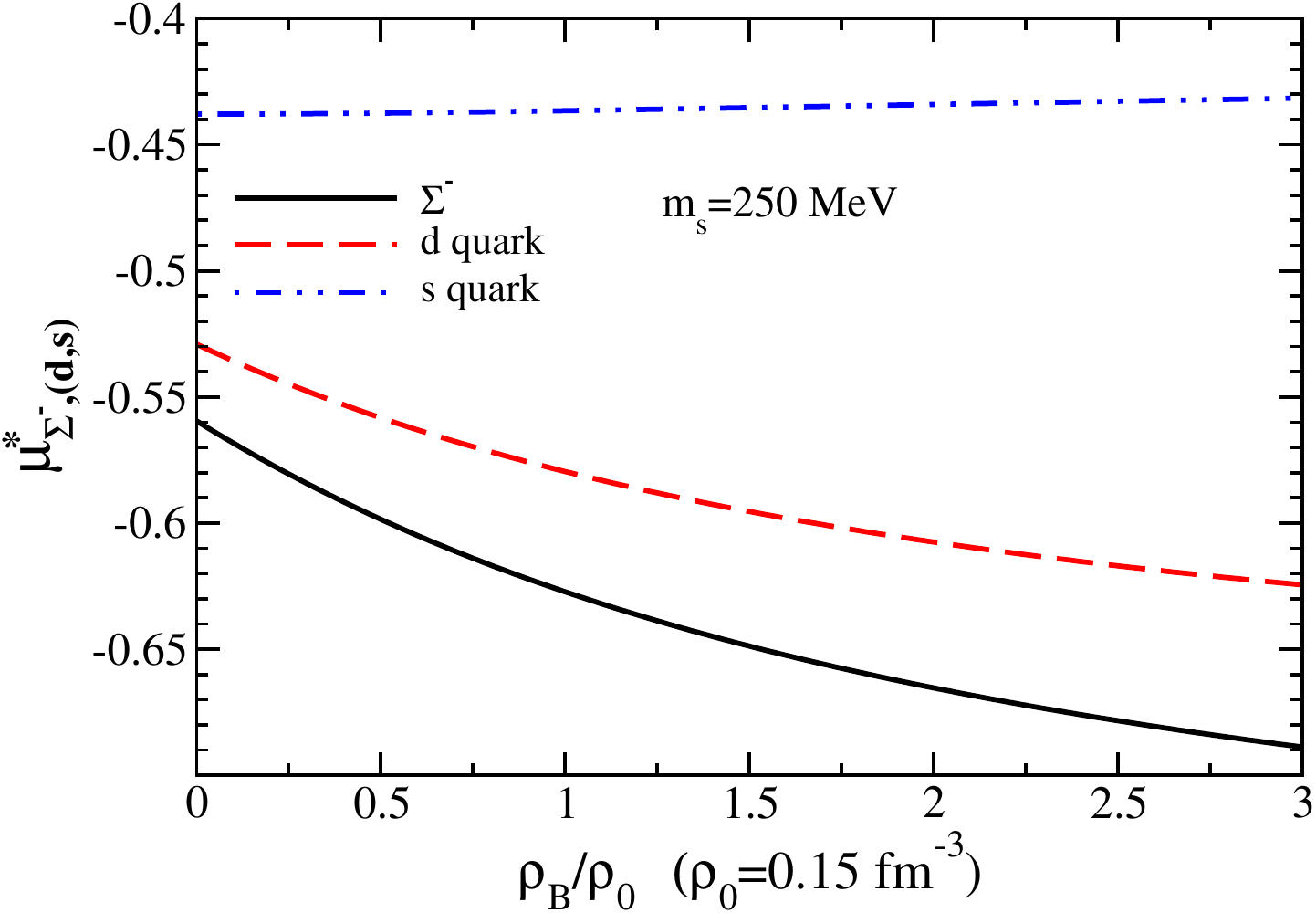}
\vspace{6ex}
\\
\includegraphics[scale=0.3]{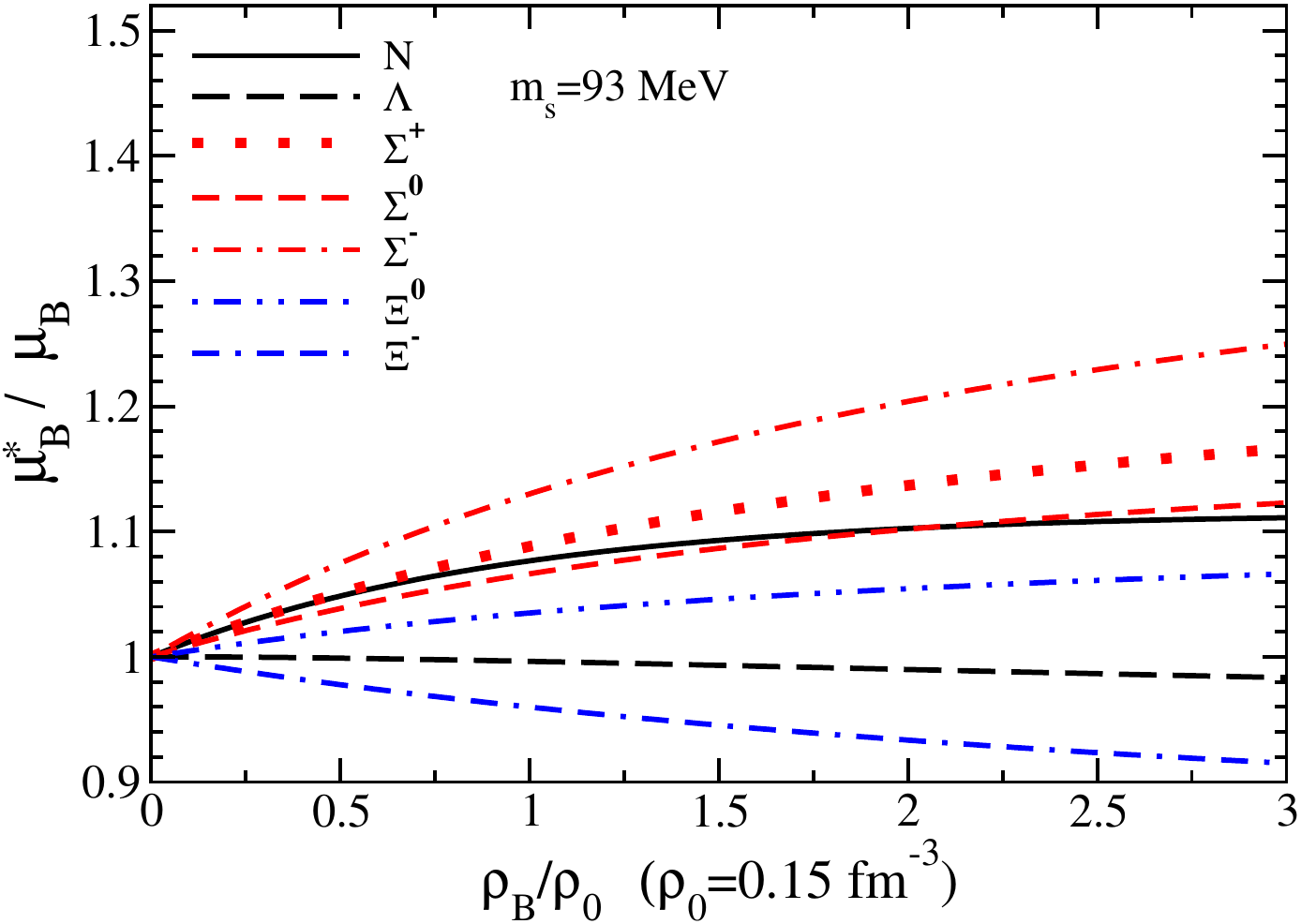}
\hspace{3ex}
\includegraphics[scale=0.3]{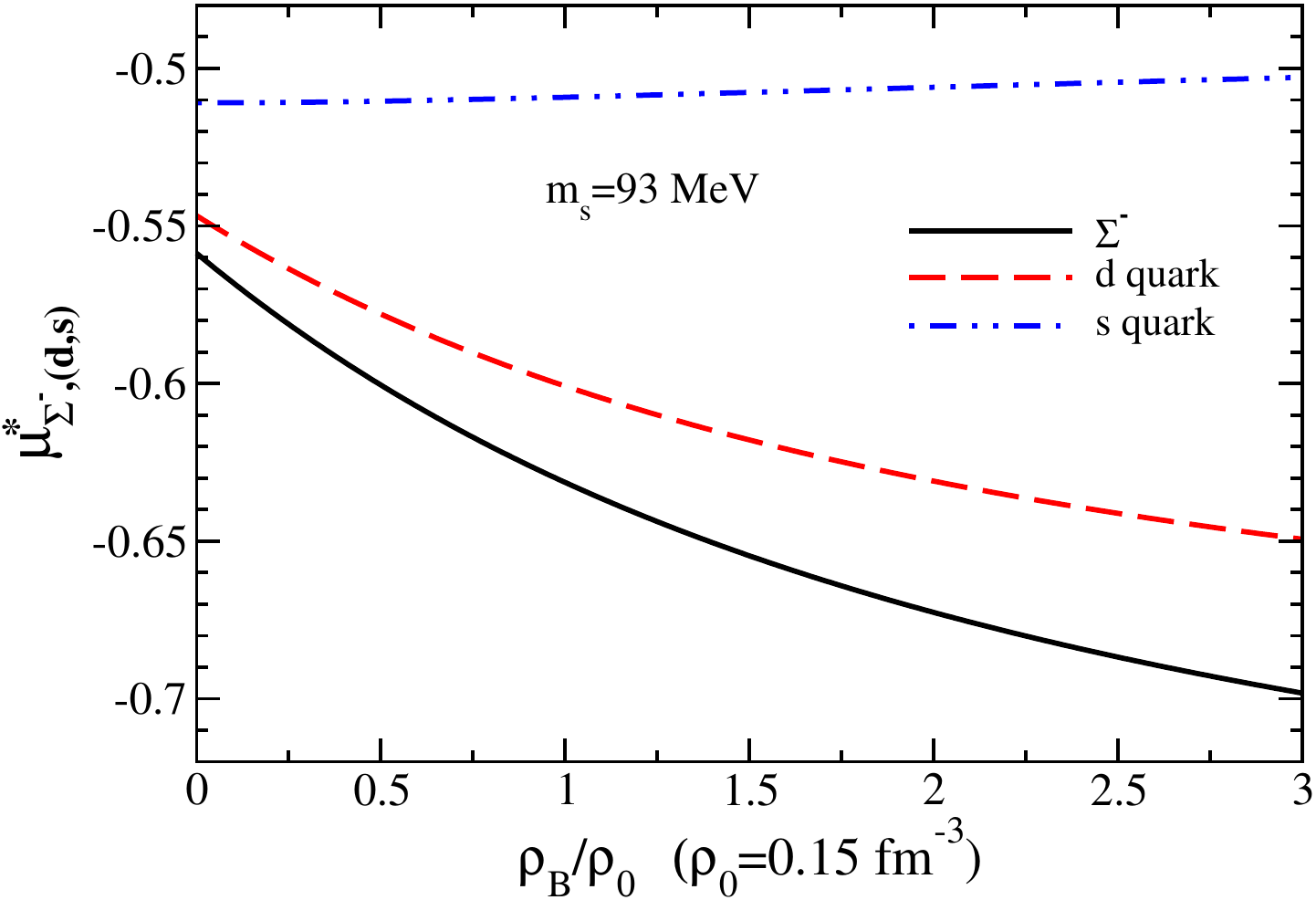}
\caption{\label{Octmag} 
Density dependence of the octet baryon magnetic moment ratios, 
the in-medium to free space (left panel), 
and the bare $\Sigma^-$ magnetic moment, 
which has the largest medium modification in the left panel, 
and the corresponding $d$ and $s$ quark magnetic moments (right panel),  
respectively calculated by the Set I (upper panel) and Set II (lower panel).
}
\end{center}
\vspace{2ex}
\end{figure}

\begin{figure}[htb]
\vspace{3ex}
\begin{center}
\includegraphics[scale=0.3]{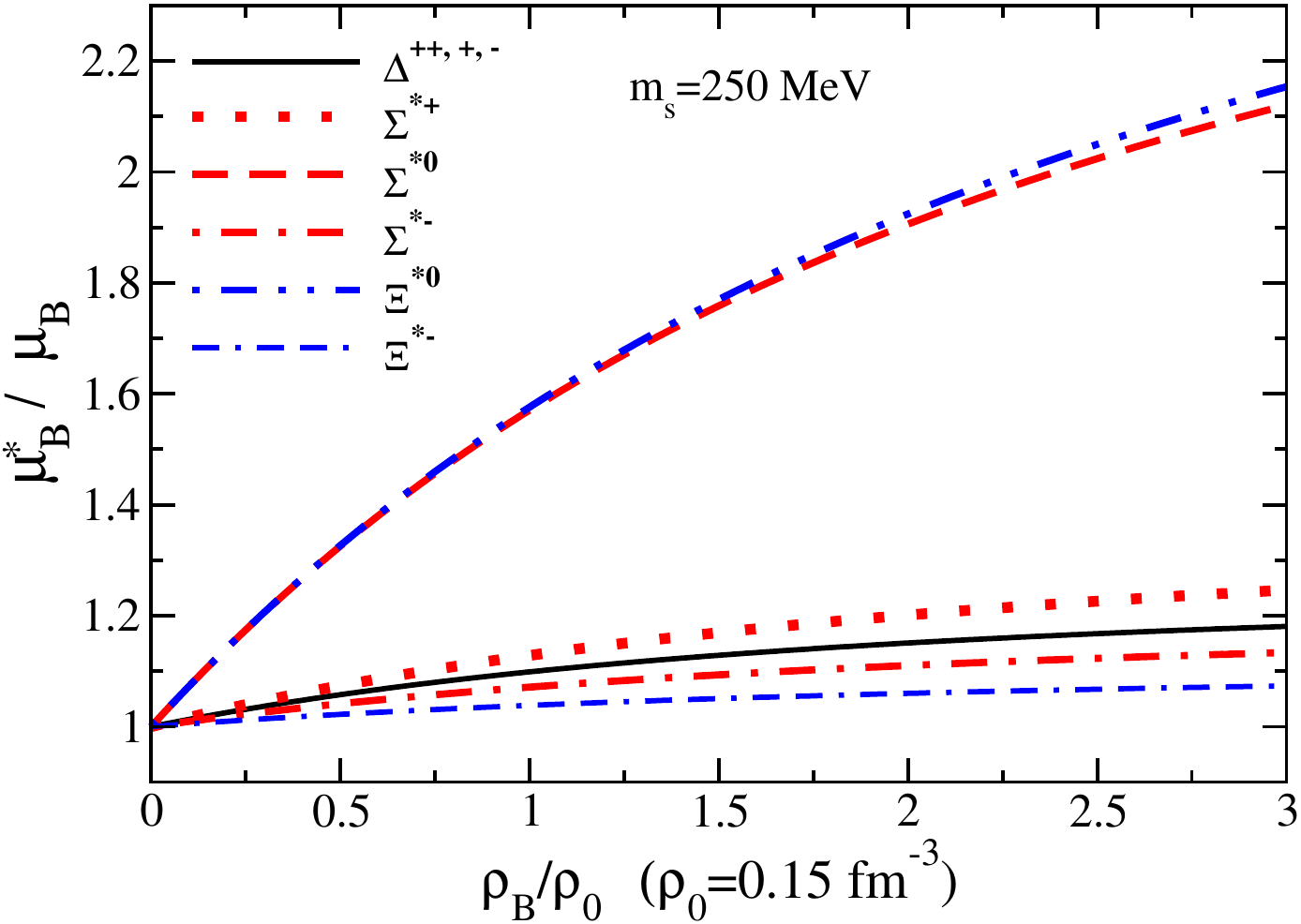}
\hspace{3ex}
\includegraphics[scale=0.3]{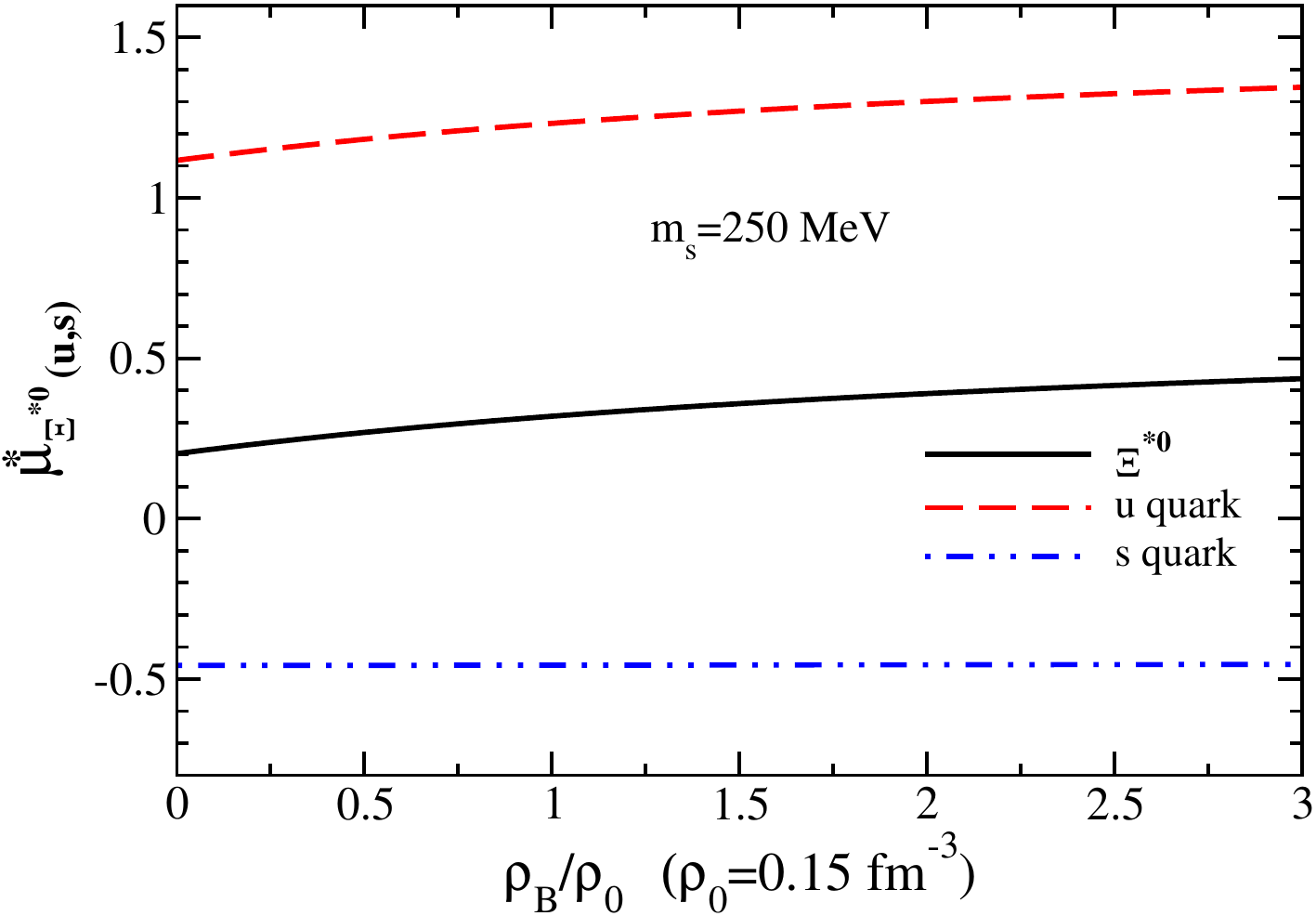}
\vspace{6ex}
\\
\includegraphics[scale=0.3]{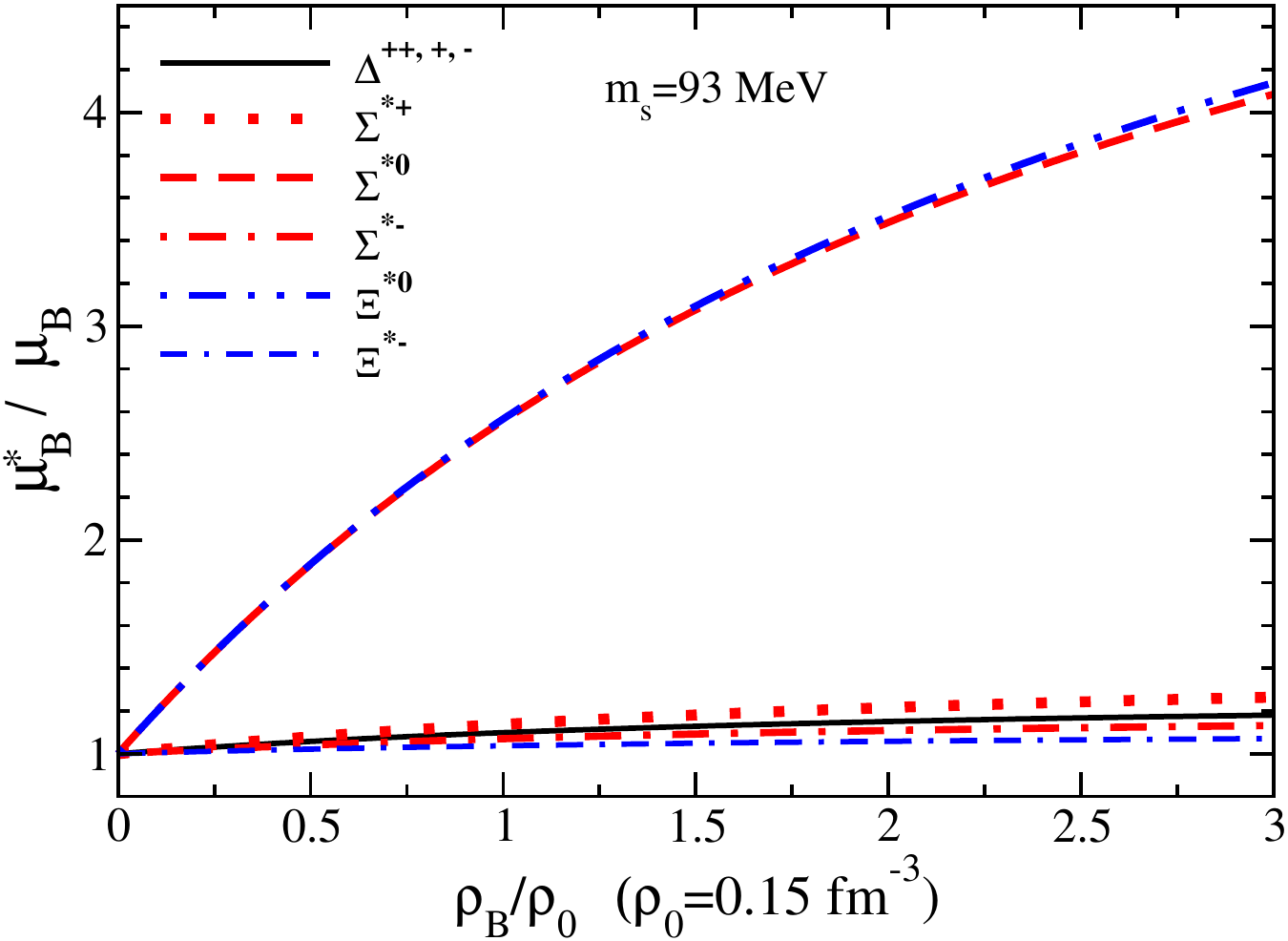}
\hspace{3ex}
\includegraphics[scale=0.3]{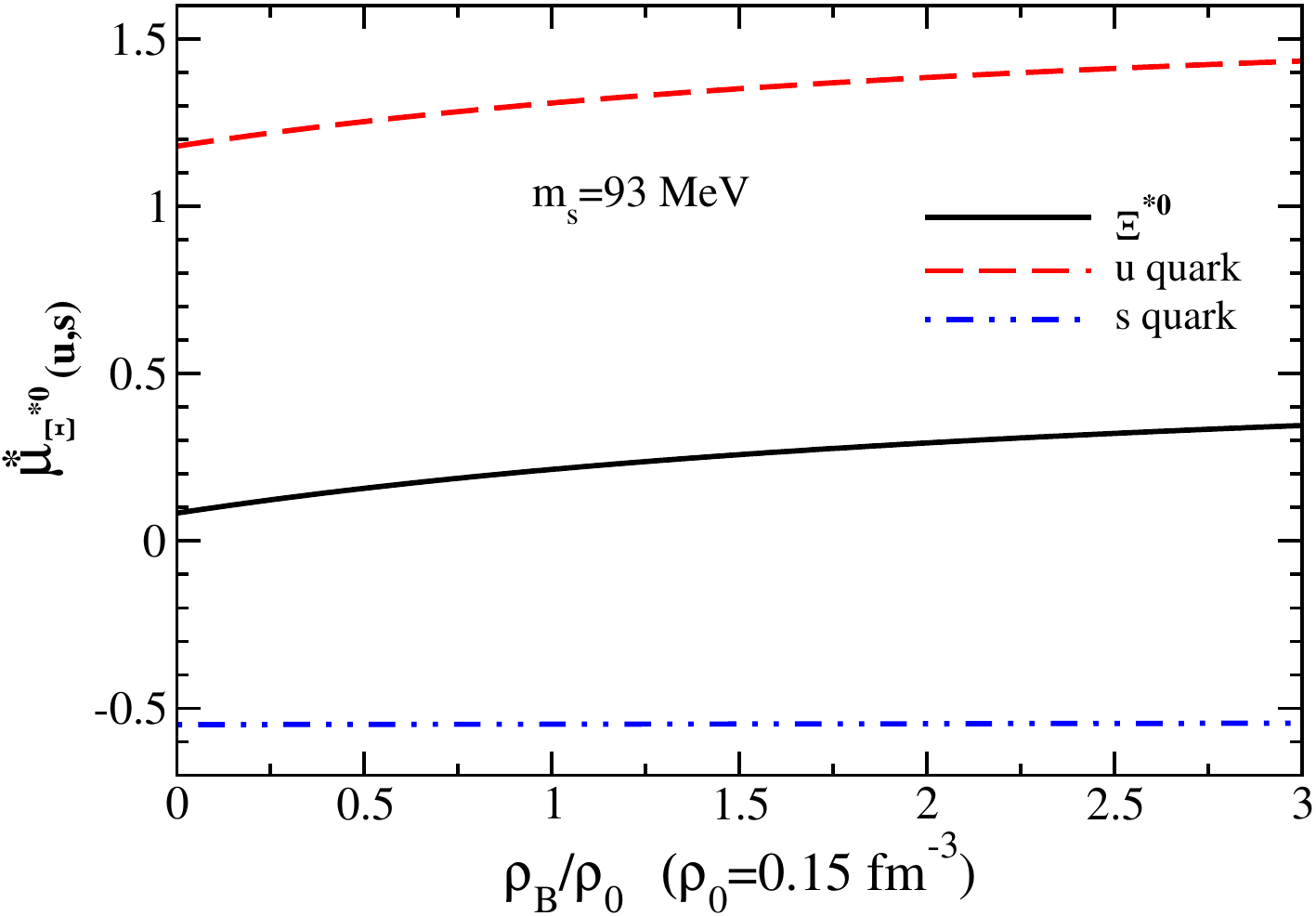}
\caption{\label{Decmag} 
Density dependence of the decuplet baryon magnetic moment ratios, 
the in-medium to free space (left panel), 
and the bare $\Xi^{*0}$ magnetic moment, 
which has the largest medium modification in the left panel, 
and the corresponding $u$ and $s$ quark magnetic moments (right panel), 
respectively calculated by the Set I (upper panel) and Set II (lower panel).
}
\end{center}
\vspace{2ex}
\end{figure}

\begin{figure}[htb]
\vspace{3ex}
\begin{center}
\includegraphics[scale=0.3]{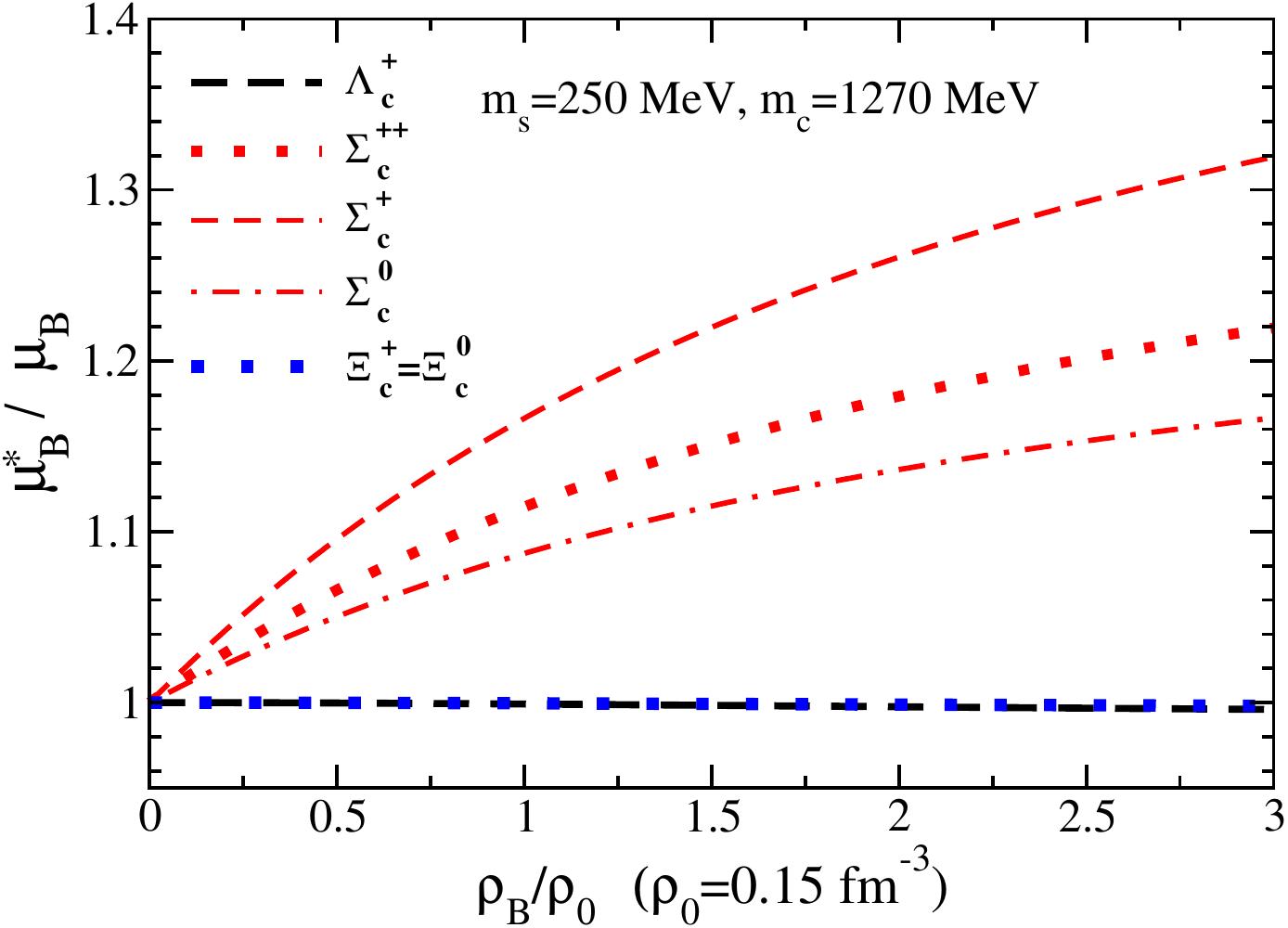}
\hspace{3ex}
\includegraphics[scale=0.3]{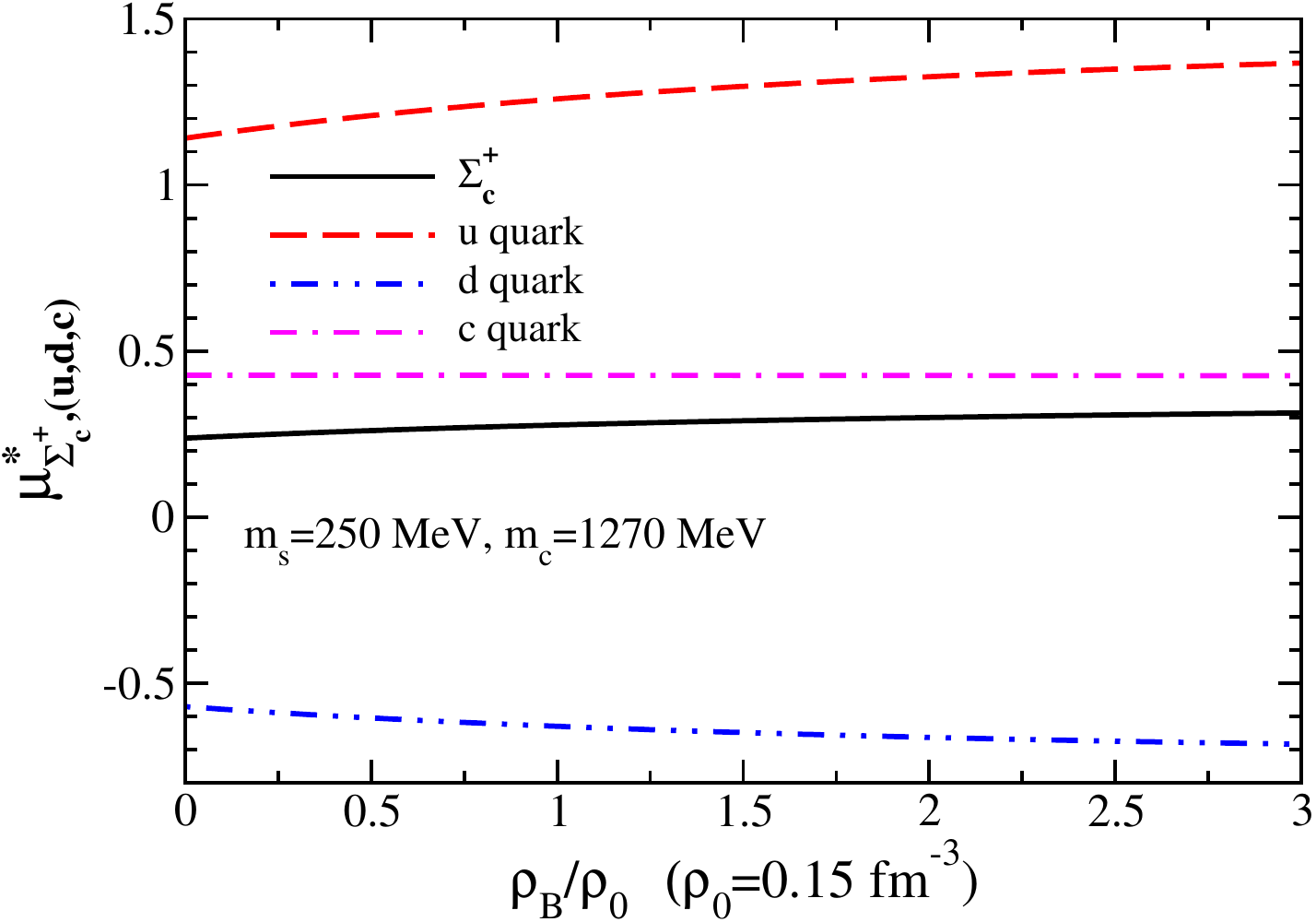}
\vspace{6ex}
\\
\includegraphics[scale=0.3]{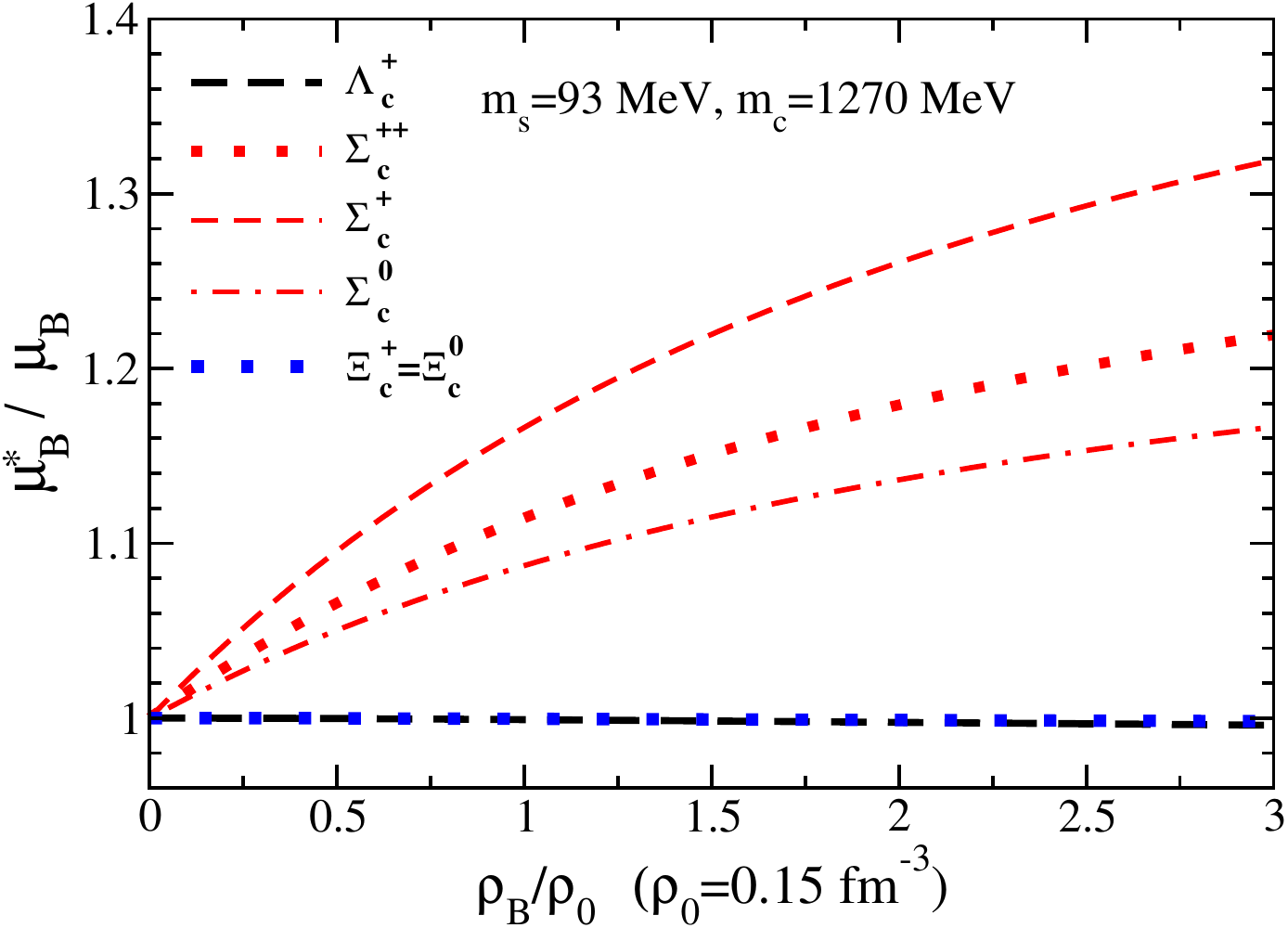}
\hspace{3ex}
\includegraphics[scale=0.3]{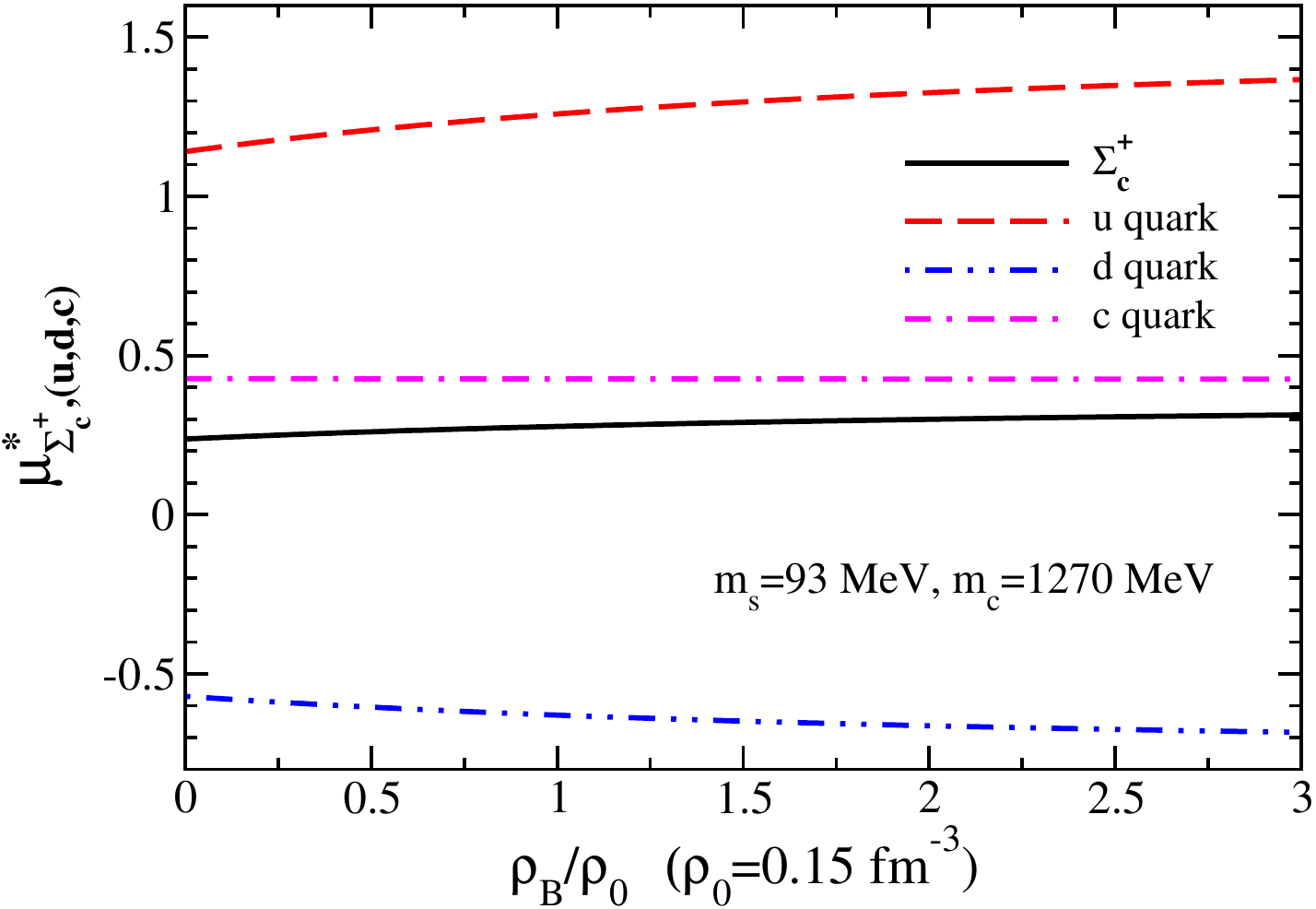}
\caption{\label{CBmag} 
Density dependence of the charm baryon magnetic moment ratios, 
the in-medium to free space (left panel), 
and the bare $\Sigma_c^+$ magnetic moment, 
which has the largest medium modification in the left panel, 
and the corresponding $u, s$, and $c$ quark magnetic moments (right panel), 
respectively calculated by the Set I (upper panel) and Set II (lower panel). 
}
\end{center}
\vspace{2ex}
\end{figure}

\begin{figure}[htb]
\vspace{3ex}
\begin{center}
\includegraphics[scale=0.3]{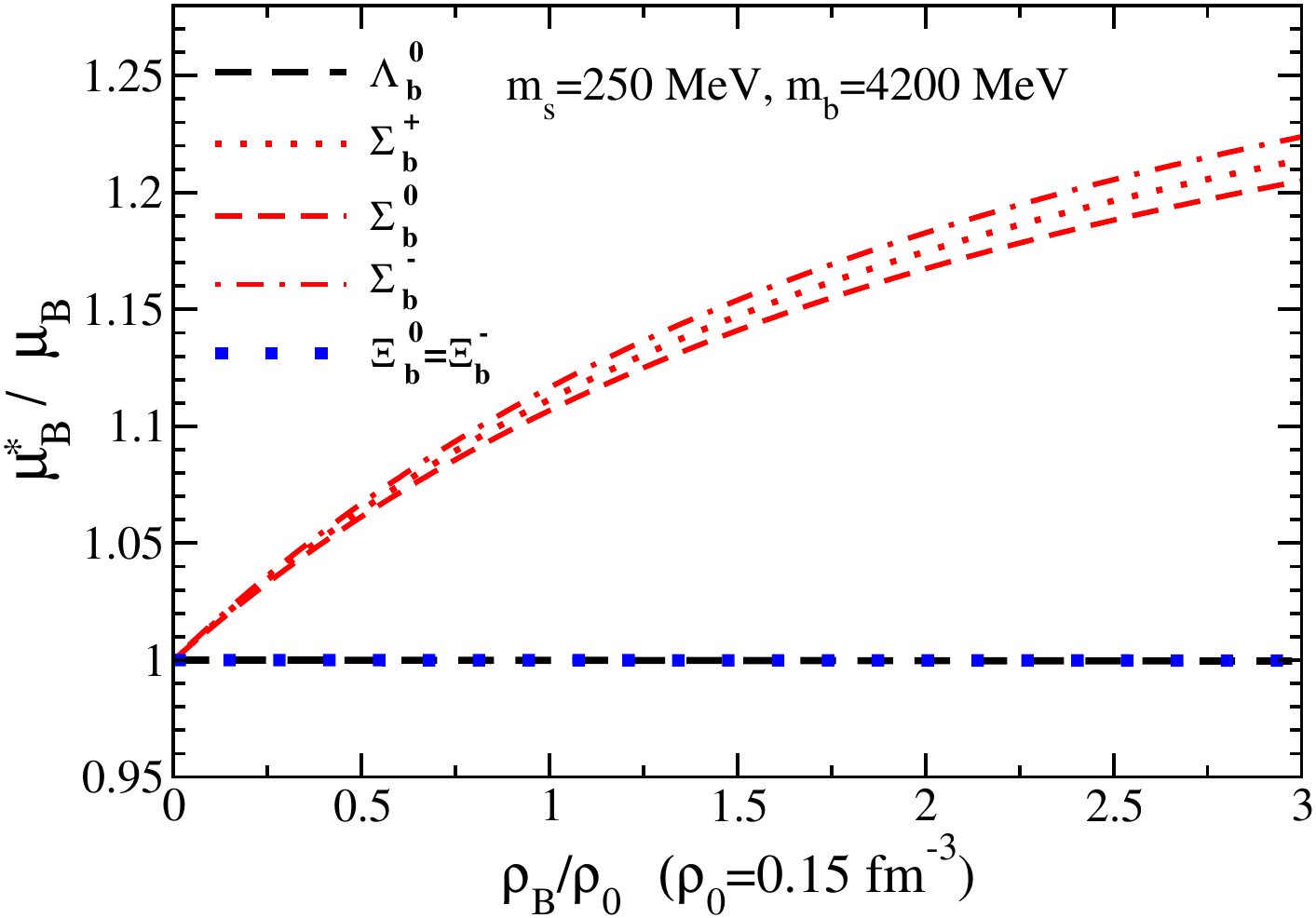}
\hspace{3ex}
\includegraphics[scale=0.3]{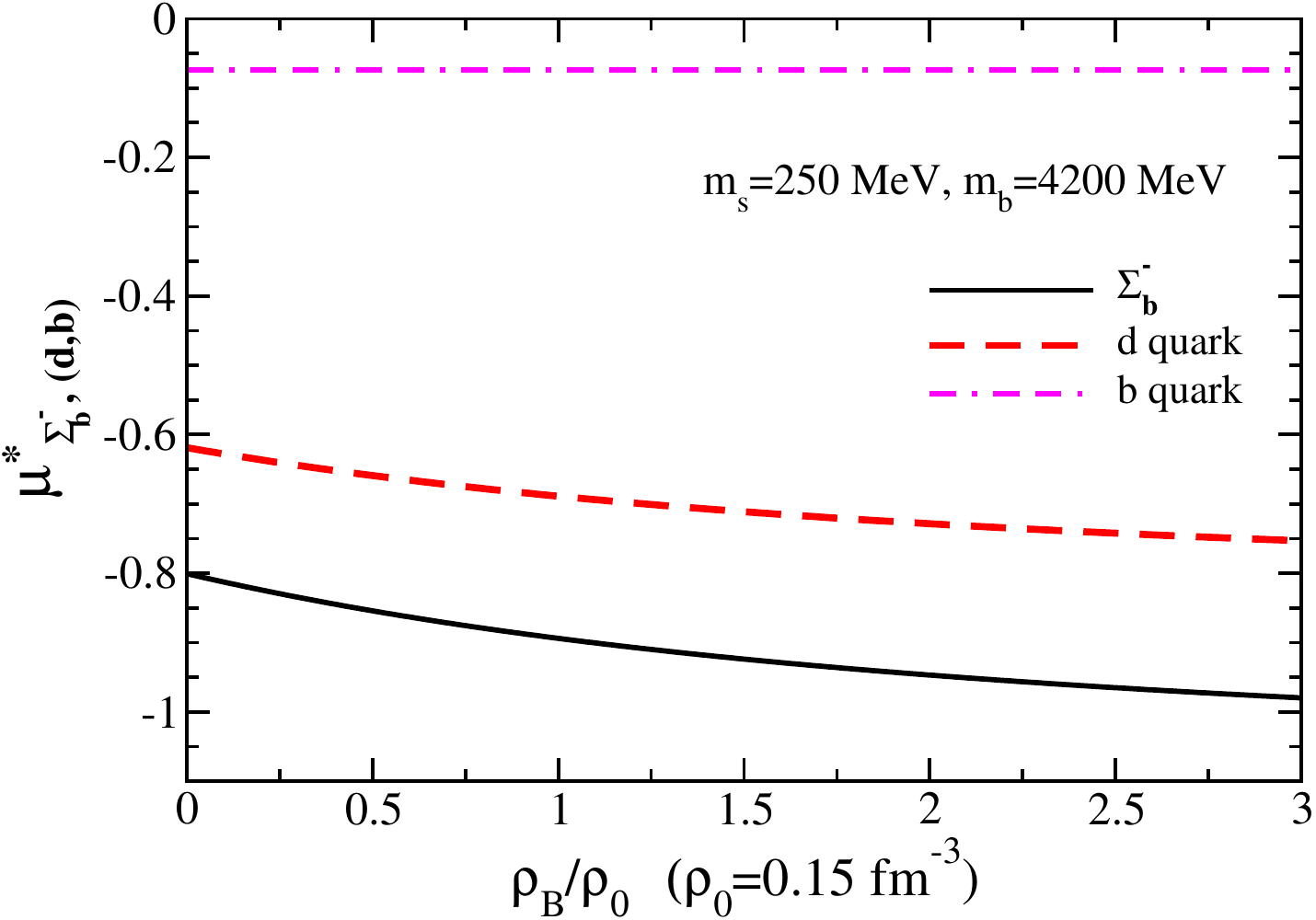}
\vspace{6ex}
\\
\includegraphics[scale=0.3]{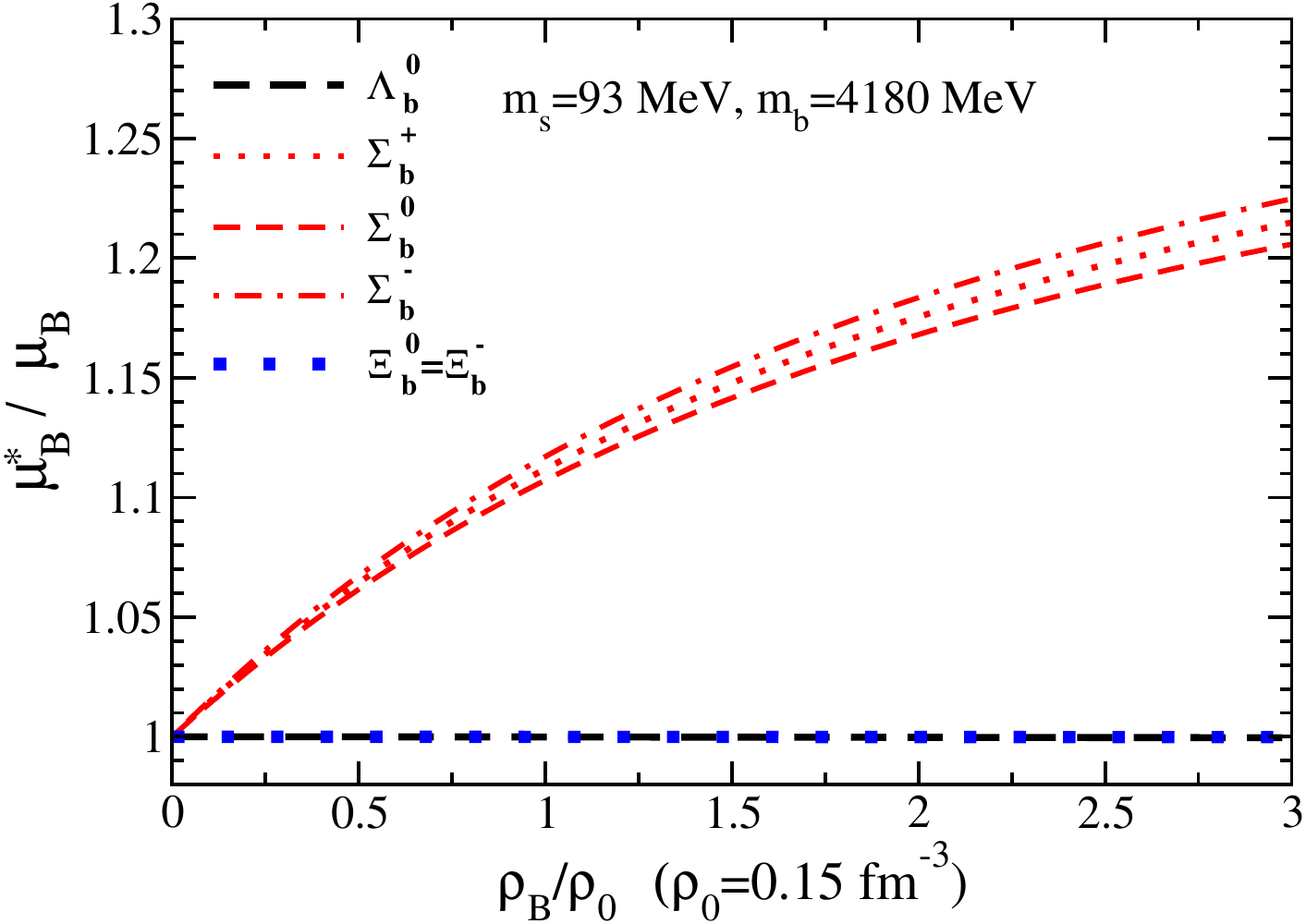}
\hspace{3ex}
\includegraphics[scale=0.3]{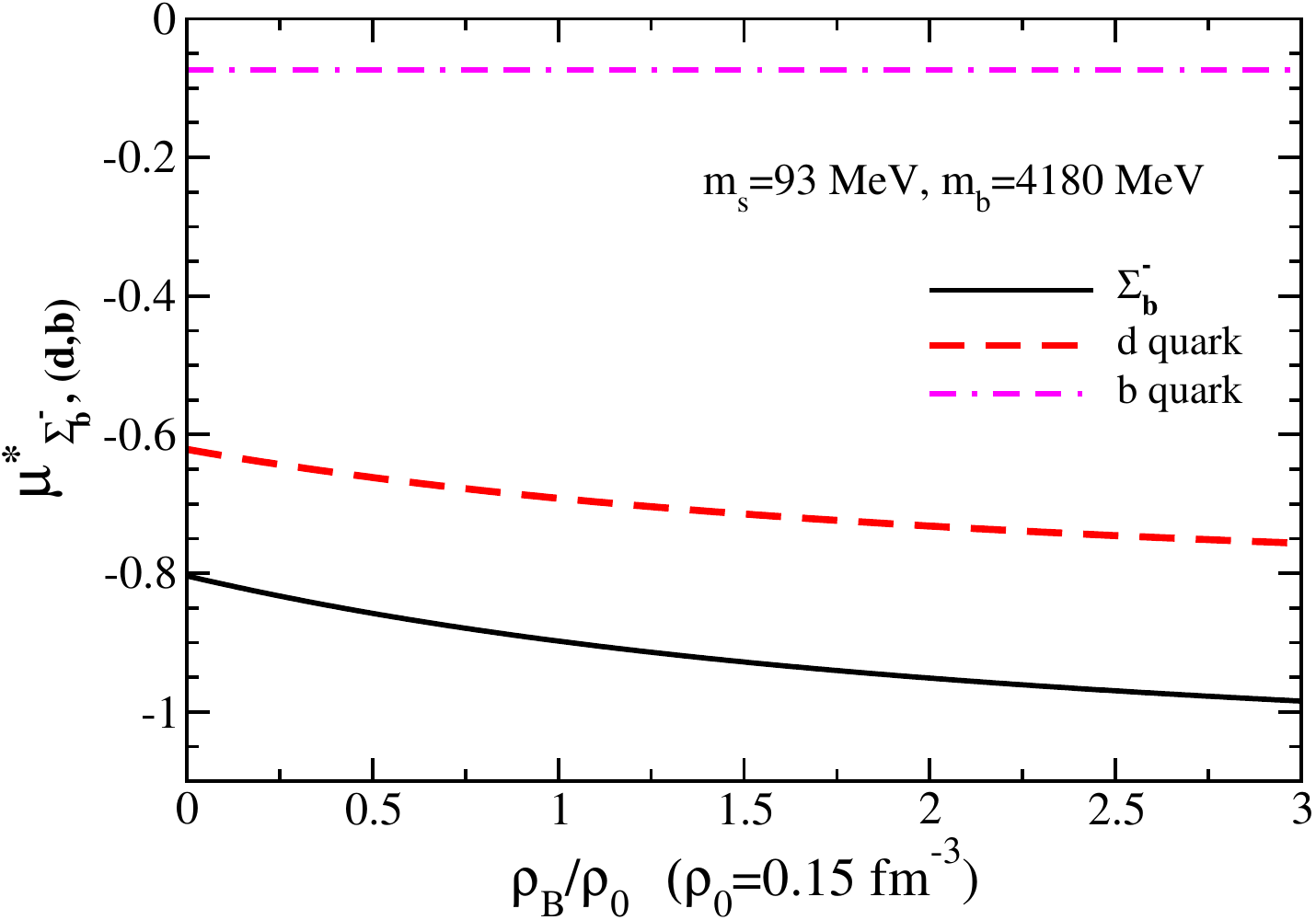}
\caption{\label{BBmag} 
Density dependence of the bottom baryon magnetic moment ratios, 
the in-medium to free space (left panel), 
and the bare $\Sigma_b^-$ magnetic moment, 
which has the largest medium modification in the left panel, 
and the corresponding $u, s$, and $b$ quark magnetic moments (right panel), 
respectively calculated by the Set I (upper panel) and Set II (lower panel).
}
\end{center}
\vspace{2ex}
\end{figure}

\begin{figure}[htb]
\vspace{3ex}
\begin{center}
\includegraphics[scale=0.3]{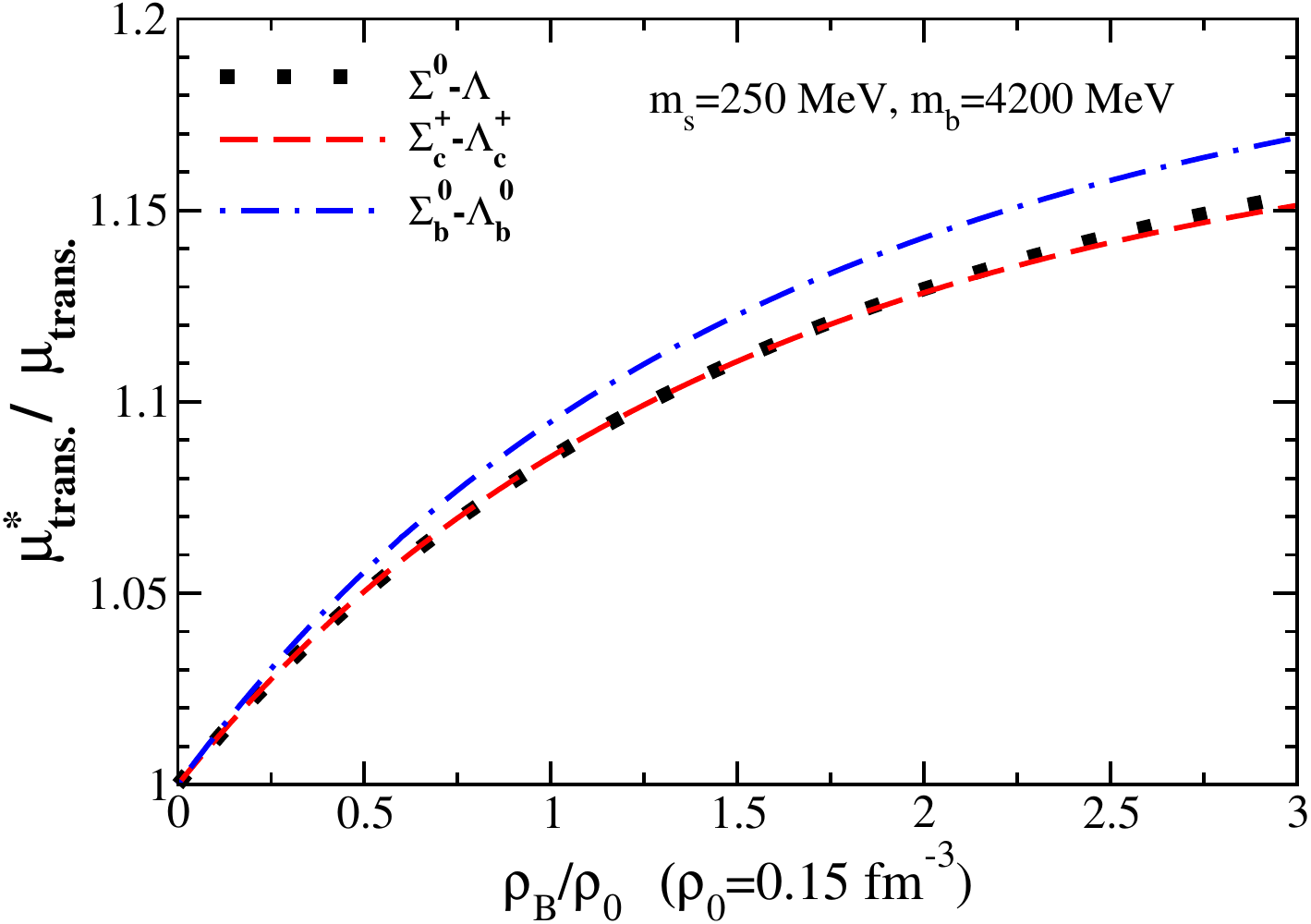}
\hspace{3ex}
\includegraphics[scale=0.3]{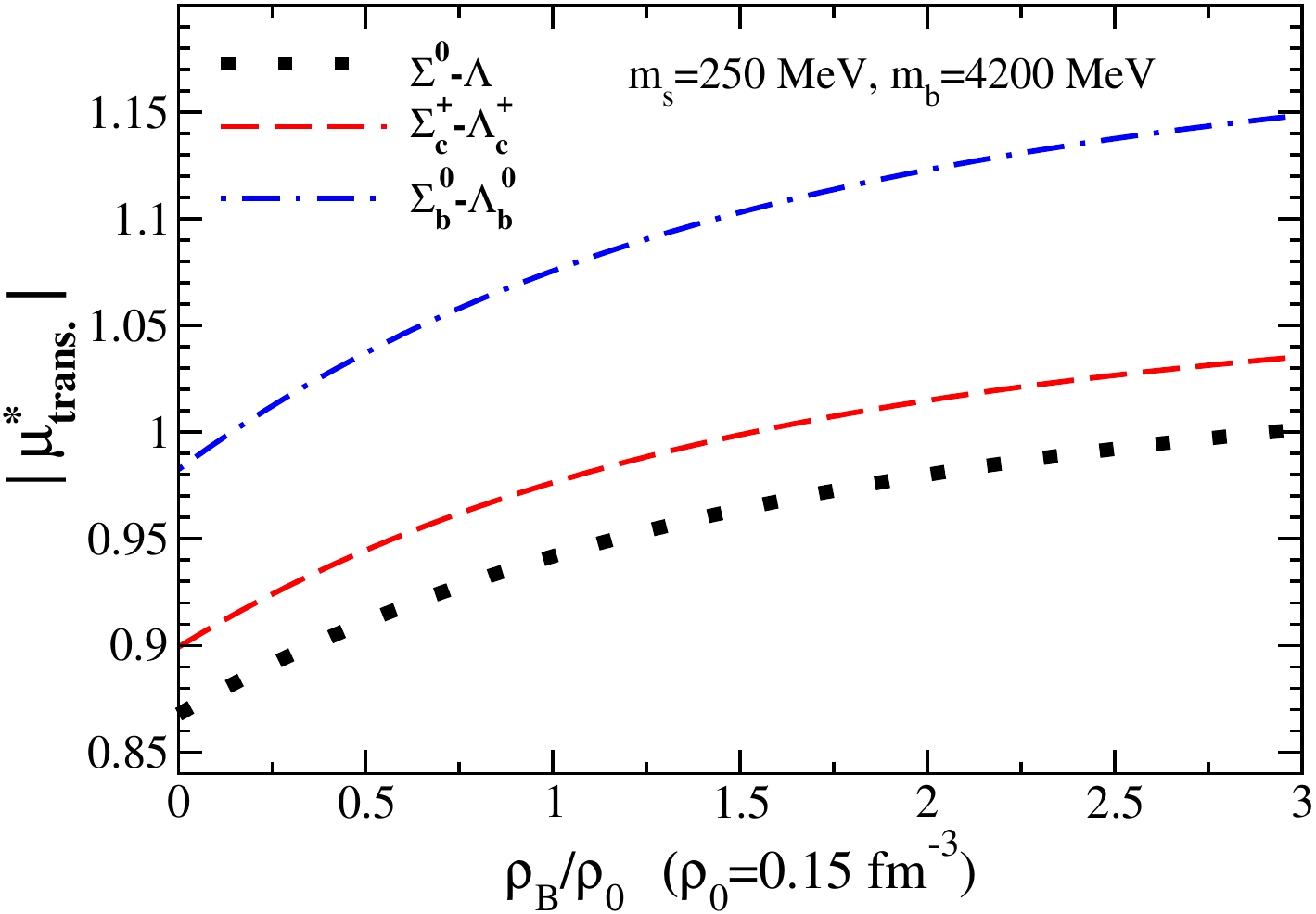}
\vspace{6ex}
\\
\includegraphics[scale=0.3]{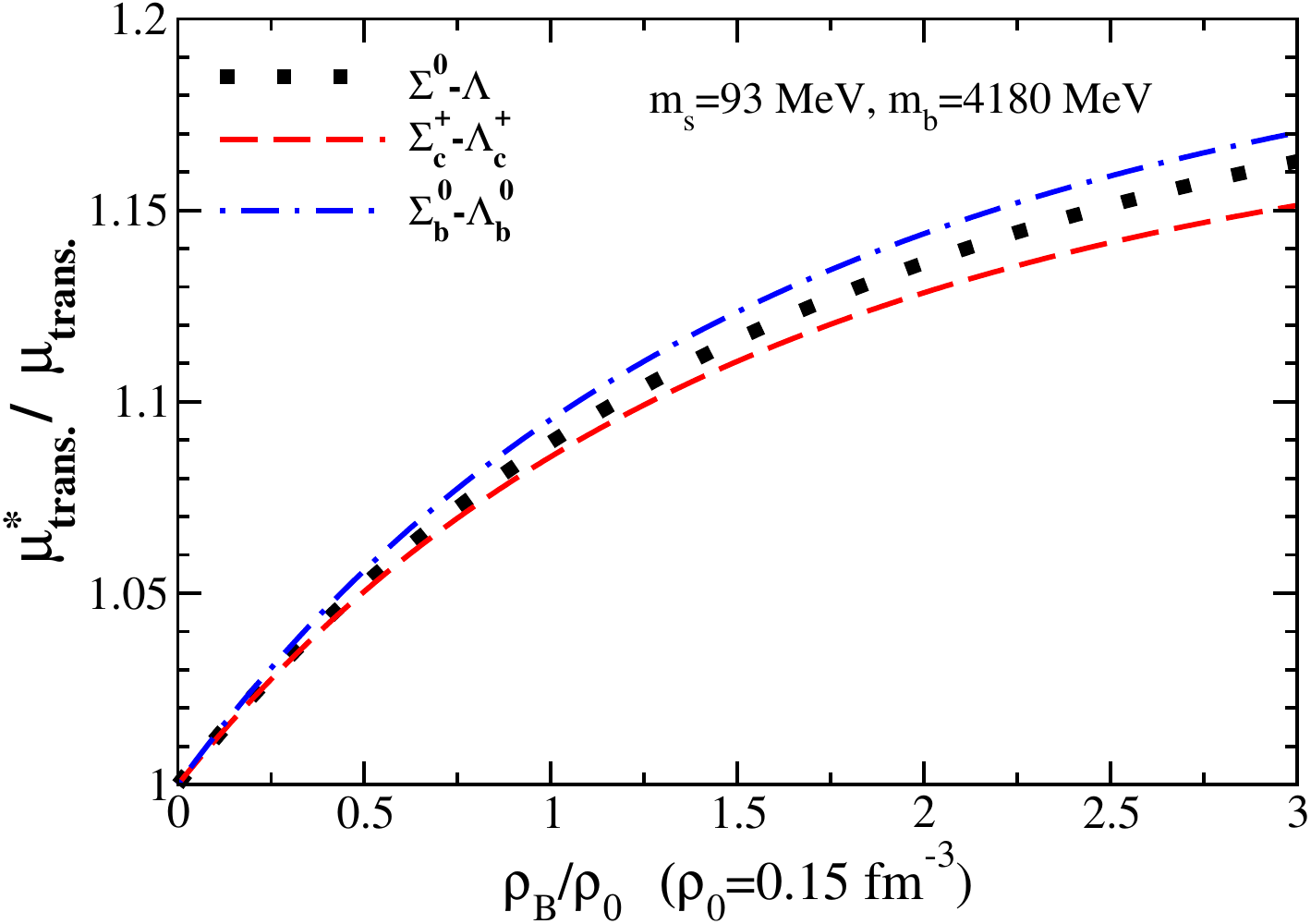}
\hspace{3ex}
\includegraphics[scale=0.3]{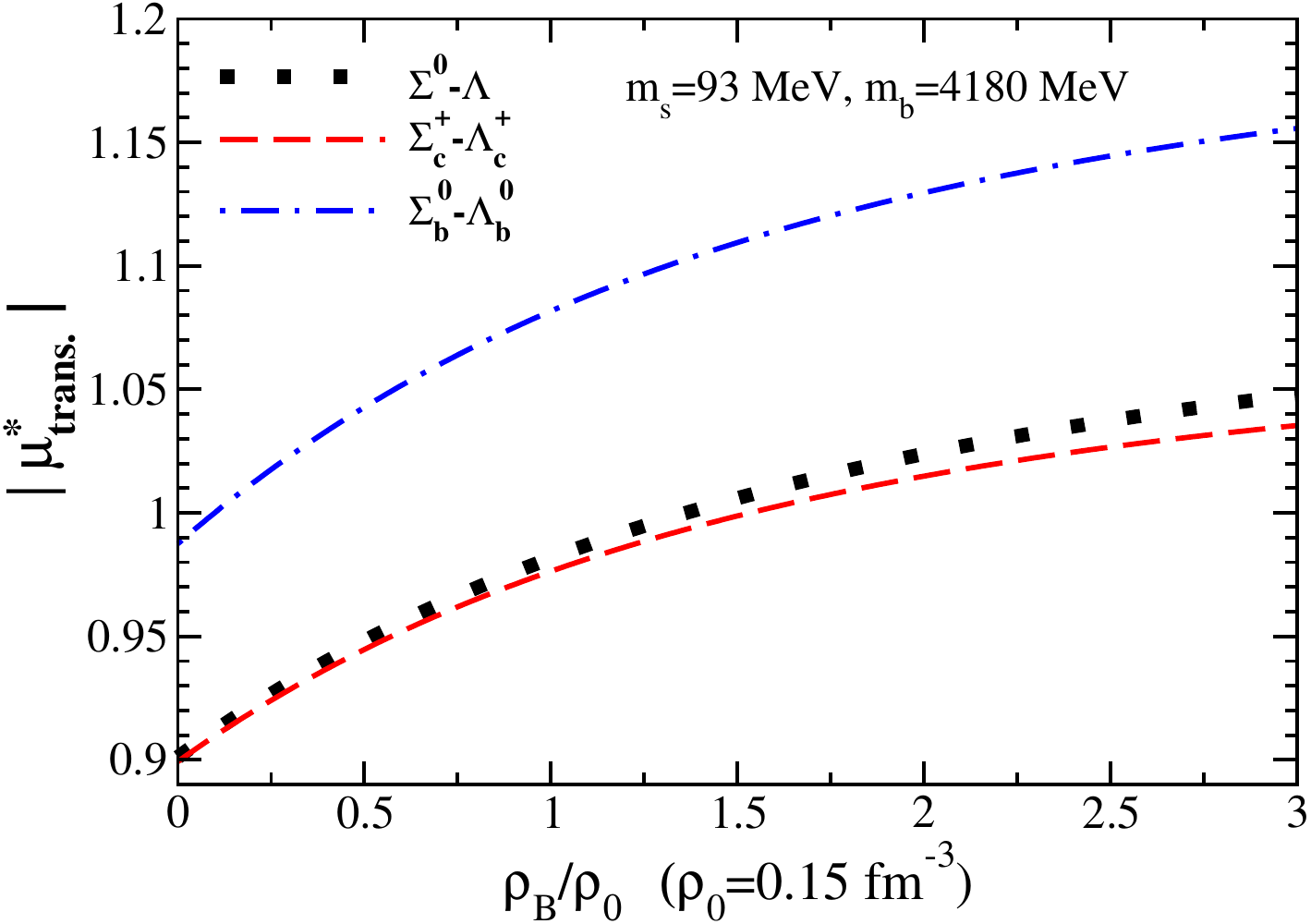}
\caption{\label{Transmag} 
Density dependence of the transition magnetic moment ratios, 
the in-medium to free space (left panel), 
and the moduli of the bare values in medium (right panel), 
respectively calculated by the Set I (upper panel) and Set II (lower panel).
}
\end{center}
\vspace{2ex}
\end{figure}

To see easier, we show the density dependence of the magnetic moments of 
the octet baryons in Fig.~\ref{Octmag}, decuplet baryons in Fig.~\ref{Decmag}, 
the low-lying charmed baryons in Fig.~\ref{CBmag}, 
the low-lying bottom baryons in Fig.~\ref{BBmag}, and the transition magnetic moments
in Fig.~\ref{Transmag}, calculated by the Set I and Set II. 
In each figure the ratios of the in-medium to free space  
are shown in the left panel for both the Set I (upper left) and Set II (lower left), 
while in the right panel the    
bare density dependence is shown for the magnetic moment that has the largest medium modification 
among all in the left panel, as well as the corresponding quark contributions 
for the Set I (upper right) and Set II (lower right).

First, we discuss the magnetic moments of the octet baryons shown 
in Fig.~\ref{Octmag}.
As is known, the MIT bag model underestimates the octet baryon magnetic moments 
in free space~\cite{DeGrand:1975cf,Allen:1975nq}, and  
the analytic expression for the magnetic moment roughly proportionals to the bag radius. 
(For example, see Table 7.1 in Ref.~\cite{Yamaguchi:1989sx} for the bag radius dependence 
on the numerically obtained octet baryon magnetic moments.) 
Thus, we estimate the medium modifications of magnetic moments by taking the ratios, 
the in-medium to free space magnetic moments. 
The calculated density dependence of the ratios (left panel) is quite different from that of  
the MQMC model shown in Ref.~\cite{Ryu:2008st}.
It is desired to take into account properly the constraint 
for the allowed maximum change derived from the $y$-scaling data analysis. 

In the right panel we show the $\Sigma^-$ magnetic moment, 
which has the largest medium modification among all in the left panel, 
as well as the corresponding quark contributions.
One can see the $s$ quark magnetic moment is only slightly modified in medium,
showing a small linear increase as density increases. 
This very small density dependence is expected,  
since the $s$ quark does not couple to any meson fields in the present model, 
and the modification comes from the change in the bag radius $R_\Sigma^*$.

Next, we show the decuplet baryon magnetic moments in Fig.~\ref{Decmag}.
The $\Sigma^{*0}$ and $\Xi^{*0}$ magnetic moments have large enhancement as baryon 
density increases for both the Set I and Set II. 
This enhancement, as already mentioned, due to their 
small magnitudes in free space.

As for the charm sector baryon magnetic moments shown in Fig.~\ref{CBmag}, 
one can easily notice that the medium modification of the 
$\Sigma_c^+$ magnetic moment is the largest in the left panel, 
while those of $\Lambda_c^+$ and $\Xi_c^{+,0}$ are negligibly modified (tiny decrease).
The $\Sigma_c^+$ magnetic moment is enhanced in the ratio by the small magnitude of 
the free space magnetic moment as shown in Tables~\ref{Bmag1} and~\ref{Bmag2}.
As can be confirmed for the $\Sigma_c^+$ in the right panel,  
the increase of the $u$ quark magnetic moment enhances the ratio. 

For the bottom baryon magnetic moments shown in Fig.~\ref{BBmag}, 
the $\Sigma_b^{\pm,0}$ magnetic moments show large medium modifications in the left panel.
Different from the case of $\Sigma_c^{++,+,0}$, the magnitudes of the free space 
magnetic moments do not reflect to the enhancement of the ratio of 
the $\Sigma_b^{\pm,0}$ magnetic moments. 
Similar to the charm sector, the $\Lambda_b$ and $\Xi_b^{+,0}$ are negligibly 
modified (tiny decrease) as expected.

Finally, we show in Fig.~\ref{Transmag} the transition magnetic moments 
for $B=(\Sigma^0,\Sigma_c^+,\Sigma_b^0) \to B'=(\Lambda,\Lambda_c^+,\Lambda_b^0)$, 
calculated by the Set I (upper panel) and Set II (lower panel).
The density dependence of the ratios, the in-medium to free space is shown in the left panel,  
while the moduli of the bare values are shown in the right panel. 
The density dependence seems to be similar for all the three cases in the left panel, 
but $\Sigma_b^0 \to \Lambda_b$ case shows larger enhanced in both the Set I and Set II.
The larger medium modification of $|\mu_{\Sigma_b^0\Lambda_b^0}|$, 
which can be seen in the left panel, may be attributed to the larger bag radii 
for the $(\Sigma_b^0 \to \Lambda_b^0)$, or the larger integral upper limit (common bag radius)     
$R_{\Lambda_b}^*$, than those of $R_{\Lambda}^*$ for $(\Sigma^0 \to \Lambda)$ and 
$R_{\Lambda_c^+}^*$ for $(\Sigma_c^+ \to \Lambda_c^+)$.
In addition, the effect due to the different $m_s$ values can be seen 
in the enhancement of the $\Sigma^0 \to \Lambda$ transition magnetic moment 
in the bottom right. In this case, the results show that the 
$m_s = 93$ MeV (Set II) yields larger bag radii for the $\Sigma$ and $\Lambda$ 
than those corresponding to $m_s = 250$ MeV (Set I), 
giving the larger transition magnetic moment, since 
the transition magnetic moments are also roughly proportional to the bag radii 
of the initial and final baryons, or common bag radius.
Since this feature may be associated with the MIT bag model artifact,
we should focus on the ratios, the in-medium to free space shown in the left panel, 
as will be discussed next. As already mentioned, ratios are also 
used to extract the useful and meaningful experimental data to reduce 
ambiguities.

Let us discuss here some aspects of partial restoration of dynamical chiral symmetry (PRDCS) 
on the in-medium magnetic moments obtained. 
As already mentioned, dynamical chiral symmetry (DCS) and its partial restoration concern the light 
quarks---the faster reductions of light quark condensates. The reductions  
result to decrease the (effective) masses of baryons (hadrons) which contain light quarks. 
Thus, the largest effect is expected to appear in the nucleon magnetic moments, 
likewise the mass reductions of the nucleons are the largest among the baryons treated in this 
study, since nucleons are composed of purely light quarks. 
Or, the magnitude of mass reduction (attractive Lorentz scalar potential) appears in proportional 
to the number of light quarks in the baryon, namely from the lager to smaller order, 
$N$, $\Lambda_{s,c,b} \simeq \Sigma_{s,c,b}$, and $\Xi_{s,c,b}$ 
($\Lambda_s \equiv \Lambda$). But for the case of the baryon magnetic moments, 
the story is a bit different, since the $\Lambda_{s,c,b}$ magnetic moments 
are respectively equal to $\mu_{s,c,b}$ without the light quark contributions. 
Note that, the effects of PRDCS are for the ''net'' modifications  
of the light quark magnetic moments $\mu_u$ and $\mu_d$. 
For the transition magnetic moments, since all the three transitions treated 
in this study are proportional to $|(\mu_u - \mu_d)/\sqrt{3}|$, the transition 
magnetic moments reflect directly the PRDCS. 
In the ratios shown in Figs.~\ref{Octmag}~-~\ref{Transmag}, the modifications of the 
$\mu_u$ and $\mu_d$ are modulated by the other contributions from $s, c$ and $b$ quark 
magnetic moments respectively $\mu_{s,c,b}$ for the sum for each baryon magnetic moment 
except for those of $\Lambda_{s,c,b}$.
In addition, small but nonnegligible modification of the in-medium bag 
radius of each baryon influences the in-medium magnetic moment small amount.  
(In the MIT bag model the magnetic moment of a baryon roughly proportional 
to the bag radius). Although the results shown in Figs.~\ref{Octmag} - \ref{Transmag} 
contain these mixed effects in the ratios, the clear conclusion is that, 
the light quark in-medium to free magnetic moment ratios
become enhanced to be $|\mu^*_{u,d}/\mu_{u,d}| > 1$, and these inequalities 
are valid for all the relevant baryons in medium, although for $\Lambda_{s,c,b}$ 
and $\Xi_{c,b}$ they do not explicitly contribute and affect the results.
The magnetic moments of $\mu_{s,c,b}$ are not modified in the first order effects of PRDCS 
in the QMC model. These features can be clearly seen from the right panels 
in Figs.~\ref{Octmag}~-~\ref{Transmag}.

Next, we discuss the ratios of the in-medium 
to free space transition magnetic moments, since the ratios are expected to reduce 
the possible ambiguities originated from the MIT bag model artifact as follows. 
Let us denote the true values in medium as $\mu_{BB'}^{\rm *true}(\rho_B)$ 
and free space as $\mu_{BB'}^{\rm true}(0)$, respectively, and     
the corresponding errors by $\epsilon^*(\rho_B)$ and $\epsilon(0)$. 
Then, the ratio of the in-medium to free space can be estimated by, 
\bea
\frac{\mu_{BB'}^*(\rho_B)}{\mu_{BB'}(0)} 
&=&\frac{\mu_{BB'}^{\rm * true}(\rho_B)(1 \pm \epsilon^*(\rho_B))}
{\mu_{BB'}^{\rm true}(0)(1 \pm \epsilon(0))},  
\nn\\
&\simeq& \frac{\mu_{BB'}^{\rm * true}(\rho_B)}{\mu_{BB'}^{\rm true}(0)}  
(1 \pm \epsilon^*(\rho_B) \mp \epsilon(0))
= \frac{\mu_{BB'}^{\rm * true}(\rho_B)}{\mu_{BB'}^{\rm true}(0)}  
[ 1 \pm (\epsilon^*(\rho_B) - \epsilon(0)) ],  
\label{error}
\eea
where $0 < \epsilon^*(\rho_B), \epsilon(0) << 1$ are assumed. 
The signs in front of them in the first line in Eq.~(\ref{error}) 
are expected to be the same, since $\epsilon^*(\rho_B)$ varies smoothly as $\rho_B \to 0$,    
and $\epsilon^*(\rho_B) \to \epsilon(0)$ to give the ratio unity.
Then, $|(\epsilon^*(\rho_B) - \epsilon(0))|$ becomes smaller in Eq.~(\ref{error}).

To have a better idea on the bag radius difference, we estimate the differences in free 
space and at $\rho_0$~\cite{Tsushima:2018goq} by the Set I and Set II results, and obtain,  
\bea
\left(\frac{R_{B}(0)-R_{B'}(0)}{R_{B'}(0)}, 
\frac{R_{B}^*(\rho_0)-R_{B'}^*(\rho_0)}{R_{B'}^*(\rho_0)}\right)  
< (0.045, 0.045),
\label{rdiff}
\eea
where, inequality holds for all the three cases, $B=(\Sigma^0,\Sigma_c^+,\Sigma_b^0) \to 
B'=(\Lambda,\Lambda_c^+,\Lambda_b^0)$ 
with $R_{B}(0)-R_{B'}(0)>0$ and $R_{B}^*(\rho_0)-R_{B'}^*(\rho_0)>0$, 
for the both sets (Tables~\ref{coupcc1} and~\ref{coupcc2}).  
Furthermore, since the bag model result for the diagonal magnetic moment $\mu_B$ ($\mu_B^*$)   
roughly proportionals to the bag radius $R_B$ (and $R_B^*$)~\cite{DeGrand:1975cf}, 
we can expect the size of the ambiguity is the same order as in Eq.~(\ref{rdiff}).

In fact, the octet baron magnetic moments and transition magnetic 
moments were studied in the U(3) symmetry model~\cite{Okubo:1961jc}
(equivalent to the SU(3) symmetry model of Refs.~\cite{Neeman:1961jhl,GellMann:1962xb} 
for this case), by including up to the first order of mass splitting interaction. 
Then, the following relations were obtained,  
\bea
\mu_{\Sigma^0} &=& \frac{1}{2} \left[ \mu_{\Sigma^+} + \mu_{\Sigma^-} \right],
\label{musig0}
\\
\mu_{\Sigma^0\Lambda} &=& \mu_{\Lambda\Sigma^0} 
= \frac{1}{2\sqrt{3}} \left[ \mu_{\Sigma^0} 
+ 3 \mu_{\Lambda} -2 \mu_{\Xi^0} -2 \mu_n \right].
\label{muls0trans}
\eea
We examine the above relations in the present 
results of the MIT bag model in free space.
For Eq.~(\ref{musig0}), we get (l.h.s., r.h.s.) = (0.499, 0.499) [(0.457,0.457)] 
using the values in Table~\ref{Bmag1} (Set I) [Table~\ref{Bmag2} (Set II)], 
while for Eq.~(\ref{muls0trans}) we get 
(l.h.s, r.h.s.) = (0.868, 0.900) [(0.901,0.914)]. 
For the latter, the deviation may be estimated as 
(0.900-0.868)/0.868 = 0.037 [(0.914-0.901)/0.901 = 0.014]. 
On the other hand, using the experimental data for the octet baryon magnetic
moments and $|\mu_{\Sigma^0 \to \Lambda}^{\rm Expt.}| = 1.61$~\cite{PDG} 
with $\mu_{\Sigma^0} \equiv (1/2)[\mu_{\Sigma^+}^{\rm Expt.}+\mu_{\Sigma^-}^{\rm Expt.}]$, 
we get, (l.h.s, r.h.s.) = (1.61, 1.483) for Eq.~(\ref{muls0trans}). 
The deviation is estimated by (1.61-1.483)/1.483 = 0.086.
Thus, including the SU(3) symmetry breaking mass splitting interaction up 
to the first order, the deviation from the relation is larger when we use  
the experimental results, than that using the MIT bag model results 
for both the Set I and Set II. 
Based on these estimates, the ambiguities arising from the possible 
MIT bag model artifact such as the bag radius difference, 
are not expected to affect the estimated ratios   
for the transition magnetic moments within  
the present status of the experimental precision. 

To make the analysis complete, we also study the Coleman-Glashow relations~\cite{Coleman:1961jn} 
that were derived in a unitary symmetry scheme by defining   
the ratios $R_i (i=1,...,8)$:
\bea
\mu_{\Sigma^+} &=& \mu_p, 
\hspace{14ex}\to R_1 \equiv \left| \frac{\mu_{\Sigma^+}-\mu_p}{\mu_p} \right|, 
\label{CG1}\\
\mu_{\Lambda} &=& \frac{1}{2} \mu_n, 
\hspace{12ex}\to R_2 \equiv \left| \frac{\mu_{\Lambda}-\frac{1}{2} \mu_n}{\frac{1}{2}\mu_n}\right|,
\label{CG2}\\
\mu_{\Xi^0} &=& \mu_n, 
\hspace{14ex}\to R_3 \equiv\left|\frac{\mu_{\Xi^0}-\mu_n}{\mu_n}\right|,
\label{CG3}\\
\mu_{\Sigma^-} &=& - \left[\mu_p+\mu_n\right],
\hspace{5ex}\to R_4 \equiv \left| 
\frac{\mu_{\Sigma^-}+\left[\mu_p + \mu_n \right]}{-\left[\mu_p+\mu_n\right]} \right|,
\label{CG4-1}\\
\mu_{\Xi^-} &=& - \left[ \mu_p + \mu_n \right],
\hspace{5ex}\to R_5 \equiv \left| 
\frac{\mu_{\Xi^-}+\left[\mu_p + \mu_n \right]}{-\left[\mu_p+\mu_n\right]} \right|,
\label{CG4-2}\\
\mu_{\Xi^-} &=& \mu_{\Sigma^-},
\hspace{13ex}\to R_6 \equiv \left| 
\frac{\mu_{\Xi^-}-\mu_{\Sigma^-}}{\mu_{\Sigma^-}} \right|,
\label{CG4-3}\\
\mu_{\Sigma^0} &=& - \frac{1}{2}\mu_n,
\hspace{11ex}\to R_7 \equiv \left| \frac{\mu_{\Sigma^0}+\frac{1}{2}\mu_n}{-\frac{1}{2}\mu_n} 
\right|,
\label{CG5}\\
|\mu_{\Sigma^0 \Lambda}| &=& |\frac{1}{2}\sqrt{3}\mu_n|, 
\hspace{10ex}\to R_8 \equiv \left| \frac{|\mu_{\Sigma^0 \Lambda}| 
-|\frac{1}{2}\sqrt{3}\mu_n|}{\frac{1}{2}\sqrt{3}\mu_n} \right|.
\label{CG6}
\eea
In the above $R_i (i=1,...,8)$ will be evaluated by the experimental data (Expt.), Set I, and 
Set II results, where we have already analyzed the relation, 
$\mu_{\Sigma^0}({\rm Expt.}) = \frac{1}{2} \left[ \mu_{\Sigma^+}({\rm Expt.})
+\mu_{\Sigma^-}({\rm Expt.}) \right]=0.649$ by Eq.~(\ref{musig0}), that is independent of the 
unitary symmetry scheme, and we will use this value as the $\mu_{\Sigma^0}{\rm (Expt.)}$ in this 
analysis. 
We also take into account the unknown sign in the experiment by $|\mu_{\Sigma^0\Lambda}|$.
The obtained ratios $R_i = (i=1,...,8)$ for the Set I, Set II and Expt. are given in 
Table~\ref{CGratios}. In addition, we give the ratios, $\mu_B/\mu_B({\rm Expt.})$, for 
convenience.

\begin{table}[htb]
\begin{center}
\caption{
The ratios $R_i (i=1,...,8)$ Eqs.~(\ref{CG1})~-~(\ref{CG6}) calculated by the Set I, Set II, 
and experimental data (Expt.), corresponding to the Coleman-Glashow 
relations~\cite{Coleman:1961jn}, and the ratios by experimental data, $\mu_B/\mu_B({\rm Expt.})$, 
where  $\mu_{\Sigma^0}({\rm Expt.}) = \frac{1}{2} \left[ \mu_{\Sigma^+}({\rm Expt.})
+\mu_{\Sigma^-}({\rm Expt.}) \right]=0.649$ is used.
}
\label{CGratios}
\vspace{1ex}
\begin{tabular}{l|c|c|c}
\hline
\hline
Ratio     &Set I  &Set II  &Expt.\\
\hline
$R_1$     &0.014  &0.061 &0.120\\
$R_2$     &0.163  &0.022 &0.359\\
$R_3$     &0.092  &0.044 &0.347\\
$R_4$     &0.094  &0.092 &0.318\\
$R_5$     &0.209  &0.006 &0.207\\
$R_6$     &0.277  &0.089 &0.439\\
$R_7$     &0.024  &0.046 &0.322\\
$R_8$     &0.020  &0.017 &0.028\\
\hline
\hline
$\mu_p/\mu_p({\rm Expt.})$                     &0.550 &0.550 &1\\
$\mu_n/\mu_n({\rm Expt.})$                     &0.535 &0.535 &1\\
$\mu_\Lambda/\mu_\Lambda({\rm Expt.})$         &0.700 &0.816 &1\\
$\mu_{\Sigma^+}/\mu_{\Sigma^+}({\rm Expt.})$   &0.633 &0.662 &1\\
$\mu_{\Sigma^0}/\mu_{\Sigma^0}({\rm Expt.})$   &0.769 &0.828 &1\\
$\mu_{\Sigma^-}/\mu_{\Sigma^-}({\rm Expt.})$   &0.483 &0.481 &1\\
$\mu_{\Xi^0}/\mu_{\Xi^0}({\rm Expt.})$         &0.743 &0.854 &1\\
$\mu_{\Xi^-}/\mu_{\Xi^-}({\rm Expt.})$         &0.622 &0.782 &1\\
\hline
\hline
\end{tabular}
\end{center}
\end{table}

The results for $R_i (i=1,...,8)$ given in Table~\ref{CGratios} show that 
the Coleman-Glashow relations~\cite{Coleman:1961jn} 
are broken up to about 45\% in data, and up to about 30\% (70\%) in 
the Set I (Set II) results for the ratios $R_i (i=1,...,8)$. 
The Coleman-Glashow relations are not well realized in nature, 
as well as in the MIT bag model results obtained in the Set I and Set II. 
In particular, the values for $R_8$ in Eq.~(\ref{CG6}) associated with 
$\mu_{\Sigma^0\Lambda}$, are obtained respectively as 
(Set I, Set II, Expt.) = (0.020, 0.017, 0.028), and the experimental data give 
the larger deviation from the Coleman-Glashow relation.   
Thus, the ambiguity due to the possible MIT bag model artifact (the bag radii difference etc.) 
are not expected to affect significantly the conclusions   
for the transition magnetic moments in medium.

The estimated medium modifications of the magnetic moments and 
the transition magnetic moments at nuclear matter saturation density ($\rho_0$) 
may not directly be reflected on some experimental results.
In the future experiment, although it is expected to be very difficult, 
if the charmed/bottom hypernuclei can be formed for example, 
it might not be impossible to measure the magnetic 
moments of the bound charmed/bottom baryons for those treated in this study.
For example, similarly to the proton knockout reaction used for extracting the bound proton 
electromagnetic form factor double ratios by the polarization transfer measurement,  
it might be possible to extract the charged charm/bottom baryon 
electromagnetic form factor double ratios at very small momentum transfer,  
where the charge form factor at zero momentum transfer gives the charge, 
and the information on the in-medium magnetic moment can be extracted. 
But such possibilities are, at present, very remote and speculations. 

However, the modifications in the higher density region near $3 \rho_0$, 
may affect some studies for heavy ion collisions, 
structure and reactions occur in the inner cores of magnetars, 
neutron stars, and compact stars. 
In the cores of such high density compact objects, in particular very high density compact stars,   
one may expect the appearance of charm and bottom baryons, although no charm quark-matter star is 
expected or stable~\cite{Kettner:1994zs,Glendenning,Jimenez:2019kji}, using the flat space-time 
equation of states (EOS). 
However, it was discussed that the use of 
EOS computed in the curved space-time of neutron stars, may yield 
the higher central energy densities and masses---about 16.9\% for an idealistic neutron 
star case---than that calculated using the flat-space EOS~\cite{Hossain:2020mvn}.     
The authors state that the result favors to resolve the ''hyperon puzzle`` of neutron star.
We would like to comment that, within the QMC model treatment, 
a possibility of resolving the ''hyperon puzzle'' due to the 
quark structure of nucleons and hyperons, has been 
reported in Refs.~\cite{RikovskaStone:2006ta,Katayama:2012ge,Miyatsu:2011bc,
Whittenbury:2013wma,Thomas:2013sea}.
It might be interesting to study the possibility of charm quark-matter star  
as well as very high density compact stars using the curved space-time EOS  
including the charm and bottom baryons to study whether or not such heavy baryons can 
indeed influence the structure.
In addition, dominant charm hadron contributions to neutrino 
fluence in ultrahigh-energy neutrino production by newborn magnetars 
are suggested~\cite{Carpio:2020wzg}.

\section{Summary and conclusion}
\label{summary}

We have studied the medium modifications of magnetic moments and transition magnetic 
moments of the octet, decuplet, low-lying charm, and low-lying bottom baryons with nonzero 
light quarks in symmetric nuclear matter  
using the quark-meson coupling model, which satisfies the constraint  
for the allowed maximum change (swelling) of the nucleon size in medium, 
derived from the $y$-scaling data analysis---thus satisfies the allowed maximum 
enhancement of the nucleon magnetic moment at nuclear matter saturation density.
The model has been extended to treat the decuplet baryons. 
This is the first study to estimate the in-medium  
magnetic moments and transition magnetic moments of the low-lying charm and 
low-lying bottom baryons with nonzero light quarks. 
In the estimates we have assumed the ''1-2 quark order'' for the charm and bottom  
baryon flavor-spin wave functions as is practiced for the octet baryon SU(6) wave functions, 
namely, the quarks 1 and 2 to be the closest in mass.
The issue of different quark order is concerned for the $\Xi_{c,b}$ baryons in the present study.
The ''1-2 quark order'' is supported  
as the best quark ordering for flavor-degenerate baryons for the masses.

In addition we have estimated using two sets of the current quark mass values, 
Set I (Table~\ref{coupcc1}) and Set II (Table~\ref{coupcc2}).
The obtained feature for the medium modifications is generally similar for the two sets.

The estimates of the medium modification are made by calculating 
the in-medium to free space baryon magnetic 
moment ratios, to compensate the known MIT bag deficiency in obtaining   
the magnitude of the free space octet baryon magnetic moments.
In connection with this, we have given examples that many experiments directly measure 
ratios to extract meaningful physics quantities, such as electromagnetic form factors 
of the free and bound proton---experimental data extracted at Jefferson Laboratory support the 
in-medium to free ratios predicted by the quark-meson coupling model, 
and the European Muon Collaboration effect.
The observed maximum modifications       
for the (octet, decuplet, charm, bottom) baryon sectors 
are, respectively for the $(\Sigma^-,\Xi^{*0},\Sigma_c^+,\Sigma_b^0)$ baryons, 
and the corresponding modifications 
are about $(12,158,17,12)\%$ [$(11,257,17,12)\%$]
at normal nuclear matter density ($\rho_0=0.15$ fm$^{-3}$), 
and about $(23,215,32,24)\%$ [$(25,414,32,23)\%$] at $3 \rho_0$ 
for Set I [Set II], where the large enhancement of the $\Xi^{*0}$ 
is due to the very small magnitude in free space in Set I [Set II].

As for the medium modifications of the transition magnetic moments $\mu_{BB'}$ with   
$B=(\Sigma^0,\Sigma_c^+,\Sigma_b^0) \to B'=(\Lambda,\Lambda_c^+,\Lambda_b^0)$,  
the modifications at $\rho_0$ are about $(9,9,10)\%$ [$(9,9,10)\%$], 
while at $3\rho_0$, they are about $(15,15,17)\%$ [$(16,15,17)\%$] 
for Set I [Set II].
Concerning the possible ambiguities arising from the MIT bag model artifact,  
we have discussed the difference of the initial and final baryon bag radii 
in free space as well as in symmetric nuclear matter, based on the evidence observed for 
the violation of the Ademollo-Gato theorem in the weak-interaction vector charge calculation. 
Furthermore, we have studied the possible impact of the ambiguities 
on the estimated results based on the SU(3) (U(3)) symmetry relations 
as well as the Coleman-Glashow relations using 
the calculated results and the experimental data.
It turned out that such ambiguities are not expected to affect 
the estimated ratios of this study, within the present status of 
the experimental precision.

The medium modifications of baryon magnetic moments, 
that have been estimated by the ratios in this study,     
may serve as new realistic inputs for the studies of magnetar and neutron 
star structure, in particular, for the systems with high baryon 
density with nearly zero temperature and extremely strong magnetic field.

For a practical use, we have given the explicit density dependent parametrizations 
for the vector potentials of the baryons 
and light-$(u, d)$ quarks, as well as for the effective masses 
of the low-lying baryons treated in this study, 
and of the mesons, $\omega,\rho,K,K^*,\eta,\eta',D,D^*,B$, and $B^*$. 
The parametrizations given in this study may open wider applications and studies 
for the properties of the corresponding baryons and mesons in medium.

As extensions, we plan to include the meson cloud effects in medium, and also study the 
medium modifications of the weak-interaction axial charges of the octet, decuplet, 
low-lying charm, and low-lying bottom baryons with nonzero light quarks.

\vspace{2ex}
\noindent
{\bf Acknowledgments}\\
The author was supported by the Conselho Nacional de Desenvolvimento 
Cient\'{i}fico e Tecnol\'{o}gico (CNPq) 
Process, No.~313063/2018-4, and No.~426150/2018-0, 
and Funda\c{c}\~{a}o de Amparo \`{a} Pesquisa do Estado 
de S\~{a}o Paulo (FAPESP) Process, No.~2019/00763-0, 
and was also part of the projects, Instituto Nacional de Ci\^{e}ncia e 
Tecnologia -- Nuclear Physics and Applications (INCT-FNA), Brazil, 
Process. No.~464898/2014-5, and FAPESP Tem\'{a}tico, Brazil, Process, 
No.~2017/05660-0.



\end{document}